\documentclass[aps,prd,twocolumn,preprintnumbers]{revtex4}

\usepackage{graphicx}
\usepackage{dcolumn}
\usepackage{bm}
\usepackage{amsmath}
\usepackage{amssymb}
\usepackage{amsfonts}
\usepackage{float}
\usepackage{hyperref}[11in]
\usepackage{dsfont}
\usepackage{slashed}
\usepackage{booktabs}
\usepackage{multirow}
\usepackage{subfigure}
\usepackage[sort&compress]{natbib}
\usepackage{xcolor}
\usepackage{color}
\usepackage{colordvi}
\usepackage{wasysym}
\usepackage{array,multirow}
\usepackage{booktabs}
\usepackage{bbold}

\usepackage{graphicx}
\usepackage{xfrac}

\newcommand{\be}{\begin{equation}}
\newcommand{\ee}{\end{equation}}
\newcommand{\beq}{\begin{eqnarray}}
\newcommand{\eeq}{\end{eqnarray}}

\newcommand{\tr}{\mathrm{Tr}}

\newcommand{\Op}{\mathcal{O}}

\newcommand{\eins}{\mathds{1}}

\begin{document}
\title{Nucleon axial and pseudoscalar form factors from lattice QCD at the physical point}

\author{
  C.~Alexandrou$^{1,2}$,
  S.~Bacchio$^{2}$,
  M.~Constantinou$^{3}$,
P.~Dimopoulos$^{4}$
  J. Finkenrath$^{2}$
  K.~Hadjiyiannakou$^{1,2}$,
  K.~Jansen$^{5}$,
  G.~Koutsou$^{2}$,
  B. Kostrzewa$^6$,
  T.~Leontiou$^7$,
  C. Urbach$^8$
}
\affiliation{
  $^1$Department of Physics, University of Cyprus, P.O. Box 20537, 1678 Nicosia, Cyprus\\
  $^2$Computation-based Science and Technology Research Center,~The Cyprus Institute, 20 Kavafi Str., Nicosia 2121, Cyprus \\
  $^3$Department of Physics, Temple University, 1925 N. 12th Street, Philadelphia, PA 19122-1801,
  USA\\
 $^4$ Dipartimento di Scienze Matematiche, Fisiche e Informatiche,~Universit\`a di Parma and INFN, \\Gruppo Collegato di Parma, Parco Area delle Scienze 7/a (Campus), 43124 Parma, Italy\\
$^5$NIC, DESY, Platanenallee 6, D-15738 Zeuthen, Germany\\
$^6$ High Performance Computing and Analytics Lab, University of Bonn, Endenicher Allee 19A, 53115 Bonn, Germany\\
$^7$Department of Mechanical Engineering, Frederick University, 1036 Nicosia, Cyprus\\
  $^8$Helmholtz-Institut f\"ur Strahlen- und Kernphysik, University of Bonn, 53115 Bonn, Germany and\\  Bethe Center for Theoretical Physics, University of Bonn, 53115 Bonn, Germany\
}

\begin{abstract}
  \centerline{\includegraphics[width=0.15\linewidth]{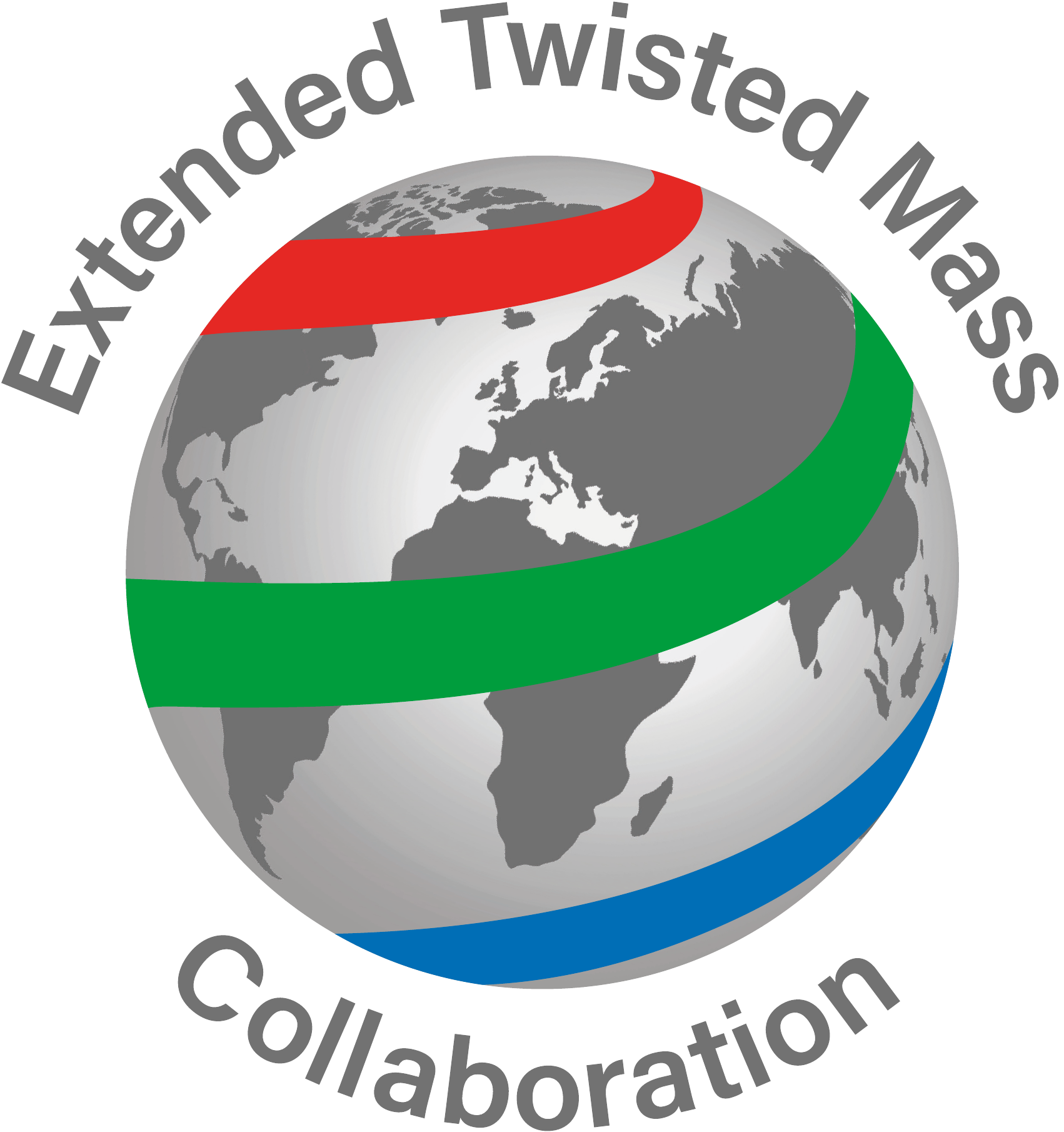}}
  \vspace*{0.3cm}
We compute the nucleon axial and induced pseudoscalar form factors using three ensembles of gauge configurations, generated with dynamical light quarks with mass tuned to approximately their physical value. One of the ensembles also includes the strange and charm quarks with their mass close to physical. The latter ensemble has large statistics and finer lattice spacing and it is used to obtain final results, while the other two  are used for assessing volume effects. The pseudoscalar form factor is also computed using these ensembles. We examine the momentum dependence of these form factors as well as relations based on pion pole dominance and the partially conserved axial-vector current hypothesis.  
\end{abstract}

\maketitle
\bibliographystyle{apsrev}

\section{Introduction}
A central aim of on-going  experimental
and theoretical studies is the
understanding of  the structure of the proton and the neutron arising from the complex nature of the  strong interactions. The electron scattering off protons is a well developed experimental approach  used in such studies. An outcome of the multi-years experimental programs in major facilities has been the precise measurement of  the electromagnetic  form factors, see e.g.~\cite{Xiong:2019umf,Yan:2018bez,Akushevich:2015toa,Smorra:2016vxa,Ablikim:2019eau,Ye:2017gyb,Haidenbauer:2014kja,Seth:2012nn}. However, despite many years of experimental effort, new features are being revealed by performing new more precise experiments as, for example, the measurement of the proton charge radius~\cite{Pohl:2010zza,Pohl:2014bpa,Kolachevsky:2018few}. Experimental efforts are accompanied by theoretical computations of such quantities 
~\cite{Alexandrou:2020aja,Alarcon:2020kcz,Hammer:2019uab,Trinhammer:2019aey,Xiong:2019umf,Bezginov:2019mdi}. However, the theoretical extraction of such form factors is difficult due to their non-perturbative nature. The  lattice formulation of Quantum Chromodynamics (QCD) provides the non-perturbative framework for computing non-perturbative quantities from first principles.   Lattice QCD computations using simulations at physical parameters of the theory of Electromagnetic form factors is a major recent achievement~\cite{Alexandrou:2018sjm,Jang:2019jkn,Alexandrou:2017ypw,Shintani:2018ozy,Ishikawa:2018rew}.

While the electromagnetic form factors are well measured and are being used to benchmark theoretical approaches, the nucleon axial form factors are less well  known. The axial form factors are important quantities for
weak interactions, neutrino scattering  and parity violation experiments. Neutrinos can interact with nucleons via the neutral current of weak interactions, exchanging a $Z^0$ boson or via the charged current of weak interactions exchanging a $W^\pm$ boson. The nucleon matrix element of the  isovector axial-vector  current $A_\mu $  is written in terms of two form factors,  the axial, $G_A(Q^2)$, and the induced pseudoscalar  $G_P(Q^2)$. The axial form factor, $G_A(Q^2)$, is experimentally determined 
from elastic scattering of neutrinos with protons, $\nu_\mu + p \rightarrow \mu^+ + n$~\cite{Ahrens:1988rr,Meyer:2016oeg,Bodek:2007vi}, while   $G_P(Q^2)$ from the longitudinal cross section in pion electro-production~\cite{Choi:1993vt,Bernard:1994pk,Fuchs:2003vw}. At zero momentum transfer the axial form factor gives the axial charge $g_A \equiv G_A(0)$, which is measured in high precision from $\beta$-decay experiments~\cite{Brown:2017mhw,Darius:2017arh,Mendenhall:2012tz,Mund:2012fq}. The induced pseudoscalar coupling $g_P^*$  can be determined via
the ordinary muon capture process $\mu^- + p \rightarrow n + \nu_\mu$ from the singlet state of the muonic hydrogen atom at the muon capture point, which corresponds to momentum transfer squared of $Q^2=0.88 m_\mu^2$~\cite{Castro:1977ep,Bernard:1998rs,Bernard:2000et,Andreev:2012fj,Andreev:2007wg}, where $m_\mu$ is the muon mass.

Besides experimental extractions,  phenomenological approaches are  being applied to study the axial form factors. Chiral perturbation theory provides a non-perturbative framework suitable for low  values of  $Q^2$ up to about $0.4$~GeV$^2$~\cite{Schindler:2006jq,Schindler:2006it,Fuchs:2003vw}.  Other  models used include  the perturbative chiral quark model~\cite{Khosonthongkee:2004qm}, the chiral
constituent quark model~\cite{Glozman:2001zc} and light-cone sum rules~\cite{Anikin:2016teg}.

As already mentioned, lattice QCD  provides the {\it ab initio} non-perturbative framework  for computing  such quantities  using directly the QCD Lagrangian. Early studies of the nucleon axial form factors  were done within the quenched approximation~\cite{Liu:1991nk,Liu:1992ab}, as well as, using  dynamical fermion simulations at heavier than
 physical pion masses~\cite{Alexandrou:2009vqd}. Only recently,  several groups are  computing the axial form factors using simulations generated directly at the physical value of the pion mass~\cite{Alexandrou:2017hac,Jang:2019vkm,Rajan:2017lxk,Bali:2018qus,Bali:2019yiy,Shintani:2018ozy,Ishikawa:2018rew}. 
 Such simulations at the physical pion mass can check  important phenomenological relations, such  as the partially conserved axial-vector current (PCAC) relation that at form factor level connects $G_A(Q^2)$ and $G_P(Q^2)$  with the pseudoscalar $G_5(Q^2)$ form factor. 
 At low $Q^2$ and assuming pion pole dominance (PPD) one can further relate $G_A(Q^2)$ to  $G_P(Q^2)$ and derive the Goldberger-Treiman relation. These relations have been studied  within lattice QCD and will be  discussed in this paper.
 The computation of the form factors is performed using one ensemble of  mass degenerate up and down quarks, and a strange and a charm quark  ($N_f=2+1+1$)  with masses tuned to their physical values, referred to as physical point.  In addition, we present results for two ensembles of   $N_f=2$ light quarks tuned to the physical pion mass. They have the same lattice spacing $a$ but different volumes in order to check for finite size effects. Final results are given for the $N_f=2+1+1$ ensemble where high statistics are used and systematic errors due to excited states are better controlled. 

 The remainder of this paper is organized as follows: In Section~\ref{sec:AP_ME} we discuss the PCAC and  PPD  relations and  in Sec.~\ref{sec:Q2Fit} the parameterization of the $Q^2$ dependence. In Sec.~\ref{sec:LatM}, we explain in detail the lattice
 methodology  to extract the axial and pseudoscalar form factors. The renormalization of the operators is discussed in Sec.~\ref{sec:Renormalization}. In Sec.\ref{sec:A0fit}, we explain how we extract the energy of the excited state and in Secs.~\ref{sec:G5_Effs} and \ref{sec:GAGP_Effs} we show results for the nucleon state matrix elements of the axial-vector and pseudoscalar currents. We compare our results of the three ensembles in Sec.~\ref{sec:AllEnsFFs} and present the final results in Sec.~\ref{sec:FinalRes}. A comparison with other studies is undertaken in Sec.~\ref{sec:CompOther}. Finally, we conclude in Sec.~\ref{sec:Concl}.

\section{Decomposition of the  nucleon axial-vector and pseudoscalar matrix elements into the form factors and their relations}\label{sec:AP_ME}
In this work we will consider the isovector axial-vector operator given by
\begin{equation}
    A_\mu = \bar{u} \gamma_\mu \gamma_5  u -\bar{d}\gamma_\mu \gamma_5  d
    \label{Eq:AVcurrent}
\end{equation}
where $u$ and $d$ is the isospin double of  the up and down quark fields. In the chiral limit, where the pion mass $m_\pi = 0$, the axial-vector current is conserved, namely $\partial^\mu A_\mu = 0$. 
For a non-zero pion mass the spontaneous breaking of chiral symmetry relates the axial-vector current to the pion field $\psi_\pi$, through the relation
\be
\partial^\mu A_\mu = F_\pi m_\pi^2 \psi_\pi.
\ee
We use the convention $F_\pi = 92$~MeV 
for the pion decay constant.   the In QCD  the axial Ward-Takahashi 
identity leads to the  partial conservation of the axial-vector current (PCAC) 
\begin{equation}
  \partial^\mu A_\mu= 2 m_q P,
  \label{Eq:PCAC}
\end{equation}
where $m_q=m_u=m_d$ is the light quark mass for degenerate  up and down quarks. Using the PCAC relation it then follows that  the pion field can be expressed as
\begin{equation}
    \psi_\pi = \frac{2 m_q P}{F_\pi m_\pi^2}.
    \label{Eq:piToP}
\end{equation}

The nucleon  matrix element of the  the axial-vector current
of Eq.~(\ref{Eq:AVcurrent}) can be written in terms of the axial, $G_A(Q^2)$, and induced pseudoscalar, $G_P(Q^2)$, form factors as
\begin{eqnarray}
    &&\langle N(p',s') \vert A_\mu\vert N(p,s) \rangle = \bar{u}_N(p',s')  \nonumber \\
    && \bigg[\gamma_\mu G_A(Q^2) - \frac{Q_\mu}{2 m_N} G_P(Q^2)\bigg] \gamma_5 u_N(p,s),
    \label{Eq:DecompA}
\end{eqnarray}
where $u_N$ is the nucleon spinor with initial (final) momentum $p(p')$ and spin $s(s')$, $q=p'-p$ the momentum transfer and $q^2=-Q^2$. The nucleon pseudoscalar matrix element is given by
\begin{equation}
    \langle N(p',s') \vert P_5 \vert N(p,s) \rangle =  G_5(Q^2) \bar{u}_N(p',s')\gamma_5 u_N(p,s).
\end{equation}
where $P_5=\bar{u} \gamma_5  u - \bar{d}  \gamma_5 d$ is the isovector pseudoscalar current.
 The PCAC relation at the form factors level relates the axial and induced pseudoscalar form factors to the pseudoscalar form factor via the relation
\begin{equation}
    G_A(Q^2) - \frac{Q^2}{4 m_N^2} G_P(Q^2) = \frac{m_q}{m_N} G_5(Q^2),
    \label{Eq:PCAC_FFs}
\end{equation}
Making use of Eq.~(\ref{Eq:piToP}) one can connect the pseudoscalar form factor to the pion-nucleon form factor $G_{\pi NN}(Q^2)$ as follows
\begin{equation}
    G_5(Q^2) = \frac{F_\pi m_\pi^2}{m_q} \frac{G_{\pi NN}(Q^2)}{m_\pi^2 + Q^2}.
    \label{Eq:PtoPiNN}
\end{equation}
Eq.~(\ref{Eq:PtoPiNN}) is written so that it illustrates the pole structure of $G_5(Q^2)$.  Substituting $G_5(Q^2)$ in Eq.~(\ref{Eq:PCAC_FFs}),  one obtains the Goldberger-Treiman relation~\cite{Alexandrou:2007hr,Alexandrou:2009vqd}
\be
G_A(Q^2)-\frac{Q^2}{4m_N^2} G_P(Q^2)=\frac{1}{m_N}\frac{G_{\pi NN}(Q^2)F_\pi m_\pi^2}{m^2_\pi+Q^2}.
  \label{GT}
  \ee
  The pion-nucleon form factor $G_{\pi NN}(Q^2)$ at the pion pole gives the pion-nucleon coupling $g_{\pi NN} \equiv G_{\pi NN}(Q^2=-m_\pi^2)$.
In the limit $Q^2 \rightarrow -m_\pi^2$, the pole on the right hand side of Eq.~(\ref{GT}) must be compensated by a similar
one in $G_P(Q^2)$, since $G_A(-m_\pi^2)$ is finite. Therefore, if we multiply Eq.~(\ref{GT}) by $(Q^2+m_\pi^2)$ and take the limit towards the pion pole we have
\begin{equation}
    \lim_{Q^2 \rightarrow -m_\pi^2} (Q^2+m_\pi^2) G_P(Q^2) = 4 m_N F_\pi g_{\pi NN}
    \label{Eq:gpiNN}
\end{equation}
and, thus, one can extract $g_{\pi NN}$ from the induced pseudoscalar form factor too. Close to the pole,  pion pole dominance means that $G_P(Q^2)=4m_N F_\pi G_{\pi NN}(Q^2)/(m^2_\pi+Q^2)$. Inserting it in Eq.~(\ref{GT}) we obtain
the well known relation~\cite{Goldberger:1958vp}
\be
m_N G_A(Q^2)=F_\pi G_{\pi NN}(Q^2),
   \label{Eq:GTRQ2}
\ee
 which means that $G_P(Q^2)$ can be expressed as~\cite{Scadron:1991ep}
\begin{equation}
    G_P(Q^2) = \frac{4 m_N^2}{Q^2+m_\pi^2} G_A(Q^2).
    \label{Eq:PPD}
\end{equation}
From Eq.~(\ref{Eq:GTRQ2}), the pion-nucleon coupling can be expressed as $g_{\pi NN} = m_N G_A(-m_\pi^2)/ F_\pi$. In the chiral limit, $\displaystyle \lim_{m_\pi\rightarrow 0} G_A(-m_\pi^2) \rightarrow g_A$ and we have that
\begin{equation}
    g_{\pi NN} = \frac{m_N}{F_\pi} g_A.
    \label{Eq:GTR}
\end{equation}
The deviation from Eq.~(\ref{Eq:GTR}) due to the finite pion mass is known
as the Goldberger-Treiman discrepancy, namely
\begin{equation}
    \Delta_{GT} = 1 - \frac{g_A m_N}{g_{\pi NN} F_\pi}
    \label{Eq:GTD}
\end{equation}
and it is estimated to be at the 2\% level~\cite{Nagy:2004tp}.

\section{$Q^2$-dependence of the axial and pseudoscalar form factors} \label{sec:Q2Fit}
For the parameterization of the $Q^2$-dependence of the axial and pseudoscalar form factors typically two functional forms are employed, the  dipole Ansatz and the model
independent z-expansion~\cite{Hill:2010yb,Bhattacharya:2011ah}.

The dipole Ansatz is given by
\begin{equation}
    G(Q^2) = \frac{G(0)}{(1+\frac{Q^2}{m^2})^2},
    \label{Eq:dipole}
\end{equation}
with $m$ the dipole mass. In the case of the axial form factor $G_A(Q^2)$, its
value for $Q^2=0$, gives the axial charge $g_A \equiv G_A(0)$ and the dipole mass $m$ is the axial mass $m_A$.

Customarily, one characterizes the size of a hadron probed by a given current  by the  root mean square radius (r.m.s) defined as $\sqrt{\langle r^2 \rangle}$. The radius of the form factors can be extracted 
from their slope  as $Q^2\rightarrow 0$, namely
\begin{equation}
    \langle r^2 \rangle = - \frac{6}{ G(0)} \frac{d G(Q^2)}{dQ^2} \bigg \vert_{Q^2 \rightarrow 0}.
    \label{Eq:radius}
\end{equation}
Combining Eq.~(\ref{Eq:dipole}) and Eq.~(\ref{Eq:radius}) one can show that the radius
is connected to the dipole mass as
\begin{equation}
    \langle r^2 \rangle = \frac{12}{m^2}.
    \label{Eq:rToM}
\end{equation}
 In the case of the z-expansion the form factor is expanded as,
\begin{equation}
    G(Q^2) = \sum_{k=0}^{k_{\rm max}} a_k\; z^k(Q^2), 
    \label{Eq:zExp}
\end{equation}
where 
\begin{equation}
    z(Q^2) = \frac{\sqrt{t_{\rm cut} + Q^2} - \sqrt{t_{\rm cut}} }{ \sqrt{t_{\rm cut} + Q^2} + \sqrt{t_{\rm cut}} }
\end{equation}
imposing analyticity constrains, with $t_{\rm cut}$ the particle production threshold.  For  $t_{\rm cut}$, we use the three-pion production threshold, namely   $t_{\rm cut}=\left(3 m_\pi\right)^2$~\cite{Bhattacharya:2011ah}. The coefficients $a_k$ should be bounded in size for the series to converge and convergence is demonstrated by increasing $k_{\rm max}$. Since the possible large values of the $a_k$ for $k>1$ can lead to instabilities, we use Gaussian priors centered around zero with standard deviation $w\max(|a_0|,|a_1|)~$\cite{Green:2017keo}, where $w$ controls the width of the prior. The value of the form factor at zero momentum is $G(0)=a_0$, while the radius is given by
\begin{equation}
    \langle r^2 \rangle = - \frac{3 a_1}{2 a_0 t_{\rm cut}}.
    \label{Eq:r2_zExp}
\end{equation}
In the case of the axial form factor, $a_0$ and $a_1$ should have opposite signs leading to positive radii.
By comparing Eq.~(\ref{Eq:r2_zExp}) to  Eq.~(\ref{Eq:rToM}),  we define the corresponding mass determined in the z-expansion to be
  \be
  m=\sqrt{-\frac{8a_0t_{\rm cut}}{a_1}}.
  \label{Eq:m_zExp}
  \ee
In the case of $G_P(Q^2)$ and $G_5(Q^2)$, the pion pole is first factored out and thus $(Q^2 + m_\pi^2)\; G_{P,5}(Q^2)$ could be fitted using the dipole and z-expansion functions.

\section{Lattice methodology}\label{sec:LatM}
In this section we describe the lattice QCD methodology  to extract the form factors, presenting  the construction of the appropriate  three- and two-point correlation functions, the procedure to
isolate the ground state and the details about the ensembles used.

\subsection{Correlation functions} \label{ssec:CorrFuncs}
The extraction of the nucleon matrix elements involves the computation of both three- and two-point Euclidean correlation functions. The two-point function is given by

\begin{eqnarray}
C(\Gamma_0,\vec{p};t_s,t_0) &&{=}  \sum_{\vec{x}_s} \hspace{-0.1cm} e^{{-}i (\vec{x}_s{-}\vec{x}_0) \cdot \vec{p}} \times \nonumber \\
&&\tr \left[ \Gamma_0 {\langle}{\cal J}_N(t_s,\vec{x}_s) \bar{\cal J}_N(t_0,\vec{x}_0) {\rangle} \right],
\label{Eq:2pf}
\end{eqnarray}
where with $x_0$ is the source and $x_s$ the sink positions on the lattice where states with the quantum numbers of the
nucleon are created and destroyed, respectively. The interpolating field is
\begin{equation}
    {\cal J}_N(t,\vec{x}) = \epsilon^{abc} u^a(x) \left[ u^{b T}(x) \mathcal{C} \gamma_5 d^{c} (x) \right],
    \label{Eq:IntF}
\end{equation}
where $\mathcal{C}$ is the charge conjugation matrix and $\Gamma_0$ is the unpolarized positive parity projector $\Gamma_0 = \frac{1}{2}(1+\gamma_0)$. By inserting the unity operator in Eq.~(\ref{Eq:2pf}) in the form of a sum over states of the QCD Hamiltonian only states with the quantum numbers of the nucleon survive. The overlap terms  between the interpolating field and the nucleon state $\vert N_j \rangle$ as $\langle \Omega \vert {\cal J}_N \vert N_j \rangle$ are terms that need to be canceled to access the matrix element. It is desirable to increase the overlap with the  nucleon state  and reduce it with  excited states so that the ground state dominates for as small as possible Euclidean time separations. This is because the signal-to-noise ratio
decays exponentially with the Euclidean time evolution. To accomplish ground state dominance, we apply Gaussian smearing~\cite{Alexandrou:1992ti,Gusken:1989qx} to the quark fields entering the interpolating field 
\begin{equation}
    \tilde{q}(\vec{x}, t) = \sum_{\vec{y}} [\mathbb{1}  + a_G H(\vec{x}, \vec{y}; U(t))]^{N_G} q(\vec{y},t),
\end{equation}
where the hopping matrix is given by
\begin{eqnarray}
H(\vec{x},\vec{y};U(t)) = \sum_{i=1}^3 \left[ U_i(x) \delta_{x,y-\hat{i}} + U_i^\dag(x-\hat{i}) \delta_{x,y+\hat{i}}  \right].
\end{eqnarray}
The parameters $a_G$ and $N_G$ are tuned~\cite{Alexandrou:2018sjm,Alexandrou:2019ali} in order to approximately give a smearing radius for the nucleon of $ 0.5$~fm. For the links entering the hopping matrix we apply APE smearing~\cite{Albanese:1987ds} to reduce statistical errors due to  ultraviolet fluctuations.

For the construction of the three-point correlation function the  current  is inserted  between the time of the  creation and annihilation operators giving 
\begin{eqnarray}
 && C_{\mu}(\Gamma_k,\vec{q},\vec{p}\,';t_s,t_{\rm ins},t_0) {=}
 \hspace{-0.1cm} {\sum_{\vec{x}_{\rm ins},\vec{x}_s}} \hspace{-0.1cm} e^{i (\vec{x}_{\rm ins} {-} \vec{x}_0)  \cdot\vec{q}}  e^{-i(\vec{x}_s {-} \vec{x}_0)\cdot \vec{p}\,'} {\times} \nonumber \\
  && \hspace{1.cm} \tr \left[ \Gamma_k \langle {\cal J}_N(t_s,\vec{x}_s) A_{\mu}(t_{\rm ins},\vec{x}_{\rm ins}) \bar{\cal J}_N(t_0,\vec{x}_0) \rangle \right],
  \label{Eq:3pf}
\end{eqnarray}
where $\Gamma_k = i \Gamma_0 \gamma_5 \gamma_k$. The Euclidean momentum trasfer squared is given by $Q^2 = - q^2 = - (p' -p)^2$, and from now on we will use $\vec{p}\,'=\vec{0}$.

\subsection{Treatment of excited states contamination}\label{ssec:ExcStates}
The interpolating field in Eq.~(\ref{Eq:IntF}) creates a tower of states
with the quantum numbers of the nucleon.  Gaussian smearing helps to reduce them but we  still need to make sure that we extract the
nucleon matrix element that we are interested in and that any contribution from nucleon excited states  and/or multi-particle states are sufficiently suppressed.

In order to cancel the Euclidean time dependence of the three-point function and unknown overlaps
of the interpolating field with the  nucleon state, we construct an appropriate ratio
of three- to a combination of two-point functions~\cite{Alexandrou:2013joa,Alexandrou:2011db,Alexandrou:2006ru,Hagler:2003jd},
\begin{eqnarray}
&&  R_{\mu}(\Gamma_{k},\vec{q};t_s,t_{\rm ins}) = \frac{C_{\mu}(\Gamma_k,\vec{q};t_s,t_{\rm ins}\
)}{C(\Gamma_0,\vec{0};t_s)} \times \nonumber \\
&&  \sqrt{\frac{C(\Gamma_0,\vec{q};t_s-t_{\rm ins}) C(\Gamma_0,\vec{0};t_{\rm ins}) C(\Gamma_0,\vec{0};t_s)}{C\
(\Gamma_0,\vec{0};t_s-t_{\rm ins}) C(\Gamma_0,\vec{q};t_{\rm ins}) C(\Gamma_0,\vec{q};t_s)}}.
\label{Eq:ratio}
\end{eqnarray}
Without loss of generality, we take $t_s$ and $t_{\rm ins}$ relative to the source time $t_0$, or equivalently $t_0$ is set to zero. The ratio in Eq.~(\ref{Eq:ratio}) is constructed such that in the limit of large time separations $(t_s-t_{\rm ins}) \gg a$ and $t_{\rm ins} \gg a$, it converges to the nucleon ground state matrix element, namely
\begin{equation}
  R_{\mu}(\Gamma_k;\vec{q};t_s;t_{\rm ins})\xrightarrow[t_{\rm ins}\gg a]{t_s-t_{\rm ins}\gg a}\Pi_{\mu}(\Gamma_k;\vec{q})\,.
\end{equation} 
How fast we ensure ground state dominance depends on the smearing procedure applied on the interpolating fields, as well as on  the type of current entering the three-point function. In order to check for ground state dominance we employ three methods as summarized below:\\
\noindent\emph{Plateau method:} Keeping only the ground state  in  the correlation functions entering in Eq.~(\ref{Eq:ratio}) we obtain
    \begin{equation}
    \Pi_{\mu}(\Gamma_k;\vec{q}) + {\cal O}(e^{-\Delta E(t_s - t_{\rm ins})}) + {\cal O}(e^{-\Delta E t_{\rm ins}}),
    \label{Eq:plateau}
\end{equation}
where $\Delta E$ is the energy gap between the nucleon first excited state and the ground state. Assuming that the exponential terms in Eq.~(\ref{Eq:plateau}) are small we can extract the first term that gives the matrix element of interest  by looking for a range of $t_{\rm ins}$ for a given $t_s$ for which Eq.~(\ref{Eq:ratio}) is time-independent (plateau region) and fit to a constant (plateau value). We  then increase $t_s$ until the plateau values  converge. The converged plateau values determine the ground state nucleon matrix element of the  current considered.\\
\noindent\emph{Summation method:} The insertion time, $t_{\rm ins}$, of the ratio in Eq.~(\ref{Eq:ratio}) can be summed leading to~\cite{Maiani:1987by,Capitani:2012gj}
\begin{align}
  R_{\mu}^{\rm summ}(\Gamma_k;\vec{q};t_s) &= \sum_{t_{\rm ins}=a}^{t_s-a} R_{\mu}(\Gamma_k;\vec{q}\
;t_s,t_{\rm ins}) = \nonumber \\
  &\hspace*{0.35cm}c + \Pi_{\mu}(\Gamma_k;\vec{q}) {\times}t_s + {\cal O}(e^{- \Delta E t_s}).
  \label{Eq:Summ} 
\end{align}
Although we also take into account only the lowest state, the contributions from excited states decay faster as compared to the plateau method. Since $t_{\rm ins}$ is  taken around $t_s/2$ the summation method may be considered equivalent to the  the plateau method with about twice $t_s$. If $e^{-\Delta Et_s}$ is sufficiently suppressed in Eq.~(\ref{Eq:Summ}) the slope gives the ground state matrix element. We probe convergence by increasing the lower value of $t_s$, denoted by $t_s^{\rm low}$ entering in the linear fit. The disadvantage of the  summation method is that one needs to do a linear fit with two parameters instead of one as for the plateau method. This leads to an increased statistical error. \\
\noindent\emph{Two-state fit method:} In this approach one considers
 explicitly the contribution of the first
 excited state.  Namely, the two-point function
is taken to be 
\begin{equation}
C(\vec{p},t_s) = c_0(\vec{p}) e^{-E_0(\vec{p}) t_s} + c_1(\vec{p}) e^{-E_1^{2\rm pt}(\vec{p}) t_s}
\label{Eq:Twp_tsf}
\end{equation}
and the three-point function 
\begin{align}
  C_{\mu}(\Gamma_k,\vec{q},t_s,t_{\rm ins}) = \nonumber \\
  &  {\cal A}^{0,0}_{\mu}(\Gamma_k,\vec{q}) e^{-m_0(t_s-t_{\rm ins})-E_0(\vec{q})t_{\rm ins}} \nonumber \\
  &+ {\cal A}^{0,1}_{\mu}(\Gamma_k,\vec{q}) e^{-m_0(t_s-t_{\rm ins})-E_1^{3 \rm pt}(\vec{q})t_{\rm ins}} \nonumber \\
  &+ {\cal A}^{1,0}_{\mu}(\Gamma_k,\vec{q}) e^{-E_1^{3 \rm pt}(0)(t_s-t_{\rm ins})-E_0(\vec{q})t_{\rm ins}} \nonumber \\
  &+ {\cal A}^{1,1}_{\mu}(\Gamma_k,\vec{q}) e^{-E_1^{3 \rm pt}(t_s-t_{\rm ins})-E_1^{3\rm pt}(\vec{q})t_{\rm ins}},
\label{Eq:Thrp_tsf}
\end{align} 
where contributions from states beyond the first excited state are  neglected. As will be discussed in detail in the following sections, we allow the first excited state in the three-point function to be in general different from that of the two-point function.
The coefficients of the exponential terms of the two-point function in Eq.~(\ref{Eq:Twp_tsf}) are overlap terms given by 
\begin{equation}
c_i(\vec{p}) = \tr[ \Gamma_0 \langle \Omega \vert {\cal J}_N | N_i(\vec{p}) \rangle \langle N_i(\vec{p}) \vert \bar{{\cal J}}_N | \Omega \rangle],
\label{Eq:c2pf}
\end{equation}
where spin indices are suppressed. The $i$-index  denotes the $i^{\rm th}$ nucleon state that may also include   multi-particle states.
The terms ${\cal A}^{i,j}$ appearing in the three-point function in Eq.~(\ref{Eq:Thrp_tsf}) are given by 
\begin{align}
    &{\cal A}^{i,j}_\mu(\Gamma_k,\vec{q}) = \tr[ \Gamma_k \langle \Omega \vert {\cal J}_N | N_i(\vec{0})\rangle \langle N_i(\vec{0}) | A_\mu | N_j(\vec{p}) \rangle  \nonumber \\
    & \langle N_j(\vec{p}) | \bar{{\cal J}}_N | \Omega \rangle],
    \label{Eq:A3pf}
\end{align}
where  $\langle N_i(\vec{0}) | A_\mu | N_j(\vec{p}) \rangle$ is the matrix element between $i^{\rm th}$ and $j^{\rm th}$ nucleon states.

\begin{figure}[h!]
        \centering
        \includegraphics[width=0.7\linewidth]{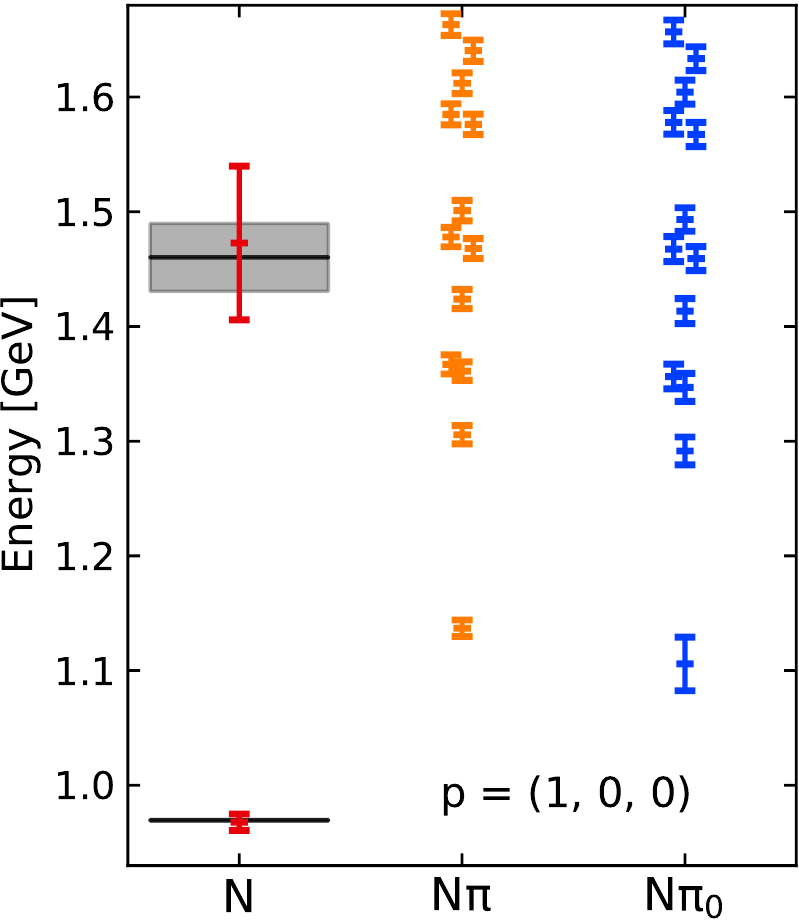}
\caption{We show the lowest two nucleon  energies (red) and the energies of the non-interacting  $\pi^+ N$ (orange) and $\pi^0 N$ (blue) for the case of the  cB211.072.64 ensemble for the smallest total momentum $\vec{p}=\{1,0,0\}$. The value of the $\pi_0$ mass is taken from Ref.~\cite{Alexandrou:2018egz}.
The nucleon energies are extracted from a three-state fit to the nucleon two-point function. The black horizontal lines with the gray bands are the 
experimental values of nucleon and Roper energies.
}
\label{fig:spectrum}
\end{figure}

\begin{figure}
        \centering
        \includegraphics[scale=0.4]{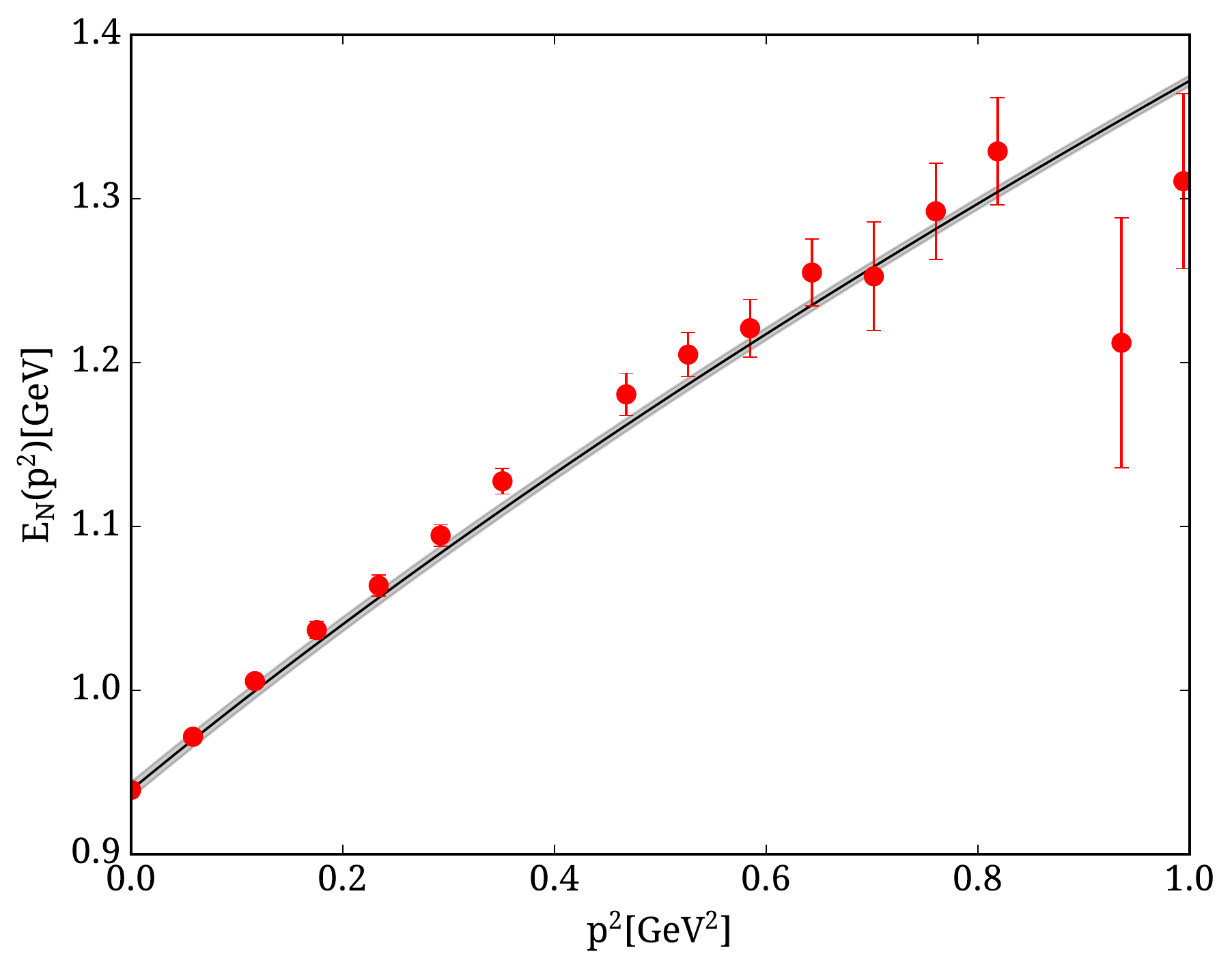}
        \caption{Red points show the energy of the nucleon $E_N(\vec{p}\,^2)$ in GeV as extracted from finite momentum two-point functions and the grey band shows the dispersion relation $E_N(\vec{p}^{~2}) = \sqrt{m_N^2 + \vec{p}^{\,2}}$ as a function of $\vec{p}^{\,2}$ in GeV$^2$. The results are from the $N_f=2+1+1$ cB211.072.64 ensemble.}
        \label{fig:dispRel}
    \end{figure}

Multi-particle states are  volume 
suppressed and are typically not observed in the two-point function. However, if they couple strongly to a current they may contribute in the three-point function. As pointed  out in Refs.~\cite{Bar:2019igf,Bar:2019zkx}, this may happen for the case of the axial-vector current considered here. 
In order to include the possibility that multi-particle states contribute to the three-point function, we perform the following types of fits:
\begin{enumerate}
    \item  [{\it M1}:]  We assume that the first excited state is the same in both the two- and three-point functions. In this case, we first fit  the two-point function extracting  $c_1(\vec{p})$ and $E_1(\vec{p})$ and then use them when fitting the ratio of Eq.~(\ref{Eq:ratio}). We also fit the zero momentum two-point function to determine the nucleon mass and then use the continuum dispersion relation $E_0(\vec{p}) = \sqrt{m_N^2 + \vec{p}^{\,2}}$ to determine the nucleon energy for a given value of momentum. The continuum dispersion relation is satisfied  for all the momenta considered in this work as can be seen in Fig.~\ref{fig:dispRel}. We will refer to this as fit \emph{M1}.
    \item [{\it M2}:]  We allow the first excited state to be different in the two- and three- point functions.   In this case, the first excited energy of the three-point function is left as a fit parameter. We will refer to this as  \emph{M2} fit.
\end{enumerate}
In Fig.~\ref{fig:spectrum} we show the energies extracted from the nucleon two-point function as well as the two-particle non-interacting   $\pi N$ energies  computed as the sum of the pion and nucleon energies. We show these energies for both the charged and neutral pions. As can been seen, the first excited state $E_1^{2\rm pt}(p)$ extracted from two-point function coincide with that of the Roper resonance with at the same momentum. The lowest two-particle states are not visible in the two-point functions, although they are much lower than the energy of the Roper. This is expected since they volume suppressed.   We note that the energies of the $\pi^+ N$ and $\pi^0 N$ system are consistent within errors.

More details on these two fit approaches are given in Sec.~\ref{sec:A0fit}.

\subsection{Extraction of the axial and induced pseudoscalar form factors}\label{ssec:SVD}
While the pseudoscalar matrix elements lead directly to the $G_5(Q^2)$ as given in Eq.~(\ref{Eq:g5_decomp}),
the matrix element of the axial-vector current in general contributes to both axial and induced pseudoscalar form factors, as given in  Eqs.~\ref{Eq:Aik_decomp}) and (\ref{Eq:A0k_decomp}.
A procedure 
to extract the two form factors is to minimize $\chi^2$ given by 

\begin{widetext}
\begin{equation}
  \chi^2(Q^2,t_s,t_{\rm ins}) = \sum_{k,\mu} \;\;\sum_{\vec{q} \; \in
    Q^2}\left[\frac{\mathcal{G}_{\mu}(\Gamma_k, \vec{q})F(Q^2;t_s,t_{\rm ins}) -
      R_{\mu}(\Gamma_k,\vec{q};t_s,t_{\rm ins})}{w_{\mu}(\Gamma_k, \vec{q};t_s,t_{\rm ins})}\right]^2,
\label{Eq:chisq_SVD}
\end{equation}
\end{widetext}
where $w_\mu(\Gamma_k, \vec{q};t_s,t_{\rm ins})$ is the statistical error of the ratio $R_{\mu}(\Gamma_k, \vec{q};t_s,t_{\rm ins})$ of Eq.~(\ref{Eq:ratio}) and  $F(Q^2;t_s,t_{\rm ins})$ is a two component vector of the axial form factors
\begin{equation}
F(Q^2;t_s,t_{\rm ins})=\begin{pmatrix}
G_A(Q^2;t_s,t_{\rm ins}) \\ G_P(Q^2;t_s,t_{\rm ins})
\end{pmatrix}. 
\label{Eq:FFAP}
\end{equation}
 The definition of the coefficient matrix $\mathcal{G}_{\mu}(\Gamma_k,\vec{q})$ that has the kinematical factors  is given in Eq.\eqref{Eq:coeffs}.
Minimization of the $\chi^2$ defined in  Eq.~(\ref{Eq:chisq_SVD})  is equivalent to a singular value decomposition (SVD), where
\begin{align}
  F (Q^2;t_s,t_{\rm ins})= \sum_{k,\mu} \;\;\sum_{\vec{q} \; \in
      Q^2}  \tilde{\mathcal{G}}^{-1}_{\mu}(\Gamma_k,\vec{q})\\
      \nonumber \times \; \tilde{R}_\mu(\Gamma_k,\vec{q};t_s,t_{\rm ins})
\end{align}
and
\begin{equation}
  \tilde{\mathcal{G}} = U\Sigma V \qquad \text{with} \qquad \tilde{\mathcal{G}}^{-1} = V \Sigma^{-1} U^\dagger
\end{equation}
where
\begin{align}
  \tilde{R}_{\mu}(\Gamma_k, \vec{q};t_s,t_{\rm ins})\equiv [w_{\mu}(\Gamma_k, \vec{q};t_s,t_{\rm ins})]^{-1} \\ 
  \nonumber \times \;R_{\mu}(\Gamma_k, \vec{q};t_s,t_{\rm ins})
\end{align}
and
\begin{align}
   \tilde{\mathcal{G}}_{\mu}(\Gamma_k,\vec{q})&\equiv [w_{\mu}(\Gamma_k, \vec{q})]^{-1}\mathcal{G}_{\mu}(\Gamma_k, \vec{q}).  
\end{align}
$U$ is a hermitian $N\times N$ matrix with $N$ being the number of
combinations of $\mu$, $k$ and components of $\vec{q}$
that contribute to the same $Q^2$. $\Sigma$ is the
pseudo-diagonal $N \times 2$ matrix of the singular values of
$\tilde{\mathcal{G}}$  and $V$  is a hermitian $2 \times 2$ matrix since we have
two form factors. Typically, $N\gg2$ for finite momenta.  
In our analysis, we use the SVD to extract the form factors since it does not need any minimization algorithm that might depend on the initial parameters. In addition, using the SVD approach for a relatively small matrix is much faster than using minimization algorithms. 

In the following sections, results are presented for the ratios of $G_A$ and $G_P$ as described by Eq.~(\ref{Eq:FFAP}).

\subsection{Parameters of the gauge configuration ensembles}\label{ssec:EnsDetails}
In this work we analyze an $N_f=2+1+1$ twisted mass clover-improved fermion ensemble. The  parameters are given in Table ~\ref{table:sim}. In addition, we analyze two $N_f=2$ ensembles with the same light quark action, referred to as cA2.09.48 and cA2.09.64 ensembles. They have the same lattice spacing and two different volumes to check for finite volume effects. The physical volume of the cB211.072.64 ensemble is in between the volume of the two $N_f=2$ ensembles. Results on the axial form factors for the cA2.09.48 ensemble have been presented in Ref.~\cite{Alexandrou:2017hac} but are reanalyzed in this study and the results are used for the volume comparison. 
  For both $N_f=2+1+1$ and $N_f=2$  ensembles the lattice spacing is determined using the nucleon mass. More details on the lattice spacing determination are given in Refs.~\cite{Alexandrou:2018egz,Alexandrou:2018sjm,Alexandrou:2017xwd,Alexandrou:2019ali}.
 
The gauge  configurations were produced by the Extended Twisted Mass Collaboration (ETMC) using the twisted mass fermion formulation~\cite{Frezzotti:2000nk,Frezzotti:2003ni} with a clover term~\cite{Sheikholeslami:1985ij} and the Iwasaki~\cite{Iwasaki:1985we} improved gauge action.  Since the simulations were carried out at maximal twist, we have automatic ${\cal O}(a)$ improvement~\cite{Frezzotti:2000nk,Frezzotti:2003ni} for the physical observables considered in this work.

 \begin{widetext}
 
   \begin{center}
 \begin{table}[ht!]
  \caption{Simulation parameters for the $N_f=2+1+1$ cB211.072.64 ensemble~\cite{Alexandrou:2018egz} and the two $N_f=2$ ensembles, cA2.09.48~\cite{Abdel-Rehim:2015pwa} and cA2.09.64. $c_{SW}$ is the value of the clover
    coefficient, $\beta=6/g$ where $g$ is the bare coupling constant, $N_f$ is the number of dynamical quark flavors in the simulation, $a$ is the lattice spacing,
    V the lattice volume in lattice units,
    $m_\pi$ the pion mass, $m_N$ the nucleon mass, and $L$  the spatial lattice length in physical units.
    The  systematic error on the determination of the lattice spacing, $a$, of  the cA2.09.48 and cA2.09.64 ensembles arises from  the slight  extrapolation  of $m_\pi$ to match the physical value~\cite{Abdel-Rehim:2015pwa}.  For the cB211.072.64 ensemble the deviation from the physical point is negligible and thus this systematic error does not enter.}
  \label{table:sim}
       \begin{tabular}{l r@{.}l r@{.}l l  r@{$\times$}l c r@{.}l ccc r@{.}l r}
    \hline\hline
    Ensemble &\multicolumn{2}{c}{$c_{\rm SW}$} & \multicolumn{2}{c}{$\beta$} & \multicolumn{1}{c}{$N_f$} & \multicolumn{2}{c}{V} & $m_\pi L$ & \multicolumn{2}{c}{$a$ [fm]} & $m_N/m_\pi$ & $a m_\pi$ & $a m_N$ & \multicolumn{2}{c}{$m_\pi$ [GeV]} &$L$ [fm] \\
    \hline
    cB211.072.64 & 1&69    & 1&778 & 2+1+1 & $64^3$&$128$ & 3.62  & 0&0801(4)    & 6.74(3)    & 0.05658(6)   & 0.3813(19)      & 0&1393(7)          &5.12(3) \\
    cA2.09.64 & 1&57551 & 2&1   & 2        & $64^3$&$128$ & 3.97  & 0&0938(3)(1) & 7.14(4)    & 0.06193(7)   & 0.4421(25)      & 0&1303(4)(2)       &6.00(2) \\
    cA2.09.48 & 1&57551 & 2&1   & 2        & $48^3$&$96$  & 2.98  & 0&0938(3)(1) & 7.15(2)    & 0.06208(2)   & 0.4436(11)      & 0&1306(4)(2)       &4.50(1) \\
    \hline\hline
  \end{tabular}

 \end{table}
  \end{center}

\end{widetext}

\subsection{Three-point functions and statistics}\label{ssec:ConnDiscStats}
Since in this work we study  only  isovector combinations, only connected contributions are needed.
For their evaluation we employ standard techniques, namely the so-called \emph{fixed-sink} method using sequential propagators through the sink. In this method, changing the  sink-source time separation $t_s$, the momentum, the projector or the interpolating field at the sink requires a new sequential inversion.   
We, thus, fix the sink momentum  $\vec{p}\,'=\vec{0}$ and use four projectors, namely the unpolarized $\Gamma_0$ and the three polarized projectors $\Gamma_k$. In the case of the cB211.072.64 ensemble, we perform the analysis using in total seven sink-source time separations, $t_s$, in the range 0.64~fm to 1.60~fm. In order to better isolate the contribution from excited states, we need  to compute the three-point functions at  similar statistical accuracy. However, the signal-to-noise ratio drops rapidly with $t_s$ and, thus, we increase statistics as  $t_s$  increases, keeping approximately the statistical error constant. The number of configurations analyzed for the $N_f=2+1+1$ ensemble is kept at 750 for all values of $t_s$.  Statistics are increased by increasing the number of source positions per gauge configuration after checking that the error continues to scale as expected for independent measurements.  The statistics used for all the three ensembles for the computation of the  connected contribution per $t_s$ are shown in Table \ref{table:StatsConn}.

\begin{table}[ht!]
  \caption{Statistics used for evaluating the three- and two-point
    functions for the three ensembles.  Columns from left to right are
    the sink-source time separation, the number of configurations
    analyzed, the number of source positions per configuration chosen
    randomly and the total number of measurements for each time
    separation. Rows with ``All'' in the first column refer to
    statistics of the two-point function, while the rest indicate
    statistics for three-point functions.}
  \label{table:StatsConn}
  \vspace{0.2cm}
  \begin{tabular}{crrr}
    \hline\hline
    $t_s/a$ & \multicolumn{1}{c}{$N_{\rm conf}$} & \multicolumn{1}{c}{$N_{\rm srcs}$} & \multicolumn{1}{c}{$N_{\rm meas}$} \\
    \hline    
    \multicolumn{4}{c}{cB211.072.64: $N_f=2+1+1$, $64^3{\times}128$} \\
    \hline
    \multicolumn{4}{c}{Three-point correlators}\\
     8 & 750 & 1 &  750 \\
    10 & 750 & 2 & 1500 \\
    12 & 750 & 4 & 3000 \\
    14 & 750 & 6 & 4500 \\
    16 & 750 & 16 & 12000 \\
    18 & 750 & 48 & 36000 \\
    20 & 750 & 64 & 48000 \\
    \hline
    \multicolumn{4}{c}{Two-point correlators}\\
    All & 750 & 264 & 198000\\
    \hline\hline
    \multicolumn{4}{c}{cA2.09.64: $N_f=2$, $64^3{\times}128$} \\
    \hline
    \multicolumn{4}{c}{Three-point correlators}\\
    12 & 333 & 16 & 5328 \\
    14 & 515 & 16 & 8240 \\
    16 & 515 & 32 & 16480 \\
    \hline
    \multicolumn{4}{c}{Two-point correlators}\\
    All & 515 & 32 & 16480 \\
    \hline\hline
    \multicolumn{4}{c}{cA2.09.48: $N_f=2$, $48^3{\times}96$} \\
    \hline
    \multicolumn{4}{c}{Three-point correlators}\\
    10,12,14 & 578 & 16 & 9248 \\
    \hline
    \multicolumn{4}{c}{Two-point correlators}\\
    All & 2153 & 100 & 215300\\
    \hline\hline
  \end{tabular}
\end{table}

\section{Renormalization functions}\label{sec:Renormalization}

Matrix elements computed in lattice QCD need to be renormalized in order to extract physical observables. The renormalization functions, or Z-factors, for the $N_f=2$ ensembles have been computed previously~\cite{Alexandrou:2017hac}. A detailed description  about our procedure can be found in Ref.~\cite{Alexandrou:2015sea}. Here we present a summary on  the evaluation of the Z-factors for the $N_f=2+1+1$ cB211.072.64 ensemble.  For this work in the twisted mass formulation, we need the renormalization functions $Z_S$ used for the renormalization of the pseudoscalar form factor $G_5(Q^2)$, $Z_P$  used for the renormalization of the bare quark mass and $Z_A$ used to renormalize the axial-vector current.

We employ the Rome-Southampton method or the so-called RI$^\prime$ scheme~\cite{Martinelli:1994ty}, and compute the quark propagators and vertex functions non-perturbatively. This scheme is mass-independent, and therefore, the Z-factors do not depend on the quark mass.  However, there might be residual cut-off effects of the form $a^2 m_q^2$~\cite{Constantinou:2010gr} and, for the scale dependent renormalization functions  $Z_S$ and $Z_P$, the RI-MOM Green functions have  also a dependence on $m_q^2/\mu^2$. This is why the RI-MOM renormalization functions must be explicitly defined in the chiral limit. If not, the scheme would not be mass-independent. To eliminate any systematic related to such effects, we extract the Z-factors using multiple degenerate-quark ensembles. We use five $N_f=4$ ensembles generated exclusively for the renormalization program at the same $\beta$ value as that of the cB211.072.64 ensemble. These are generated with quark mass which is less than half of the strange mass, in order to suppress the $m_q^2/p^2$ for to scale-dependent renormalization functions, and the  lattice artifacts ${\cal O}(a^2 m_q^2)$. These ensembles are generated at different pion masses in the range of [366-519]~MeV and a lattice volume  of $24^3 \times 48$ in lattice units. Having five pion masses enables us to perform the chiral extrapolation to eliminate from the Z-factors any residual cut-off effects. It should be noted, that the extrapolation in the Z-factors does not have an impact on the nucleon matrix elements, which are calculated directly at the physical point.

 In this study, we use the operators
\begin{eqnarray}
   \Op_S^b &= \bar \chi \tau^b \chi           &= \begin{cases} \bar \psi \tau^b          \psi   & b=1,2 \\
                                                             -i\bar \psi \gamma_5 \eins  \psi   & b=3 \end{cases} \\
   \Op_P^b &= \bar \chi \gamma_5\tau^b \chi   &= \begin{cases} \bar \psi \gamma_5 \tau^b \psi   & b=1,2 \\
                                                             -i\bar \psi          \eins  \psi   & b=3 \end{cases} \\
   \Op_V^b &= \bar \chi \gamma_\mu\tau^b \chi &= \begin{cases} \bar \psi  \gamma_5\gamma_\mu \tau^2 \psi   & b=1 \\
                                                              -\bar \psi  \gamma_5\gamma_\mu \tau^1 \psi   & b=2 \\
                                                               \bar \psi  \gamma_\mu         \tau^3 \psi   & b=3 \end{cases} 
\end{eqnarray}
written in the twisted ($\chi,\,\bar\chi$) and physical basis ($\psi,\,\bar\psi$) with $\psi$ and $\chi$ the $u$ and $d$ doublet and $\tau^b$ are the three Pauli matrices.
In the chiral limit, the renormalization functions become independent of the isospin index $b$, and can be dropped. We use the combination $\bar u \Gamma d$, which is extracted from $\tilde \tau\equiv\frac{\tau^1 +i \tau^2}{2}$. Thus, the operators $ \bar \chi \gamma_\mu\tilde \tau \chi $, $ \bar \chi \tilde \tau \chi $, $ \bar \chi \gamma_5 \tilde \tau \chi $ are used to obtain $Z_A$, $Z_S$ and $Z_P$, respectively.

We note that the PCAC relation  in the twisted basis is given by
\begin{equation}
  \partial^\mu A^{b}_\mu=2m_{\rm PCAC} P^{b}+2im_q\delta^{3b}S^0+{\cal O}(a),
    \label{twisted PCAC}
\end{equation}
where  the axial-vector current   $A^{b}_\mu=Z_A\bar{\chi}\gamma_\mu\gamma_5\tau^b\chi$, the pseudoscalar operator $P^b=Z_P\bar{\chi}\gamma_5\tau^b\chi$ and  the scalar $S^0=Z_S\bar{\chi}\chi$. For the isovector flavor combination $b=3$ and at maximal twist where the PCAC mass $m_{\rm PCAC}$ is tuned to zero, Eq.~(\ref{twisted PCAC}) reduces to
\begin{equation}
  \partial^\mu A^{3}_\mu=2im_qS^0+{\cal O}(a^2),
  \end{equation}
  where  $m_q$ the renormalized quark mass determined from the twisted light quark mass parameter $\mu$ as $m_q=\mu/Z_P$.

The aforementioned operators are renormalized multiplicatively with $Z_{\cal O}$, using the condition 
\begin{equation}
    Z_q^{-1} Z_{\cal O}\frac{1}{12} \tr \left[ (\Gamma^L(p)) \Gamma^{\rm Born-1} \right]\bigg\vert_{p^2=\mu_0^2}=1\,,
\end{equation}
where
\begin{equation}
    Z_q = \frac{1}{12} \tr \left[ (S^L(p))^{-1} S^{\rm Born}(p) \right]\bigg\vert_{p^2=\mu_0^2}\,.
\end{equation}
 $S^L(p)$ and $\Gamma^L(p)$ are the quark propagator and amputated vertex function, respectively, while $S^{\rm Born}(p)$ and $\Gamma^{\rm Born}$ are their tree-level values. The trace is taken over spin and color indices and the momentum $p$  is set to be the same as the RI$^\prime$ renormalization scale $\mu_0$. For the non-perturbative calculation of the vertex functions we use momentum sources~\cite{Gockeler:1998ye} that allow us to reach per mil statistical accuracy with ${\cal O}(10)$ configurations~\cite{Alexandrou:2010me,Alexandrou:2012mt}. High statistical precision means that one has to sufficiently suppress systematic errors. We choose momenta in a democratic manner, namely 
\begin{equation}
  (a\,p) \equiv 2\pi \left(\frac{2n_t+1}{2T/a},
\frac{n_x}{L/a},\frac{n_x}{L/a},\frac{n_x}{L/a}\right),
\end{equation}
where $n_t\in[2, 10],\,n_x\in[2, 5]$ and $T/a$($L/a$) the temporal(spatial) lattice extent. The momenta are chosen in the aforementioned ranges with the constraint $ {\sum_i p_i^4}/{(\sum_i p_i^2)^2}{<}0.3$~\cite{Constantinou:2010gr} to suppress non-Lorentz invariant contributions. These constraints are chosen to suppress ${{\cal O}(a^2)}$ terms in the perturbative expansion of the Green's function and are expected to have non-negligible contributions from higher order in perturbation theory~\cite{Alexandrou:2010me,Alexandrou:2012mt,Alexandrou:2015sea}. We subtract such finite lattice spacing effects by explicitly computing such unwanted contributions to one-loop in perturbation theory and all orders in the lattice spacing. These finite $a$ artifacts appear in both the  $S^L(p)$ and $\Gamma^L(p)$ functions. This improvement of non-perturbative estimates using perturbation theory, significantly improves our estimates, as can be seen in the plots of this section.

\begin{figure}[ht!]
 \includegraphics[scale=0.4]{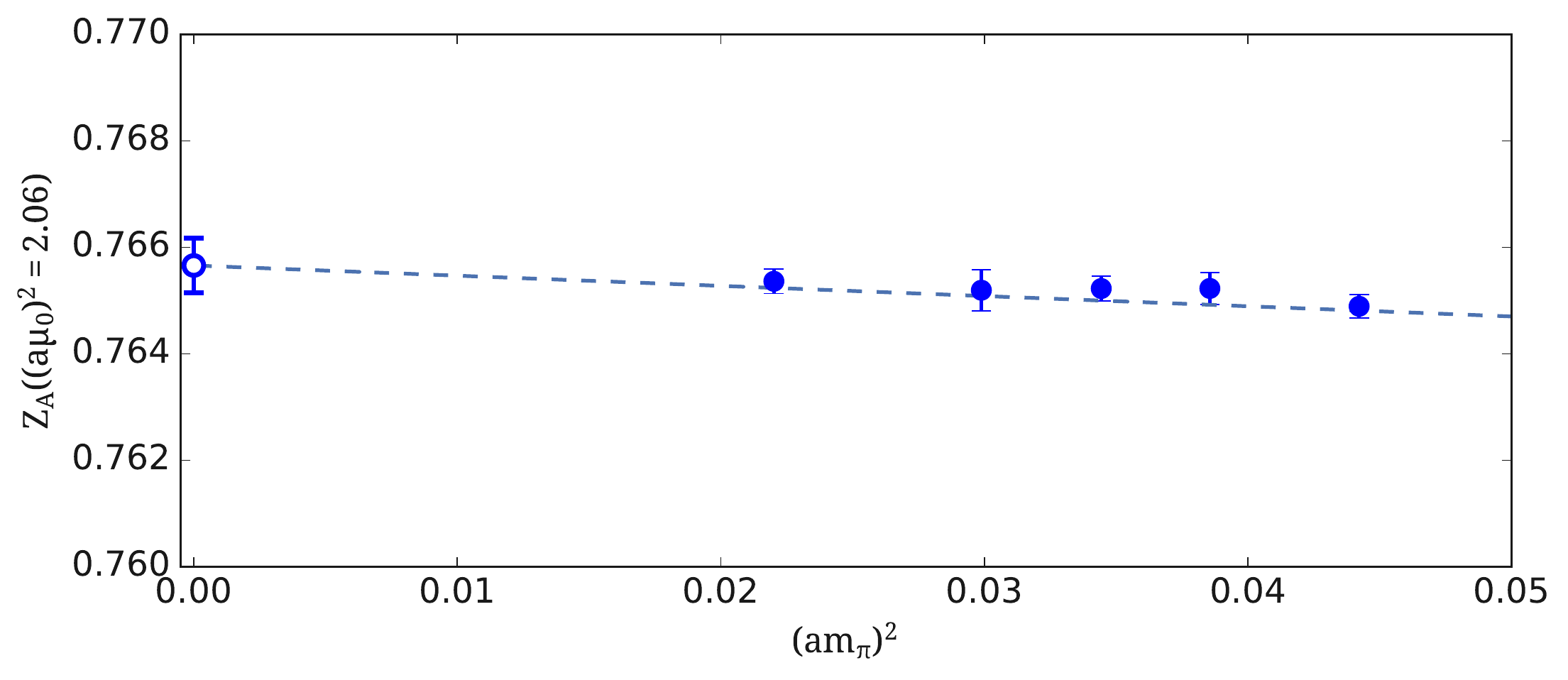}
 \caption{Chiral extrapolation of $Z_A$ for a selected value of $(a\mu_0)^2=2.06$ in the RI$^\prime$ scheme. We use a linear fit (indicated with the dashed line) with respect to $(a m_\pi)^2$, and the extrapolated value in the massless limit is given by the open blue circle.}
 \label{fig:ZA_vs_mpi}
 \end{figure}

Let us first discuss our results on  $Z_A$, which is scheme and scale independent.
In order to eliminate cut-off effects in $Z_A$, we perform a linear fit with respect to $(a m_\pi)^2$ (equivalently $a m_q$), for every value of the renormalization scale. In Fig.~\ref{fig:ZA_vs_mpi} we show the mass dependence for a specific value of the RI$^\prime$ scale. 
We find a slope that is compatible with zero, as expected from our previous studies~\cite{Alexandrou:2015sea}.

\begin{figure}[ht!]
 \includegraphics[scale=0.4]{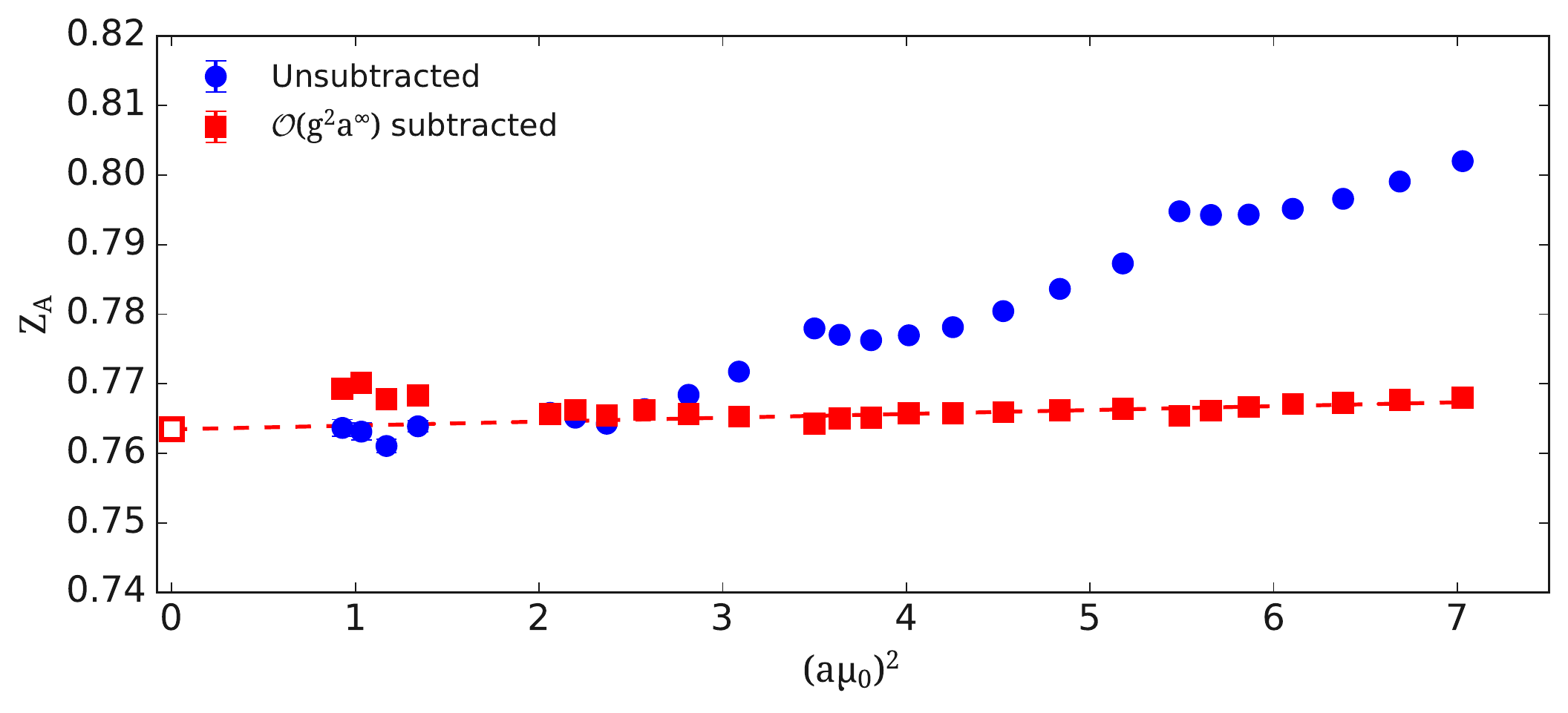}
 \caption{Results for $Z_A$ as a function of the initial renormalization scale $(a\mu_0)^2$. With blue circles are the results before the perturbative subtraction of lattice cut-off artifacts and with red squares after the subtraction of ${\cal O}(g^2 a^\infty)$ contributions. The dashed red line is a linear fit in $(a\mu_0)^2\in [2,7]$  and the open red square is the extrapolated value.}
 \label{fig:ZA_vs_ap2}
 \end{figure}

In order to eliminate the residual dependence on the initial scale due to lattice artifacts, we perform an extrapolation to $(a\mu_0)^2 \rightarrow 0$. In Fig.~\ref{fig:ZA_vs_ap2}, we show the linear extrapolation in  $(a\mu_0)^2$. In the plot we show the purely non-perturbative values of $Z_A$, as well as the improved values obtained after subtracting the lattice artifacts calculated perturbatively. Such a subtraction procedure improves greatly the estimates for Z-factors, as it captures the bulk of lattice artifacts. Indeed, a linear fit in $(a\mu_0)^2$ in the improved subtracted data yields a slope that is consistent with zero within uncertainties.  

The $Z_P$ and $Z_S$ renormalization factors are scheme and scale dependent. Therefore, after the extrapolation $(a m_\pi)^2 \to 0$, we convert to the ${\overline{\rm MS}}$-scheme, which is commonly used in experimental and phenomenological studies. The conversion procedure is applied on the Z-factors at each initial RI$'$ scale $(a\,\mu_0)$, with a simultaneous evolution to a $\overline{\rm MS}$ scale, chosen to be $\overline{\mu}{=}$2 GeV. For the conversion and evolution we employ the intermediate Renormalization Group Invariant (RGI) scheme, which is scale independent and connects the Z-factors between the two schemes:
\begin{align}
Z^{\rm RGI}_{\cal{O}} =&
Z_{\cal{O}}^{\mbox{\scriptsize RI$^{\prime}$}} (\mu_0) \, 
\Delta Z_{\cal{O}}^{\mbox{\scriptsize RI$^{\prime}$}}(\mu_0) \nonumber\\
= &
Z_{\cal{O}}^{\overline{\rm MS}} (2\,{\rm GeV}) \,
\Delta Z_{\cal{O}}^{\overline{\rm MS}} (2\,{\rm GeV})\,,
\end{align}
with ${\cal O} = {P,S}$.
Therefore, the appropriate conversion factor to multiply $Z_{\cal{O}}^{{\rm RI}'}$ is
\begin{equation}
C_{\cal{O}}^{{\rm RI}',{\overline{\rm MS}}}(\mu_0,2\,{\rm GeV}) \equiv 
\frac{Z_{\cal{O}}^{\overline{\rm MS}} (2\,{\rm GeV})}{Z_{\cal{O}}^{{\rm RI}'} (\mu_0)} = 
\frac{\Delta Z_{\cal{O}}^{\mbox{\scriptsize RI$^{\prime}$}}(\mu_0)}
     {\Delta Z_{\cal{O}}^{\overline{\rm MS}}(2\,{\rm  GeV})}\,.
     \label{eq:Conv}
\end{equation}

\begin{figure}[ht!]
 \includegraphics[scale=0.4]{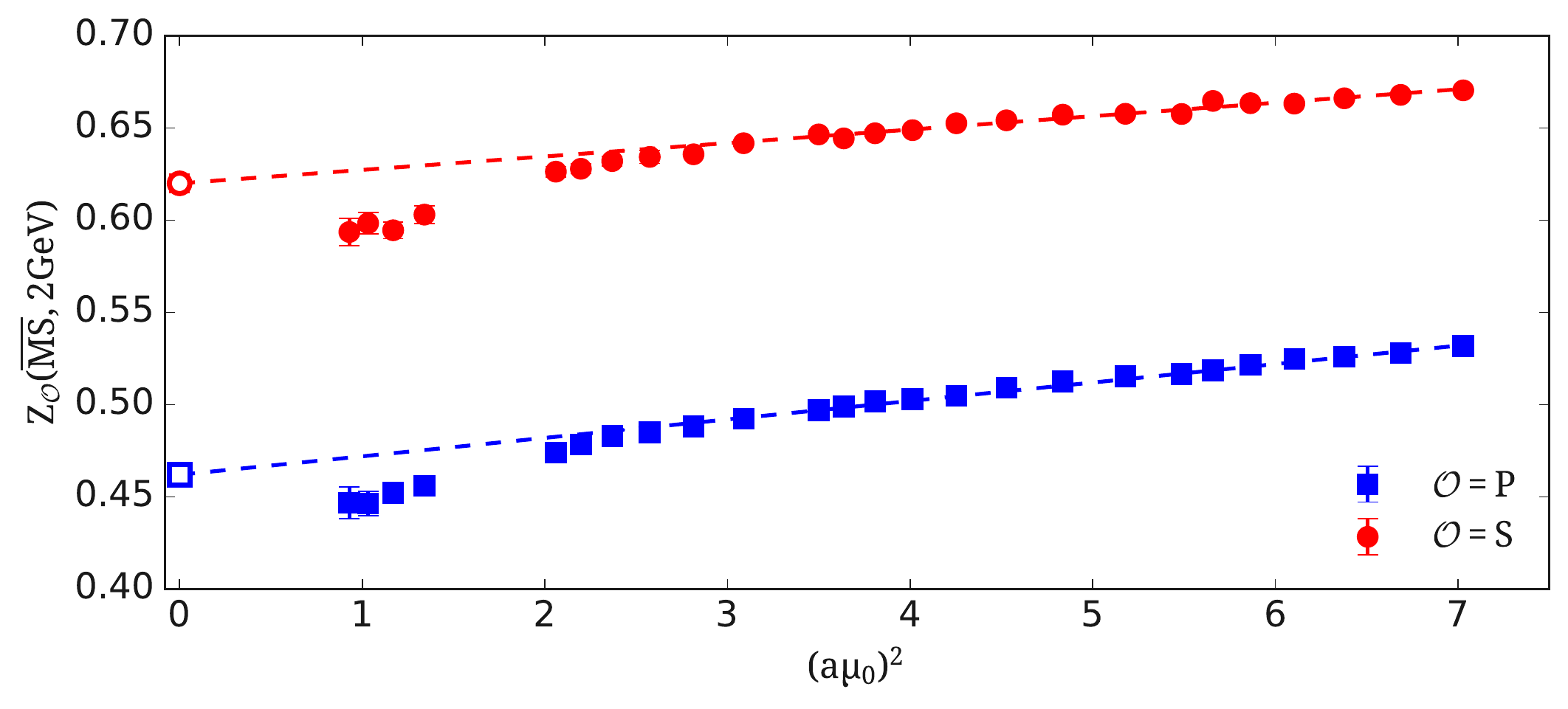}
 \caption{Results for $Z_P$ (blue squares) and $Z_S$ (red circles) as a function of the initial renormalization scale $(a\mu_0)^2$. The final scheme is the ${\rm \overline{MS}}$ scheme at scale $\bar{\mu}=2$~GeV.}
 \label{fig:ZPS_vs_ap2}
 \end{figure}

The quantity $\Delta Z_{\cal{O}}^{\mathcal S}(\mu_0)$ is expressed in terms of the $\beta$-function and the anomalous dimension, $\gamma^S$, of the operator under study
\begin{align}
\Delta Z_{\cal O}^{\mathcal S} (\mu) =&
  \left( 2 \beta_0 \frac {{g^{\mathcal S} (\mu)}^2}{16 \pi^2}\right)
^{-\frac{\gamma_0}{2 \beta_0}}\times\nonumber\\
  &\exp \left \{ \int_0^{g^{\mathcal S} (\mu)} \! \mathrm d g'
  \left( \frac{\gamma^{\mathcal S}(g')}{\beta^{\mathcal S} (g')}
   + \frac{\gamma_0}{\beta_0 \, g'} \right) \right \}\,,
\end{align}
and may be expanded to all orders of the coupling constant. The superscript $S$ denotes the scheme of choice. The expressions for the scalar and pseudoscalar operators are known to three-loops in perturbation theory and can be found in Ref.~\cite{Alexandrou:2015sea} and references therein.
In Fig.~\ref{fig:ZPS_vs_ap2} we present our results for $Z_P$ and $Z_S$. We collect our results for the renormalization functions in Table.~\ref{tab:Zfacs}. We note that the errors given are statistical. A full analysis of
      systematic errors is ongoing and will be presented in an upcoming publication. It is expected that systematic errors will mostly affect the errors on   $Z_P$ and $Z_S$ and will not have any significant effect on the results presented here.

\begin{table}[ht!]
    \centering
    \caption{Scalar and pseudoscalar renormalization functions after lattice cut-off artifacts are subtracted, the chiral limit taken  and the conversion to $\overline{\rm MS}$-scheme.  The first row has the results  for the $N_f=2+1+1$ ensemble with $\beta=1.778$, and the second row for the two $N_f=2$ ensembles with $\beta=2.1$. The errors given are statistical. }
    \begin{tabular}{c|c|c|c}
    \hline\hline 
    Ensemble       & $Z_A$   & $Z_P({\rm \overline{\rm MS}},2{\rm GeV})$ & $Z_S({\rm \overline{MS}},2 {\rm GeV})$ \\
    \hline
    cB211.072.64   & 0.763(1)& 0.462(4) & 0.620(4)    \\
    cA2.09.\{48,64\}  &  0.791(1) & 0.500(30) & 0.661(2) \\
    \hline\hline
    \end{tabular}
    \label{tab:Zfacs}
\end{table}

\section{Extraction of excited energies}\label{sec:A0fit}
In this section we discuss the details for the identification of the nucleon
matrix elements. As mentioned in Sec.~\ref{ssec:ExcStates}, we apply two procedures, referred to as  \emph{M1} and \emph{M2}. Our fit procedure is illustrated  for the case of the $N_f=2+1+1$ cB211.072.64 ensemble but the same procedure is carried out for the two $N_f=2$ ensembles.
 The two-state \emph{M1} fit has been used in previous analyses of form factors, including $G_P(Q^2)$. However, as pointed out in Ref.~\cite{Bar:2019igf}, the $\pi N$ state that is suppressed in the two-point function may become dominant in the three-point function in the case of $G_P(Q^2)$ that is dominated by the pion pole at low $Q^2$ values.
 Therefore, we allow the energy of the first excited state to be different in the two- and three-point functions, as done in the type \emph{M2} fit. As suggested in Ref.~\cite{Jang:2019vkm}, one can use the temporal component of the axial vector current, $A_0$, which is very precise, in order to determine the first excited energy. The temporal component has not been used in past studies~\cite{Shintani:2018ozy,Capitani:2017qpc,Alexandrou:2017hac,Green:2017keo,Bali:2014nma,Rajan:2017lxk}, since it has been found to suffer from large excited state contributions.

 \begin{figure}[!ht]
     \centering
     \includegraphics[scale=0.55]{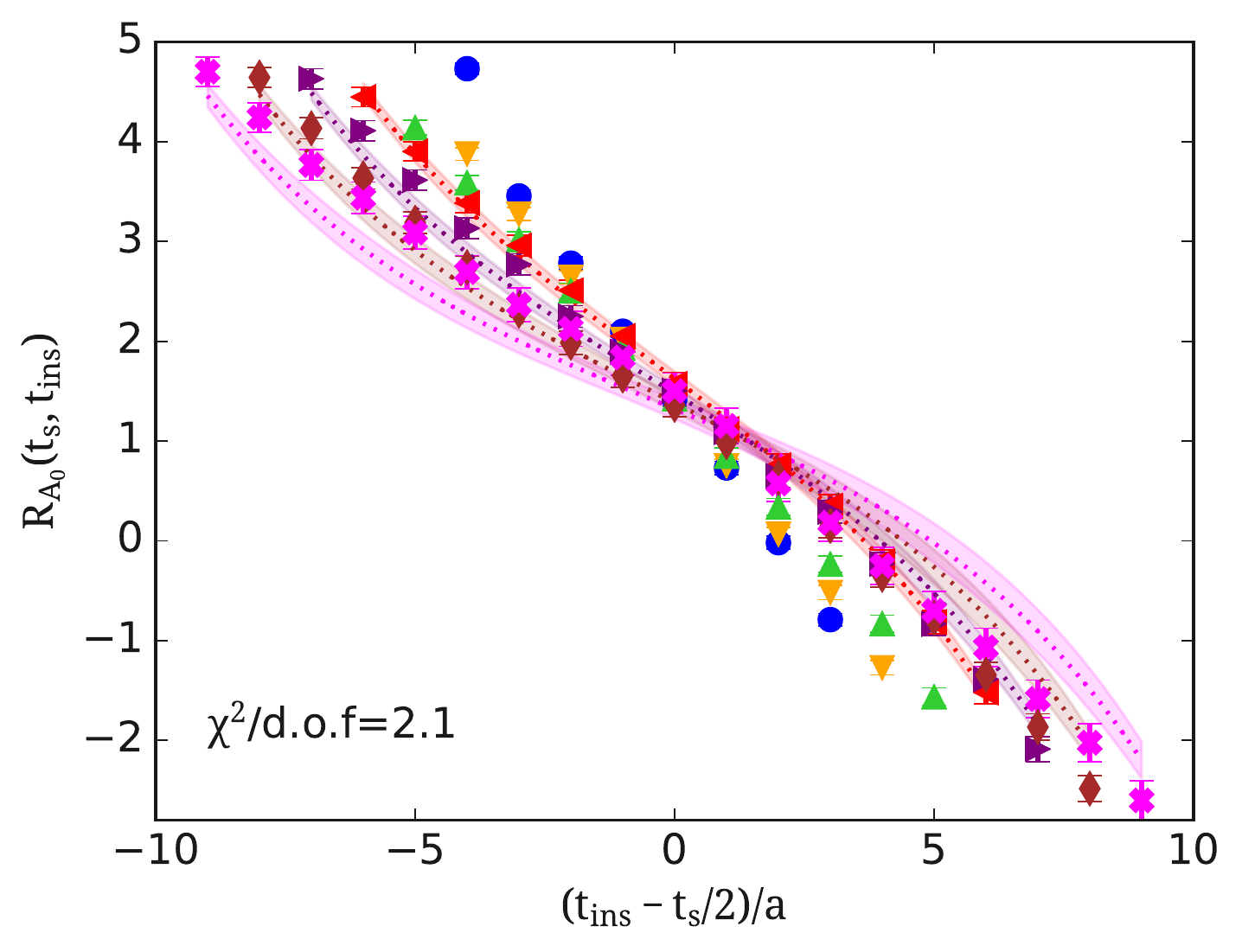}
     \caption{The ratio when using the $A_0$ current versus
       $t_{\rm ins}-t_s/2$ for the lowest non-zero $Q^2$. The sink-source time separations $t_s/a=8,10,12,14,16,18,20$ are shown with blue circles, orange down triangles, up green triangles, left red triangles, right purple triangles, brown rhombus and magenta crosses, respectively. The bands are constructed using a two-state fit where the energy gap with $\vec{p}\,'=\vec{0}$ and $\vec{p}$ is fixed from a two-state fit to the two-point function (fit type {\it M1}). The $\chi^2$/d.o.f=2.1.}
     \label{fig:RA0_M1}
 \end{figure}
   
  \begin{figure}[!ht]
     \centering
     \includegraphics[scale=0.55]{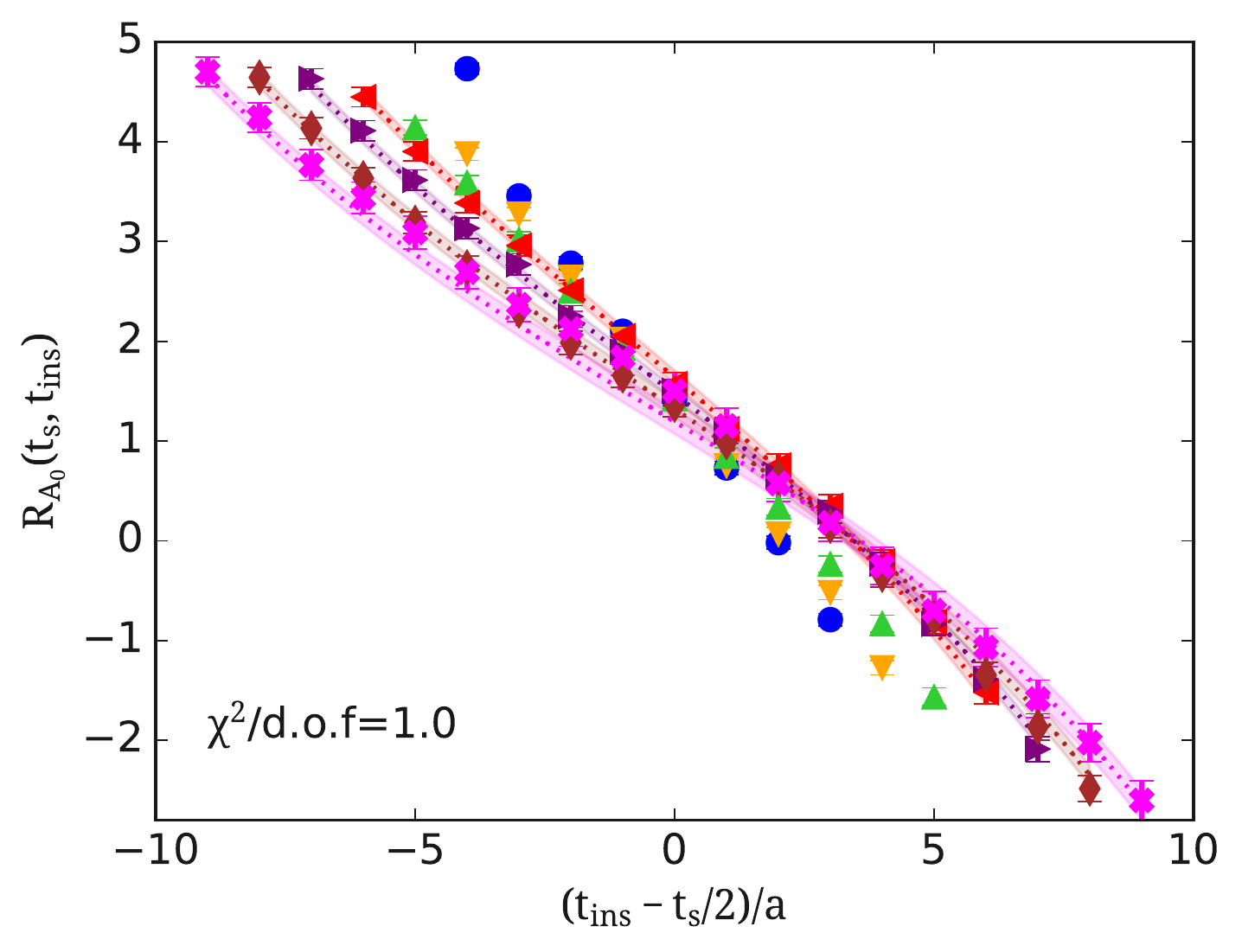}
     \caption{The same as in Fig.~\ref{fig:RA0_M1} but using the two-state approach where the energy gap at $\vec{p}\,'=\vec{0}$ and $\vec{p}$ in the three-point functions are treated as free parameters (fit type {\it M2}).  The $\chi^2$/d.o.f=1.} 
     \label{fig:RA0_M2}
 \end{figure}
 In Figs.~\ref{fig:RA0_M1} and \ref{fig:RA0_M2} we show, respectively, the results when using the two-state \emph{M1} and \emph{M2} fit types. We use the ratio constructed with the three-point function of  the temporal axial-vector current. We perform a simultaneous fit on several sink-source time separations, $t_s$,  excluding the three smallest $t_s$ to ensure no contamination from higher excited states.
 As can be seen, the \emph{M2} fit describes better the data as reflected by the better $\chi^2$/d.o.f.
 
\begin{figure}[!ht]
     \centering
     \includegraphics[scale=0.55]{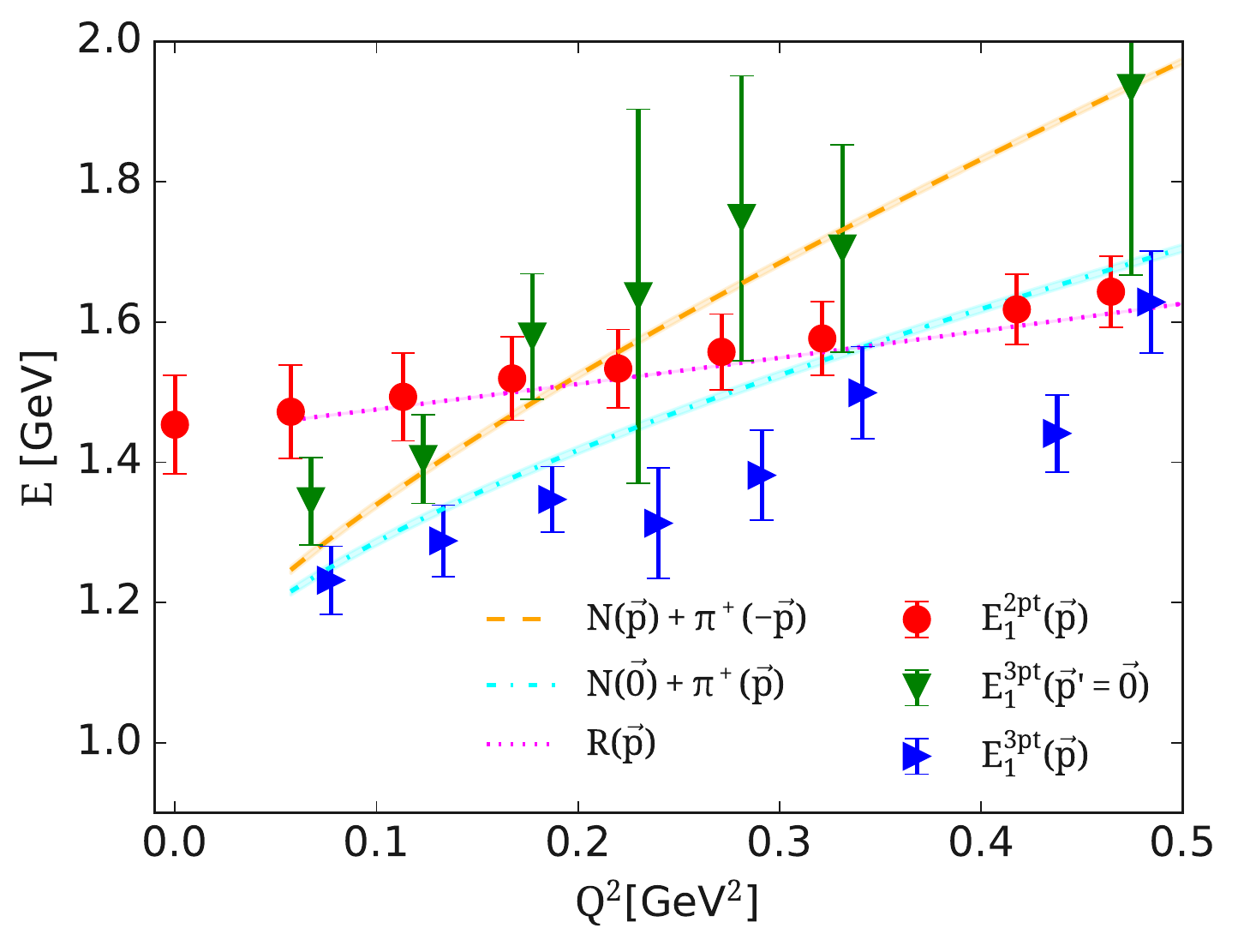}
     \caption{The energy  of the first excited state as a function of  $Q^2$. The orange dashed and cyan dashed-dotted lines are the energies of the non-interacting systems $N(\vec{p}) + \pi(-\vec{p})$ and $N(\vec{0}) + \pi(\vec{p})$, respectively, and with magenta dotted line is  the Roper energy (using as mass the one from PDG~\cite{Zyla:2020zbs}). The red circles are extracted by fitting the two-point function including one excited state. The blue right- and green down-pointing triangles are $E_1^{3\rm pt}(\vec{p})$ and $E_1^{3 \rm pt}(\vec{p}\,'=\vec{0})$, respectively, extracted from the three-point function of the temporal axial-vector current with two-state fits as given in Eq.~\eqref{Eq:Thrp_tsf}.}
     \label{fig:E_piN}
 \end{figure}
 
In Fig.~\ref{fig:E_piN} we show the energy of the first excited state extracted  from fitting the two-point and the three-point function of the temporal axial-vector current. We observe that the first excited energy as extracted from the two-point function is in agreement with the energy of the Roper. This is a different behavior from what is observed in the two recent studies~\cite{Jang:2019vkm,Bali:2019yiy}, where the first excited state extracted from the two-point function is much higher.  Moreover, the energy of the first excited state extracted from the three-point function, is in general in agreement with the  energy of the non-interacting two-particle states of $ N(0)+\pi(-\vec{p})$ and $N(\vec{p})+\pi(-\vec{p})$. We do not observe states with energies lower than the non-interacting state energies unlike what  was found in Ref.~\cite{Jang:2019vkm}.

\section{Extraction of the pseudoscalar form factor $G_5(Q^2)$ from lattice QCD correlators}\label{sec:G5_Effs}
In this section we discuss the  analysis of the correlators for the extraction of the pseudoscalar form factor, and in particular the effect of the excited states. We first consider the pseudoscalar matrix element, since it is only connected to one form factor, $G_5(Q^2)$, as described in Eq.~(\ref{Eq:g5_decomp}) and thus the simplest to extract. 
\begin{widetext}

 \begin{figure}[ht!]
 \includegraphics[scale=0.55]{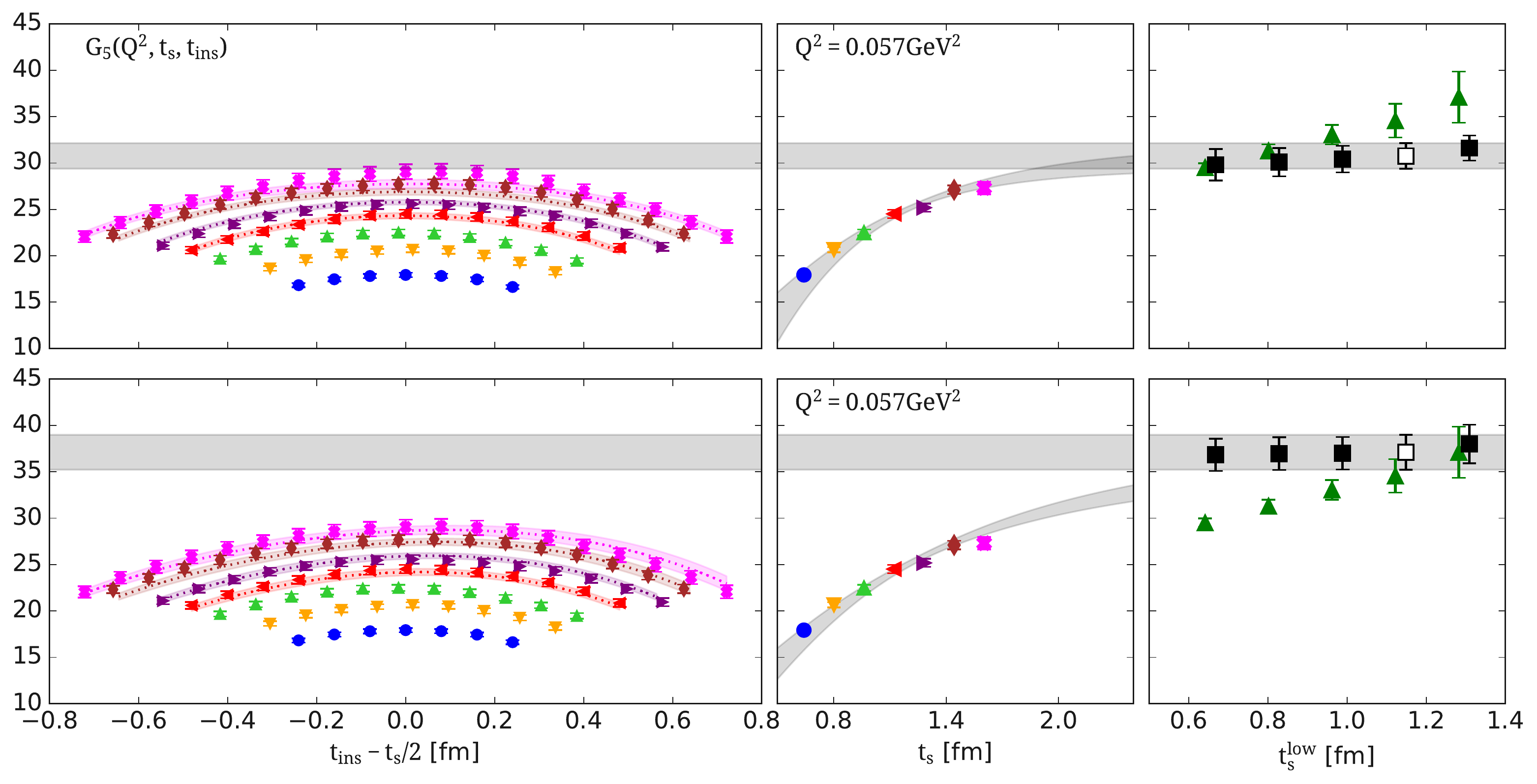}
 \caption{
 Excited states analysis for the ratio of the pseudoscalar three-point correlator for the extraction of $G_5(Q^2)$,  renormalized with $Z_S$. We show results for the first non-zero momentum transfer. In the upper panel,  we show results   when using \emph{M1} and in the second when  using \emph{M2}. In the left panel, we show the ratio given in Eq.~(\ref{Eq:ratio}), for sink-source time separations $t_s/a=8,10,12,14,16,18,20$ with blue circles, orange down-pointing triangles, up-pointing green triangles, left-pointing red triangles, right-pointing purple triangles, brown rhombus and magenta crosses, respectively. The results are shown as a function of the insertion time $t_{\rm ins}$ shifted by $t_s/2$. In the middle panel, we show  the plateau method as a function of $t_s$ using the same  symbol for each  $t_s$ as used for the ratio in the left panel. These are obtained by excluding seven time slices  away from the source and sink for  $t_s/a > 14$, while for smaller time separations, the value at the midpoint is used. In the right panel we show  summation (green triangles) and two-state fits (black squares) results as we increase the smallest time separation $t_s^{\rm low}$ used  in the fit. The open symbol is our choice of the ground state matrix element. The grey band in the middle panel is the predicted time-dependence of the ratio using the parameters extracted  from the two-state fit corresponding to the open symbol, namely when $t_s^{\rm low}=14a=1.12$~fm.
 The dotted lines and  associated error bands shown in the left panel are the resulting two-state fits using the aforementioned value of  $t_s^{\rm low}$. The $\chi^2/$d.o.f is 1.02 and 0.98 for \emph{M1} and \emph{M2}, respectively.
}
 \label{fig:G5_q1_effFFs}
 \end{figure}
 \end{widetext}

\begin{widetext}

 \begin{figure}[ht!]
 \includegraphics[scale=0.55]{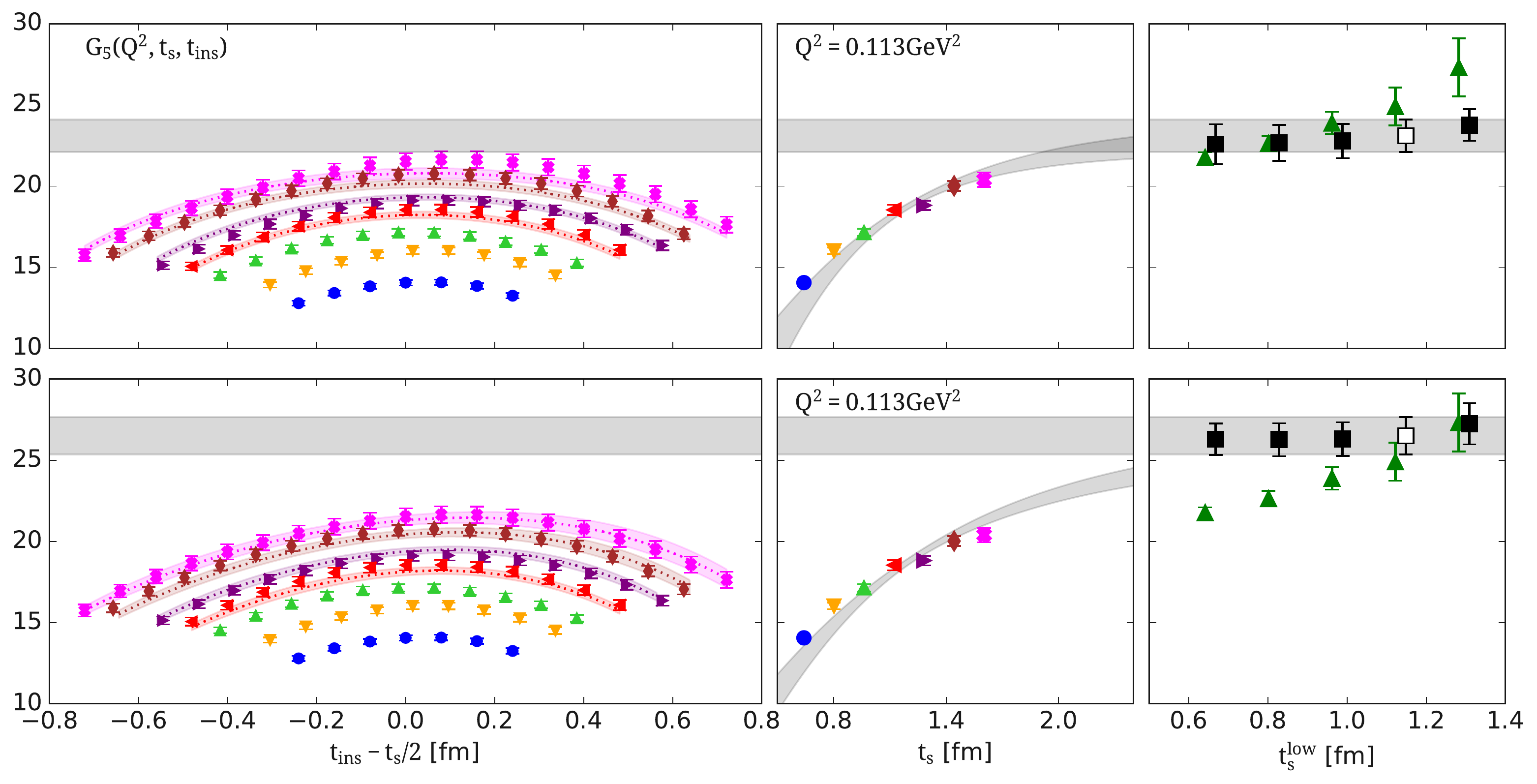}
 \caption{
Excited states analysis for the ratio of the pseudoscalar three-point correlator for the extraction of $G_5(Q^2)$ for the second smallest $Q^2$ value. The notation is the same as that in Fig.~\ref{fig:G5_q1_effFFs}. The $\chi^2/$d.o.f is 1.2 and 1 for \emph{M1} and \emph{M2} fits, respectively.
}
 \label{fig:G5_q2_effFFs}
 \end{figure}
 \end{widetext}

For the identification of the nucleon matrix element we apply the three approaches  discussed in Sec.~\ref{ssec:ExcStates} in order to analyze  contributions from excited states. In Figs.~\ref{fig:G5_q1_effFFs} and \ref{fig:G5_q2_effFFs}, we demonstrate how  excited state contributions are identified for the two smallest $Q^2$. 
In particular, we show the ratio of Eq.~(\ref{Eq:ratio}) for all the available values of $t_s$.  In the construction of the ratio we use  two-point functions computed at the same  source positions as the corresponding three-point functions to exploit their correlation that results in a reduction in the the error.  
As $t_s$ increases we see a significant increase in the values of the ratio pointing to a sizeable excited states contamination.
In the same figure, we show also the plateau values for the two largest time separations obtained by discarding 7 time slices from source and sink  or the midpoint ($t_{\rm ins}=t_s/2$) of the ratio for the smaller time separations.  

In  Fig.~\ref{fig:G5_q1_effFFs}, we include results when using both the  \emph{M1} and \emph{M2} fits for the two-state approach as well as the summation method.  We note that, while for the $G_P(Q^2)$ form factor there is a chiral perturbation theory support for the dominance  of the lowest $\pi N$ state in the three-point function~\cite{Bar:2019igf}, such a theoretical argument is not presented in the case of $G_5(Q^2)$. However, we empirically use the \emph{M2} fit in order to examine if the excited states in the lattice data can be described  with such a fit function. As can been seen, both \emph{M1} and \emph{M2} describe well the data with \emph{M2} providing a better agreement for the larger values of $t_s$. Increasing $t_s^{\rm low}$ does not change the results extracted   from the two-state fits, which shows that including an excited state captures well the time dependence of the ratio. This is unlike the summation method, for which we observe an  increase with increasing  $t_s^{\rm low}$.    We use as our final values the one determined from  the two-state fit at a value of $t_s^{\rm low}$ that  is consistent with the summation values in some range. The final value is larger for the case of \emph{M2}. This is expected since for these momentum transfer the exited energy extracted from the three-point function is lower as compared to the one extracted from the two-point function. However, this increase is not as large as observed in studies of Refs.~\cite{Bali:2018qus,Jang:2019vkm}. Comparing the behavior of the excited states at the second smallest $Q^2$  in Fig.~\ref{fig:G5_q2_effFFs} we find the same  conclusions as for the lowest $Q^2$ value. In both cases our final value is the one from the two state fit at $t_s^{\rm low}$=1.12~fm as discussed in Sec.~\ref{sec:A0fit}. This is what we use for  all the $Q^2$ values. 

\section{Extraction of the  form factors $G_A(Q^2)$ and $G_P(Q^2)$ from lattice QCD correlators }\label{sec:GAGP_Effs}
In this section we discuss the  analysis of the correlators of the axial-vector current from which
the axial and induced pseudoscalar form factors are extracted. For the determination of the  two form factors we follow the procedure discussed in Sec.~\ref{ssec:SVD}.  As explained in  Secs.~\ref{sec:A0fit} and \ref{sec:G5_Effs}, the dominance of two-particle states is expected only to enter the determination of the induced pseudoscalar form factor. For $G_A(Q^2)$ no such strong coupling is expected. Therefore, only  the \emph{M1} fit is applied for the extraction of $G_A(Q^2)$.

\begin{widetext}

 \begin{figure}[ht!]
 \includegraphics[scale=0.55]{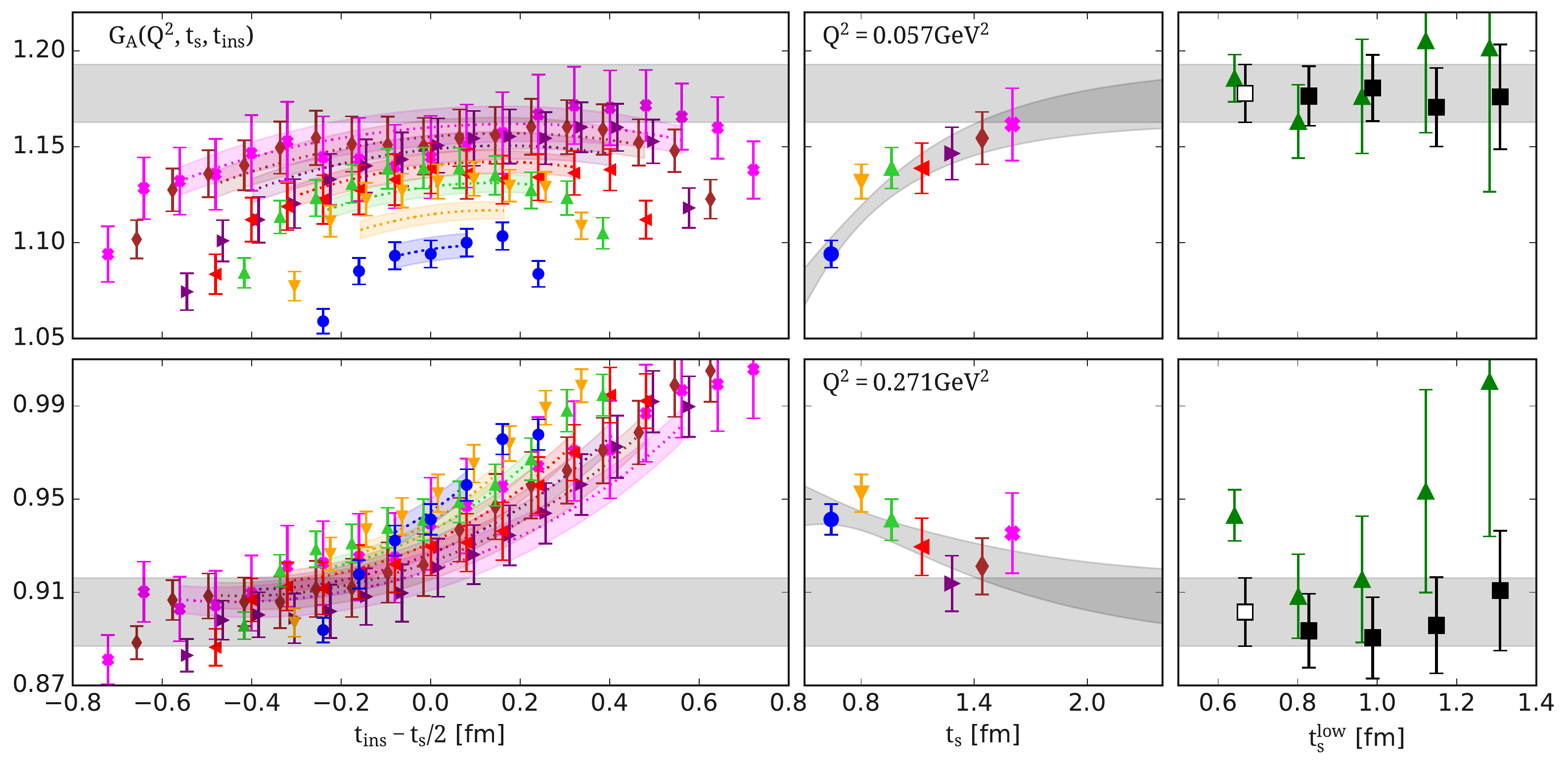}
 \caption{
  Excited states analysis for the ratio of the  three-point correlator for  the extraction of  $G_A(Q^2)$ for  $Q^2=0.057$~GeV$^2$ (top) and $Q^2=0.271$~GeV$^2$ (bottom). The notation is the same as that in Fig.~\ref{fig:G5_q1_effFFs}. For the middle panel, the plateau values are used. The two-state fit analysis is done only with the type \emph{M1} fit. In this case we use $t_s^{\rm low}/a=8$ because it does not suffer from the issues discussed for $G_5(Q^2)$. }

 \label{fig:GA_q2_effFFs}
 \end{figure}
 \end{widetext}
 In Fig.~\ref{fig:GA_q2_effFFs}, we present the analysis of the effect of excited states for the ratio leading to the axial form factor $G_A(Q^2)$. We show results at the smallest $Q^2$ value and at some  intermediate $Q^2$ value to give the general behavior as $Q^2$ increases. We observe that there is a faster convergence as compared to the case of $G_5(Q^2)$.  It is interesting that, while for the smaller values of $Q^2$ the effect of suppressing excited states is to increase the value of $G_A(Q^2)$, for  higher momenta we find that the effect is to decrease it.  Comparing the values of $G_A(Q^2)$ extracted  from the summation and the two-state fits, we find agreement. 
 \begin{widetext}
 
  \begin{figure}[ht!]
 \includegraphics[scale=0.55]{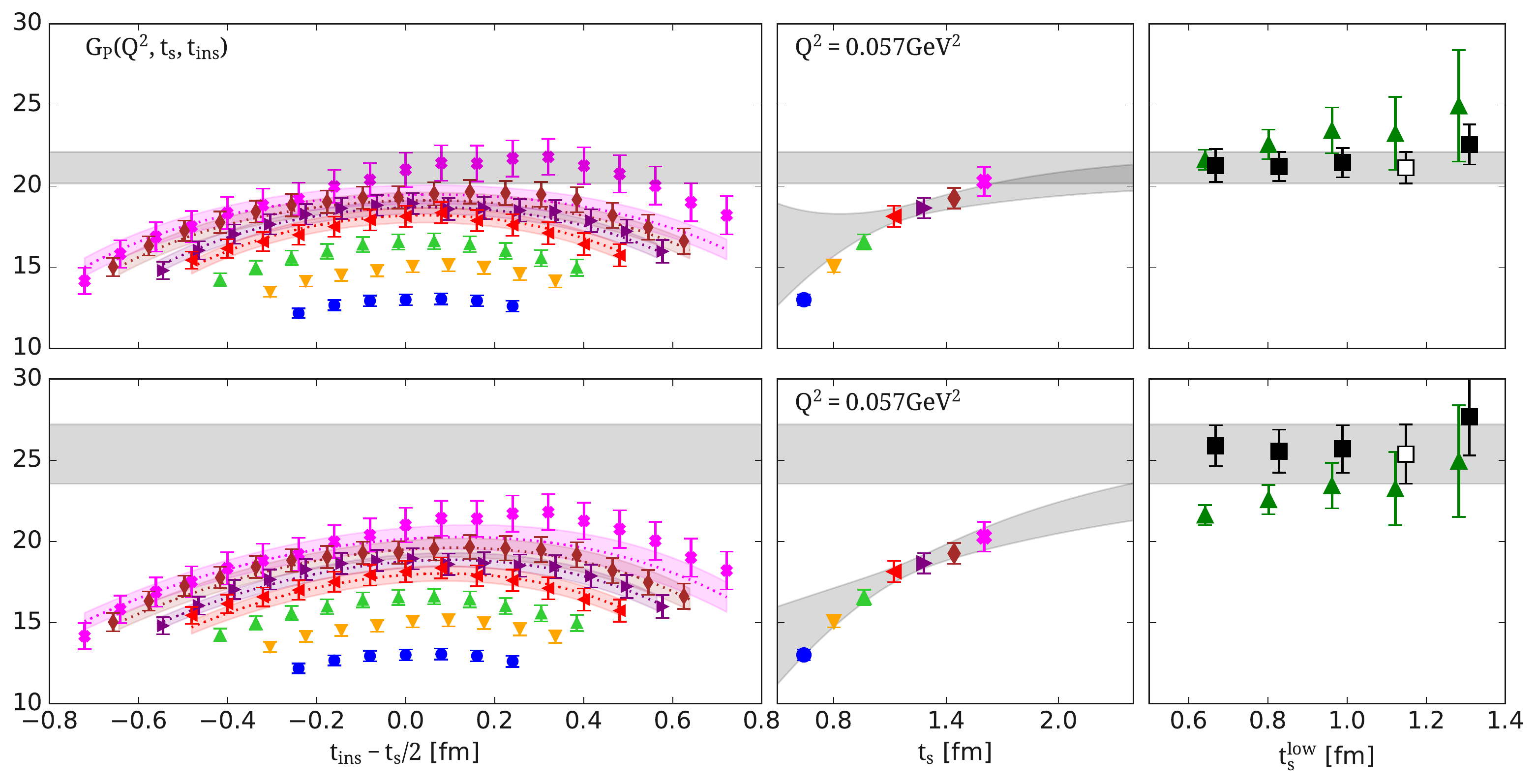}
 \caption{Excited states analysis for the ratio of the  three-point correlator for  the extraction of  $G_P(Q^2)$ for the smallest  $Q^2$. The notation is the same as that in Fig.~\ref{fig:G5_q1_effFFs}. 
}
 \label{fig:GP_q2_effFFs}
 \end{figure}
 \end{widetext}
 In Fig.~\ref{fig:GP_q2_effFFs}, we present the analysis of the effect of excited states for the ratio for   the induced pseudoscalar form factor for the smallest $Q^2$.  What we observe is that the effect of excited states is similar to what is observed in the analysis of $G_5(Q^2)$ in Fig.~\ref{fig:G5_q1_effFFs}.  $G_5(Q^2)$ and $G_P(Q^2)$ have the same pion pole behavior and therefore such similarities are expected.
 As in the case of $G_5(Q^2)$ we carry out the \emph{M2} fit in addition to \emph{M1}.

\section{Comparison of results using the  three ensembles}\label{sec:AllEnsFFs}
\begin{figure}[ht!]
 \includegraphics[scale=0.55]{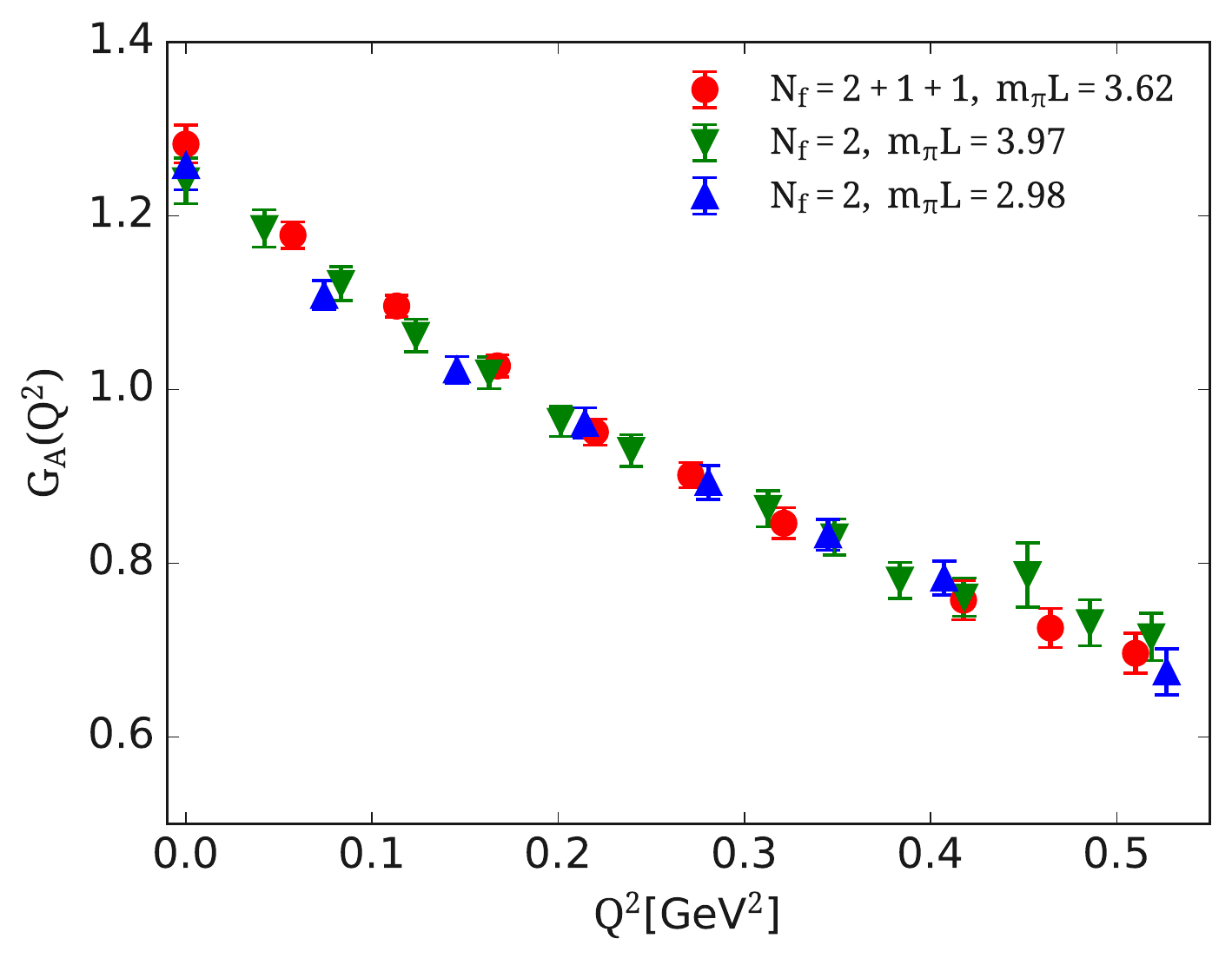}
 \caption{ Results for the $G_A(Q^2)$ form factor as a function of $Q^2$. With red circles are results from the cB211.072.64 ensemble, while with green down and blue up triangles are results from the cA2.09.64 and cA2.09.48 ensembles correspondingly. The \emph{M1} approach has been used for this case. 
 }
 \label{fig:GA_ens}
 \end{figure}

\begin{figure}[ht!]
 \includegraphics[scale=0.55]{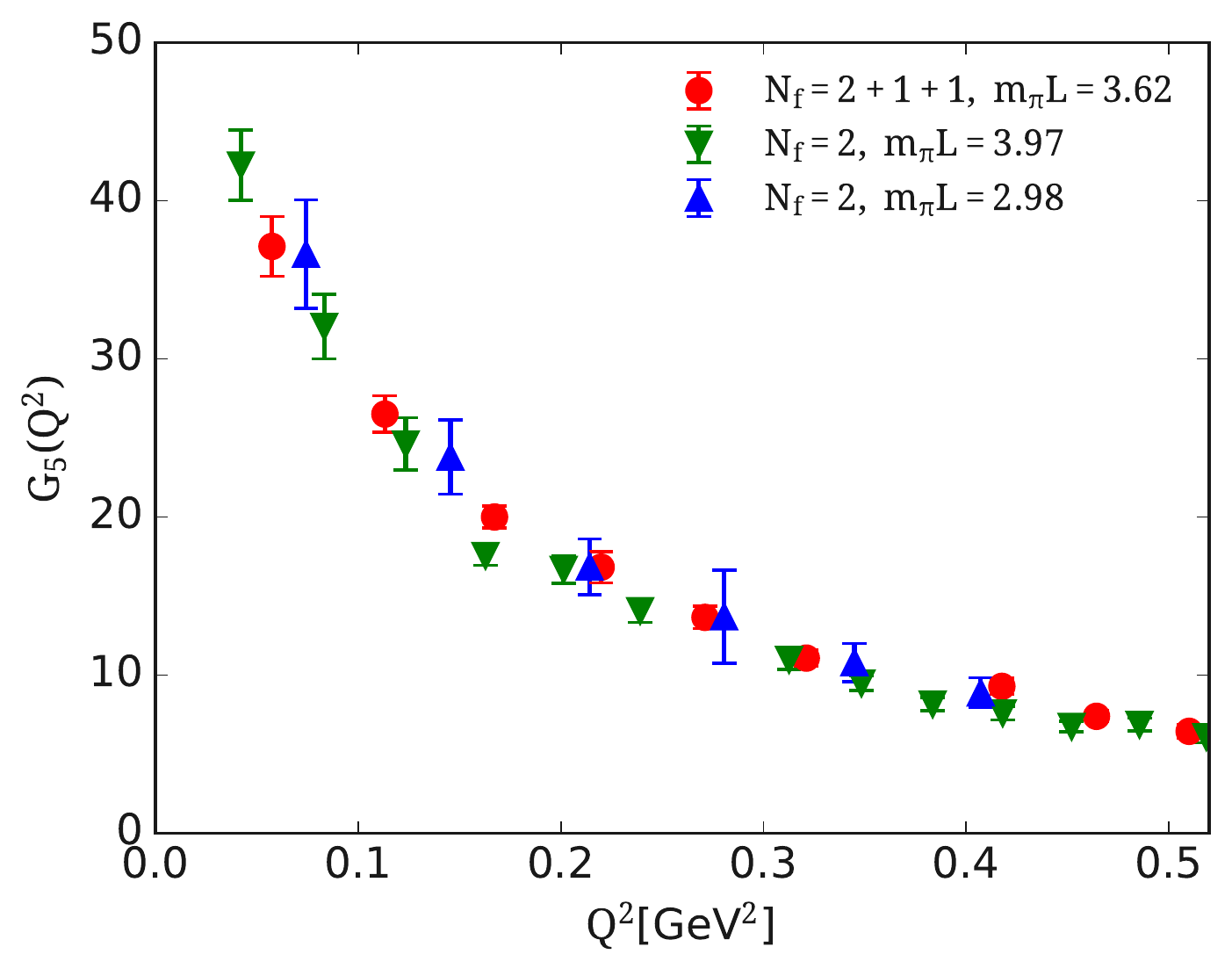}
 \caption{  Results for $G_5(Q^2)$ form factor as a function of $Q^2$. Results are shown for the \emph{M2} fit. The notation is as in Fig.~\ref{fig:GA_ens}. 
}
 \label{fig:G5_ens}
 \end{figure}

\begin{figure}[ht!]
 \includegraphics[scale=0.55]{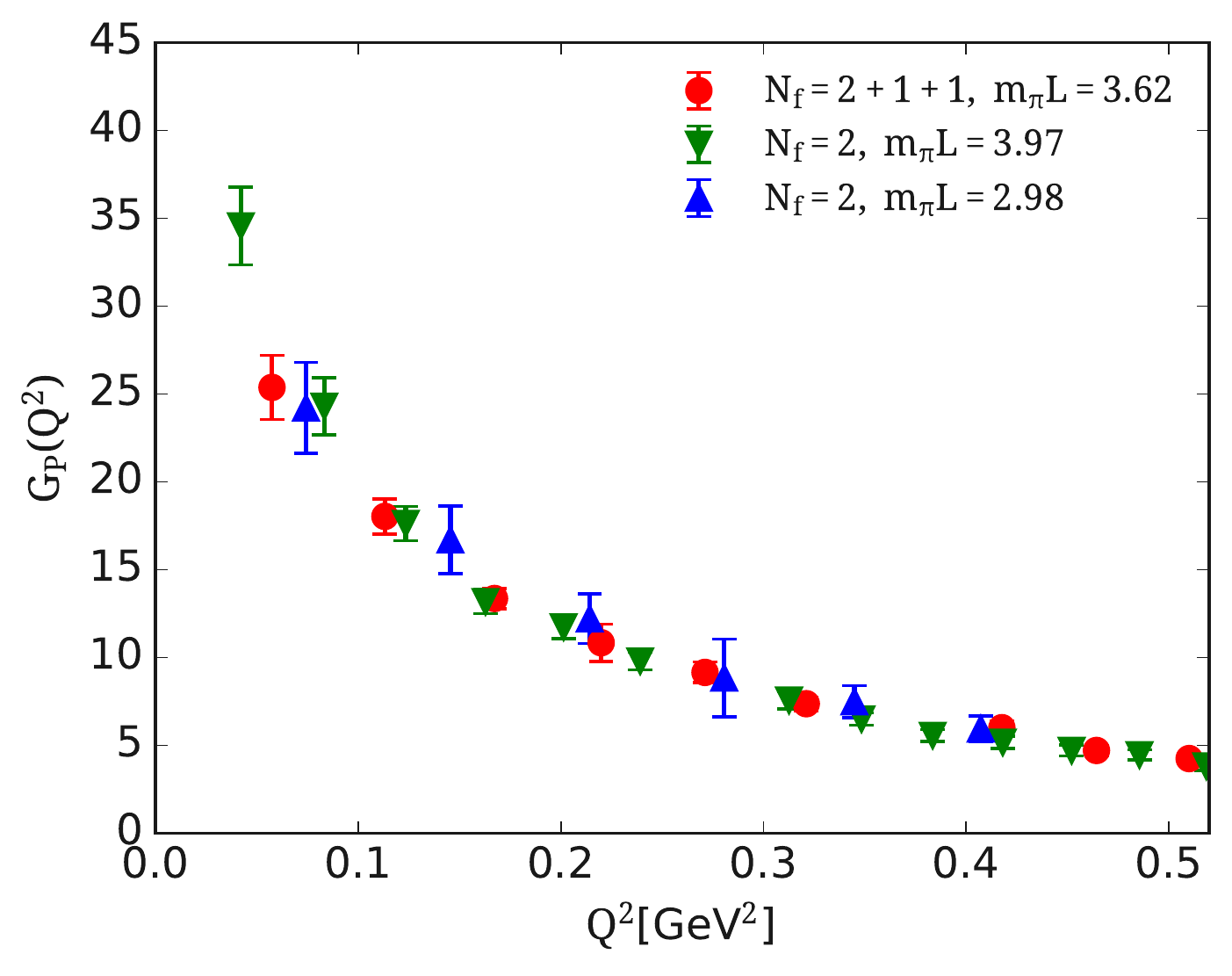}
 \caption{Results for the $G_P(Q^2)$ form factor as a function of $Q^2$. The notation is as in Fig.~\ref{fig:G5_ens}.
}
 \label{fig:GP_ens}
 \end{figure}

We perform a similar analysis as for the $N_f=2+1+1$ cB211,072.64 ensemble also for the two $N_f=2$ ensembles.
In Fig.~\ref{fig:GA_ens},  we compare results from the three ensembles for $G_A(Q^2)$. In particular, comparing the results between  the two $N_f=2$ ensembles we do not observe any finite volume effects in the range $m_\pi L \in [3,4]$.  
 In Figs.~\ref{fig:G5_ens} and \ref{fig:GP_ens} we compare our results for $G_5(Q^2)$ and $G_P(Q^2)$ for the three ensembles, using the \emph{M2} fit. We observe  a very good agreement among the results for the three ensembles. Like for the case of $G_A(Q^2)$, comparison between the results of the two $N_f=2$ ensembles does not show  any finite volume effects in the range $m_\pi L \in [3,4]$.

\begin{figure}[ht!]
 \includegraphics[scale=0.55]{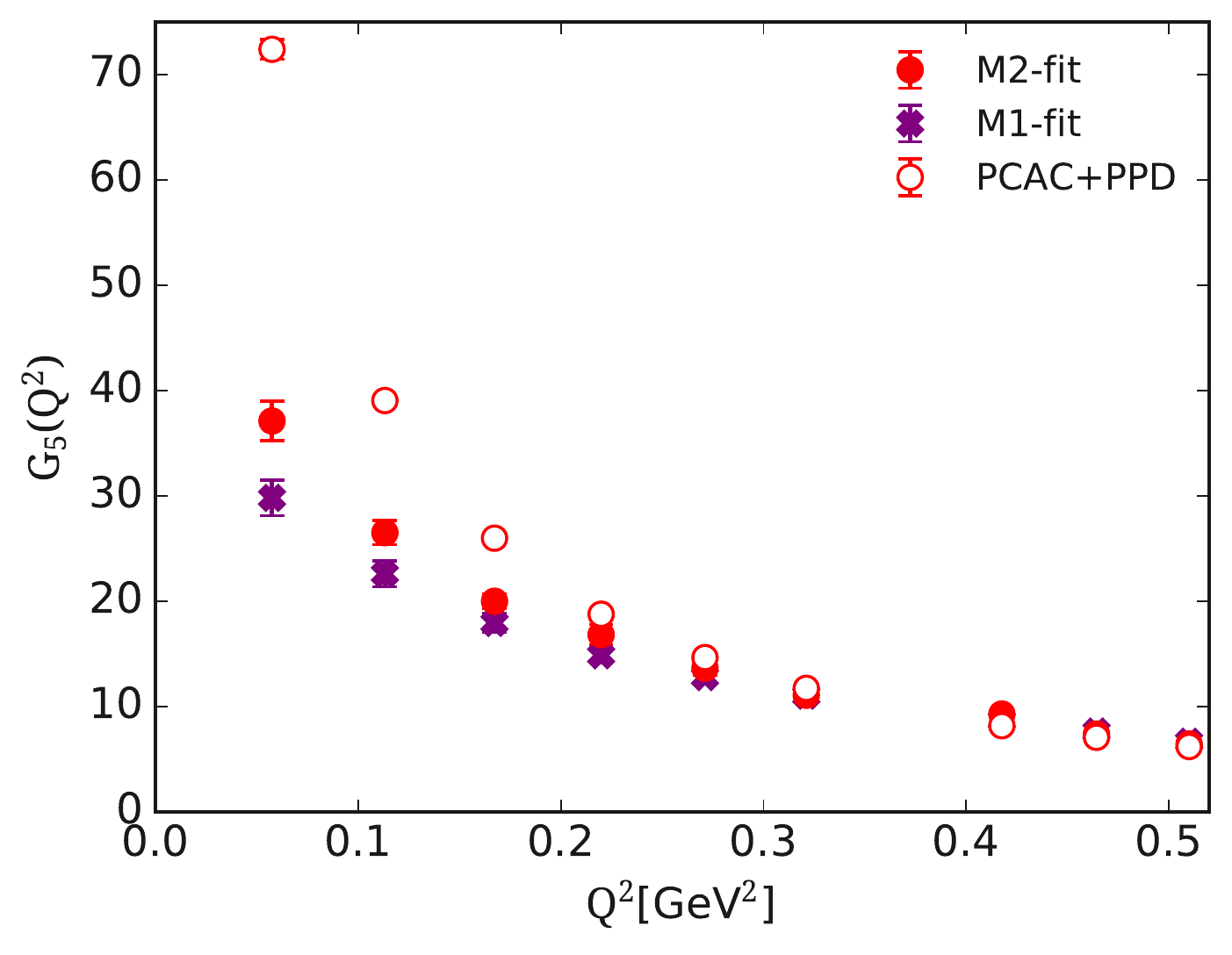}
 \caption{Results for the $G_5(Q^2)$ form factor as a function of $Q^2$ from the cB211.072.64 ensemble. Filled red circles are results using \emph{M2} approach and purple crosses using \emph{M1}. Open red circles are results from the Eq.\eqref{Eq:G5eq}.
}
 \label{fig:G5_typesM1M2}
 \end{figure}

\begin{figure}[ht!]
 \includegraphics[scale=0.55]{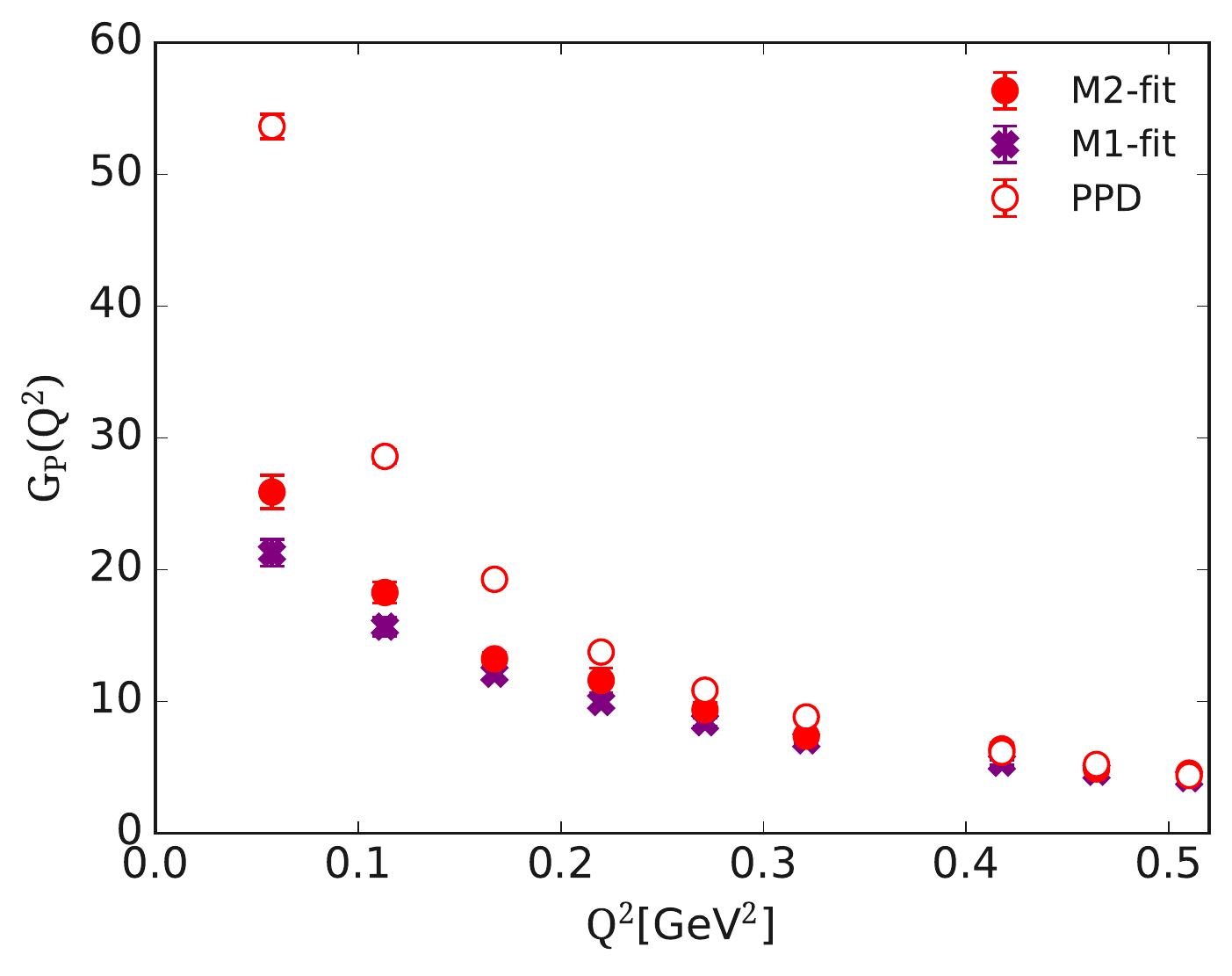}
 \caption{Results for $G_P(Q^2)$ form factor as a function of $Q^2$. The notation is as in Fig.~\ref{fig:G5_typesM1M2}. Open red circles are results from the pion pole dominance prediction. 
}
 \label{fig:GP_typesM1M2}
 \end{figure}

  In Figs.~\ref{fig:G5_typesM1M2} and \ref{fig:GP_typesM1M2} we show a comparison between the \emph{M1} and \emph{M2} fits for $G_5(Q^2)$ and $G_P(Q^2)$. For $G_5(Q^2)$ we include the prediction when using the  PCAC and PPD relations given by Eqs.~(\ref{Eq:PCAC_FFs})  and (\ref{Eq:PPD}) 
 \be
 G_5(Q^2)= \frac{m_N}{m_q} \frac{m_\pi^2}{Q^2+m_\pi^2} G_A(Q^2).
 \label{Eq:G5eq}
 \ee
   Comparing the results extracted using the two-state \emph{M1} to \emph{M2} fits to extract the nucleon matrix elements, we find that the latter approach yields higher values for $Q^2 < 0.2$~GeV$^2$. Despite the increase, however, results for $G_5(Q^2)$  predicted from PCAC  deviate significantly in the low $Q^2$ region from those extracted directly from the nucleon matrix element of the pseudoscalar operator, in contrast to what has been observed in Refs.~\cite{Jang:2019vkm,Bali:2019yiy}. This different behavior can be traced to the fact that the authors of Refs.~\cite{Jang:2019vkm,Bali:2019yiy} find a higher energy for the first excited state from their two-point functions as compared to us. Also the energy of the first excited state extracted from the three-point function of the temporal axial-vector current in Ref.~\cite{Jang:2019vkm} is lower than what we find and lower than the corresponding  non-interacting energy. The authors of Ref.~\cite{Bali:2019yiy} on the other hand find an exited state that is closer to the non-interacting energy as we do, although a direct comparison is not possible since only  results for a heavier than physical pion mass are shown. 
  These observations also hold for  $G_P(Q^2)$, as shown in Fig.~\ref{fig:GP_ens}.
  
\begin{figure}[ht!]
 \includegraphics[scale=0.55]{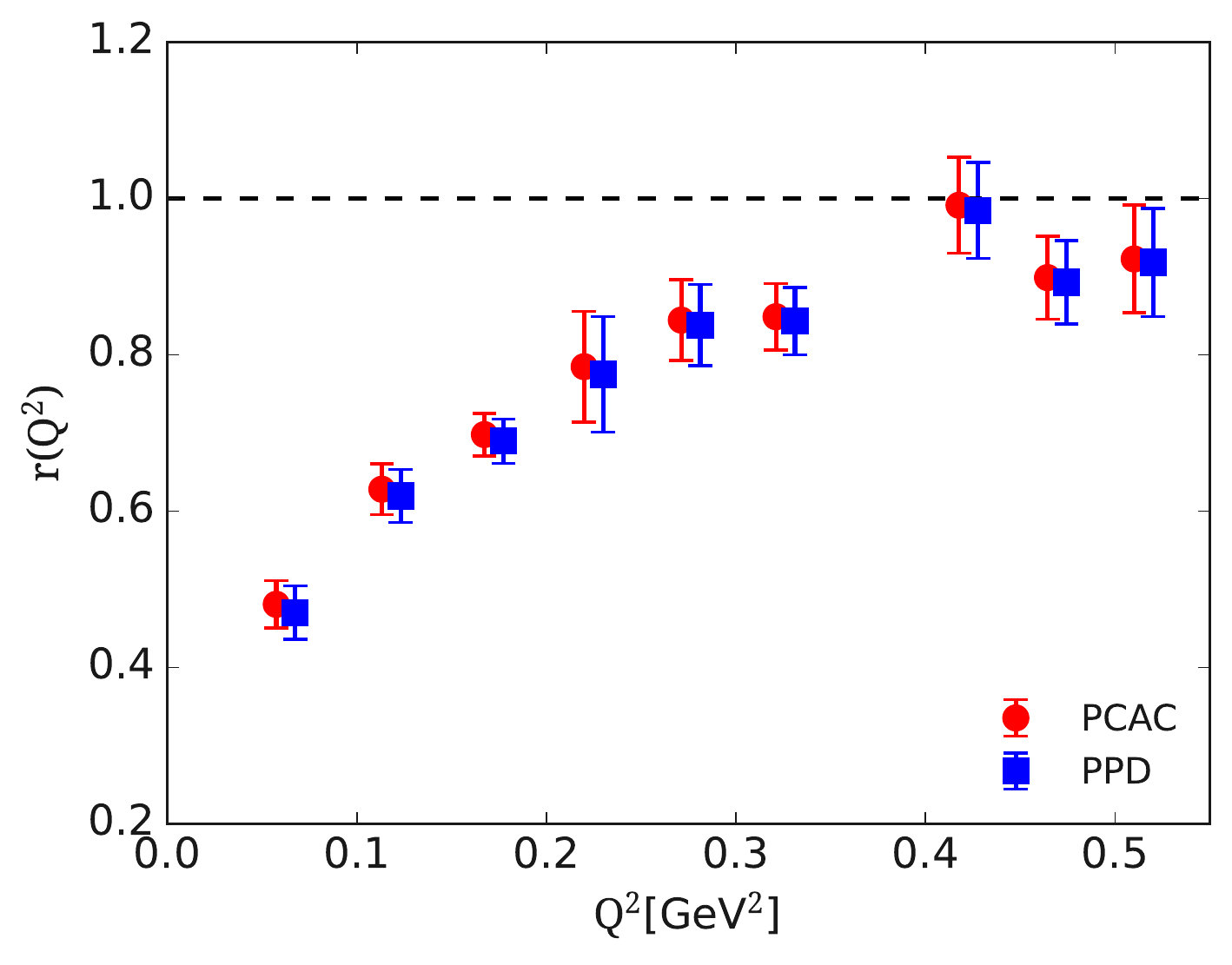}
 \caption{The breaking of PCAC (red circles) and PPD (blue squares) relations as defined in Eq.~(\ref{Eq:rPCAC}) and Eq.~\eqref{Eq:rPPD} respectively, using the cB211.072.64 ensemble. The horizontal black dashed line indicates the recovery of the two relations.  
}
 \label{fig:PCAC_PPD_ens}
 \end{figure}

 It is interesting to examine the breaking of the PCAC and PPD relations  as a function of $Q^2$.  We define two ratios, one checking the PCAC and one the PPD relation as follows
\begin{eqnarray}
r_{\rm PCAC} = \frac{\frac{m_q}{m_N} G_5(Q^2) + \frac{Q^2}{4 m_N^2} G_P(Q^2) }{G_A(Q^2)}
\label{Eq:rPCAC}
\end{eqnarray}
and 
\begin{eqnarray}
r_{\rm PPD} = \frac{G_P(Q^2)}{\frac{4 m_N^2}{m_\pi^2 + Q^2} G_A(Q^2) }.
\label{Eq:rPPD}
\end{eqnarray}
These ratios are unity if PCAC and PPD hold, respectively. In Fig.~\ref{fig:PCAC_PPD_ens}, we concentrate on the results for the cB211.072.64  ensemble since the results using the other two ensembles behave similarly. As can be seen there is a sizeable deviation for both ratios  at small $Q^2$ even though we use the \emph{M2} fit.
In our view,  further investigation is needed to understand the deviations from  the PCAC and PPD relations.
Therefore, in what follows we use the results of $G_A(Q^2)$ to extract both $G_P(Q^2)$ and $G_5(Q^2)$ from Eqs.~\eqref{Eq:PPD} and \eqref{Eq:G5eq}. Also we only discuss our results extracted using the cB211.072.64 ensemble, since they   are more precise and are computed for a lattice volume that is in between the two lattice volumes used for checking for volume effects, for which  we see no effects.

\section{Results}\label{sec:FinalRes}
\begin{figure}[ht!]
 \includegraphics[scale=0.55]{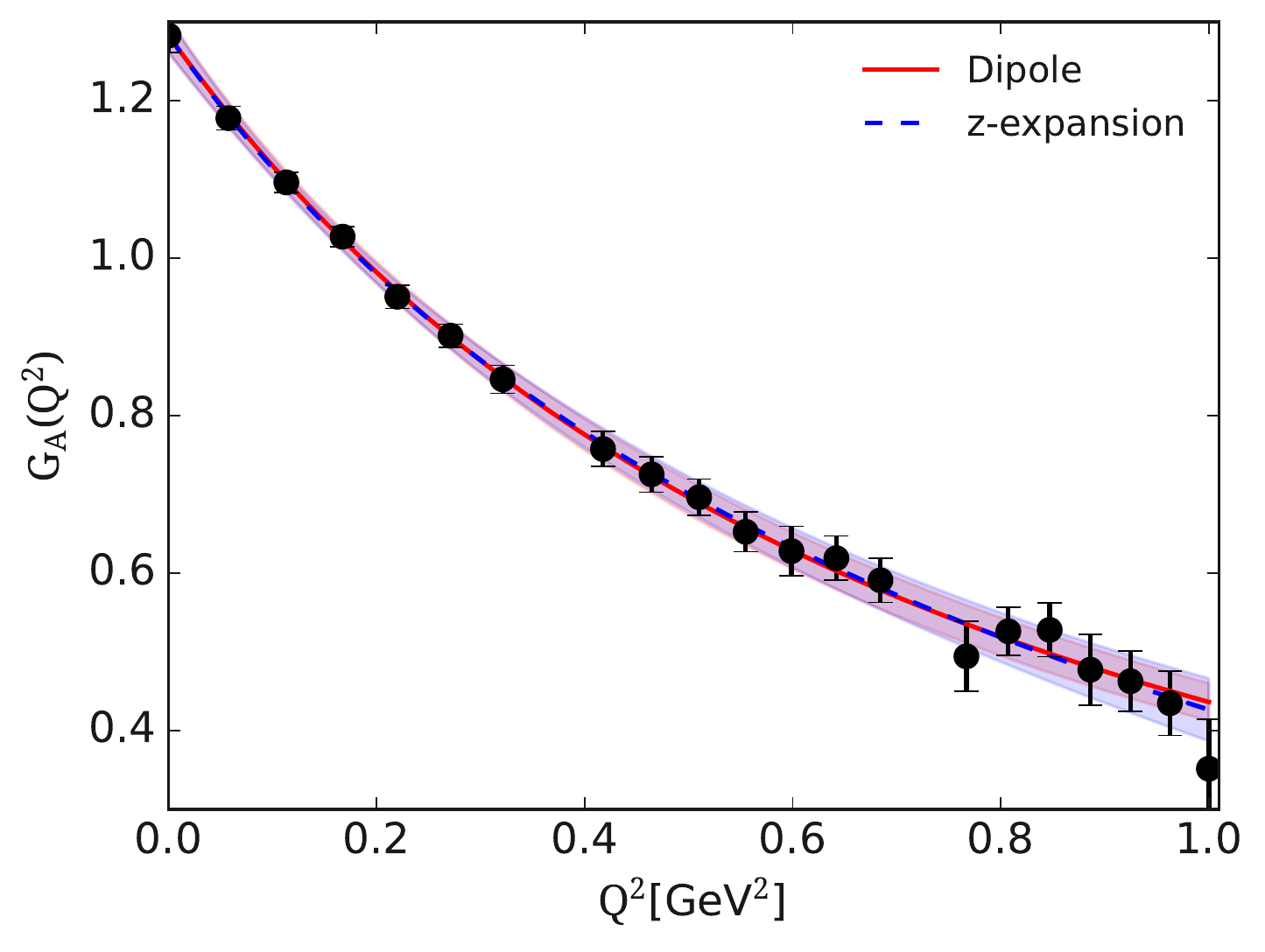}
 \caption{The axial form factor, $G_A(Q^2)$, as function of $Q^2$. The red solid line is the result of the dipole fit defined in Eq.~(\ref{Eq:dipole}) and the blue dashed line is the z-expansion of Eq.~(\ref{Eq:zExp}) with $k_{\rm max}=5$. }
 \label{fig:GA}
 \end{figure}

 All results given in this section are extracted using the $N_f=2+1+1$ cB211.072.64 ensemble. In Fig.~\ref{fig:GA} we show our results for the axial form factor. The value of the form factor at zero momentum transfer gives the axial charge, $g_A \equiv G_A(0)$. We find $g_A=1.283(22)$~\cite{Alexandrou:2019brg}. In order to fit the form factor, we use both the dipole and z-expansion (see Sec.~\ref{sec:Q2Fit} for details). Since for $G_A(Q^2)$ the value for zero momentum transfer is directly accessible, we use $G_A(0)$ in the jackknife fits. This reduces the number of fit parameters in each jackknife bin. The consequence is that the error on  the determination of the radius is smaller. In the case of the z-expansion, we use $k_{\rm max}=5$, where we check that this is large enough to ensure convergence. The width coefficient, $w$, of the Gaussian priors is chosen to be $w=5$. We  provide a systematic error taken as the difference in the mean values when using $w=5$ and  when $w=20$. Comparing the dipole Ansatz with the z-expansion we find excellent agreement for all  $Q^2$ values. Therefore, we conservatively quote as final values (Table~\ref{tab:final_res}) those from the z-expansion, since it is model independent although they typically carry larger statistical uncertainties. The axial mass and the radius are determined from the parameters of the z-expansion as given  in Eqs.~(\ref{Eq:m_zExp} and \ref{Eq:r2_zExp}) correspondingly. 
 
\begin{figure}[ht!]
 \includegraphics[scale=0.55]{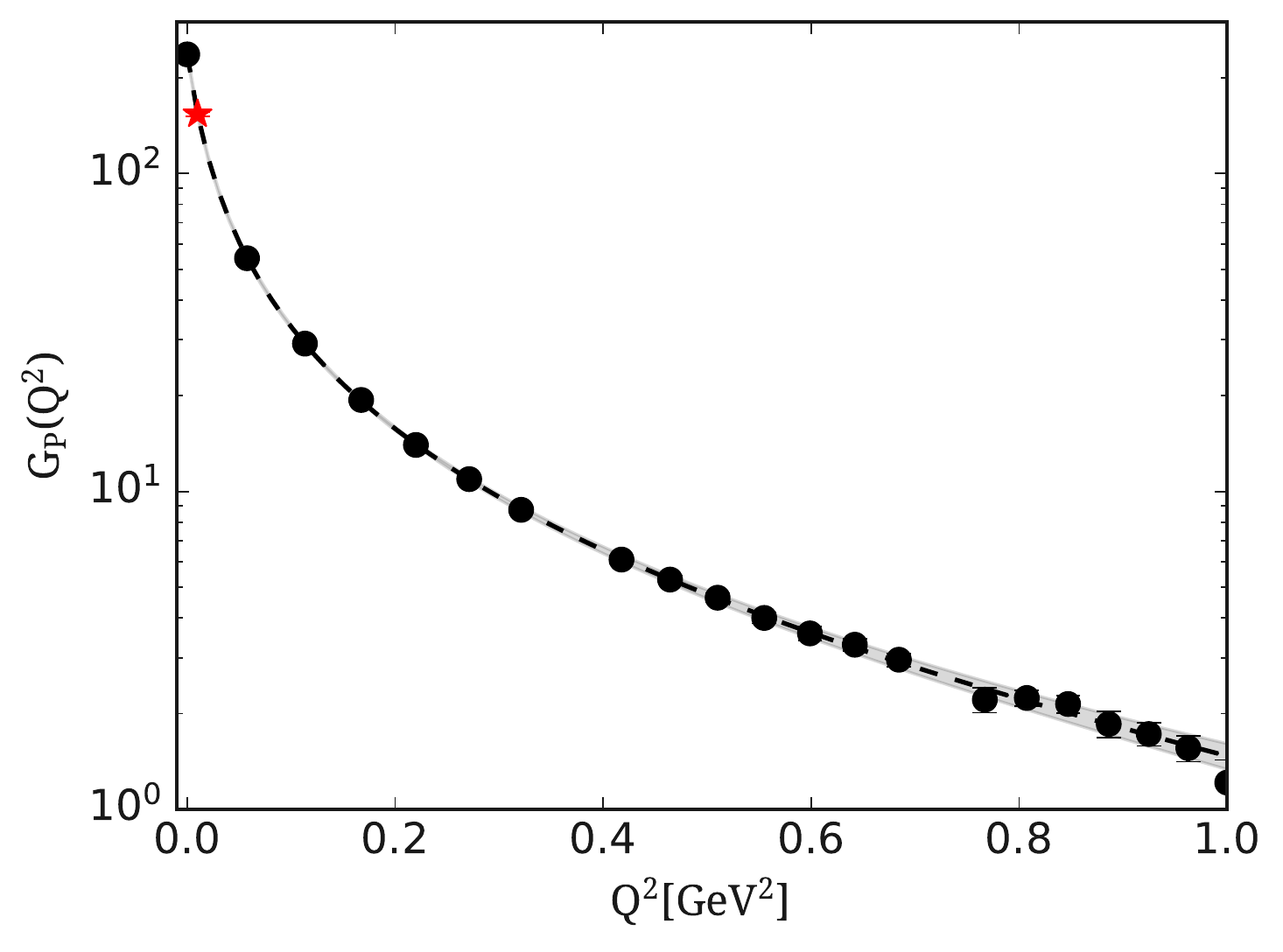}
 \caption{The induced pseudoscalar form factor, $G_P(Q^2)$, as a function of $Q^2$. The black dashed line is the result of the fit using the z-expansion. The red star is the value of the form factor at muon capture.}
 \label{fig:GP}
 \end{figure}

 In Fig.~\ref{fig:GP} we show our results for the induced pseudoscalar form factor extracted using $G_A(Q^2)$ and the PPD of Eq.~(\ref{Eq:PPD}).
The induced pseudoscalar coupling determined at the muon capture~\cite{Egger:2016hcg} is determined as
 \begin{equation}
     g_P^* = \frac{m_\mu}{2 m_N} G_P(Q^2=0.88\,m_\mu^2)
     \label{Eq:gP*}
 \end{equation}
with $m_\mu=105.6$~MeV the muon mass. The pion-nucleon coupling constant  given in Eq.~(\ref{Eq:gpiNN}), and the Goldberger-Treiman discrepancy given in Eq.~(\ref{Eq:GTD}) can also be extracted from the induced pseudoscalar form factor. We tabulate the extracted values in Table~\ref{tab:final_res}. The error on both $g_P^*$ and $g_{\pi NN}$ due to using a different fit Ansatz as well as the maximum value of   $Q^2$  used in the fits is negligible compared to the statistical error. The Goldberger-Treiman discrepancy is determined to high precision since it uses the  precise values of the axial form factor.

\begin{figure}[ht!]
 \includegraphics[scale=0.55]{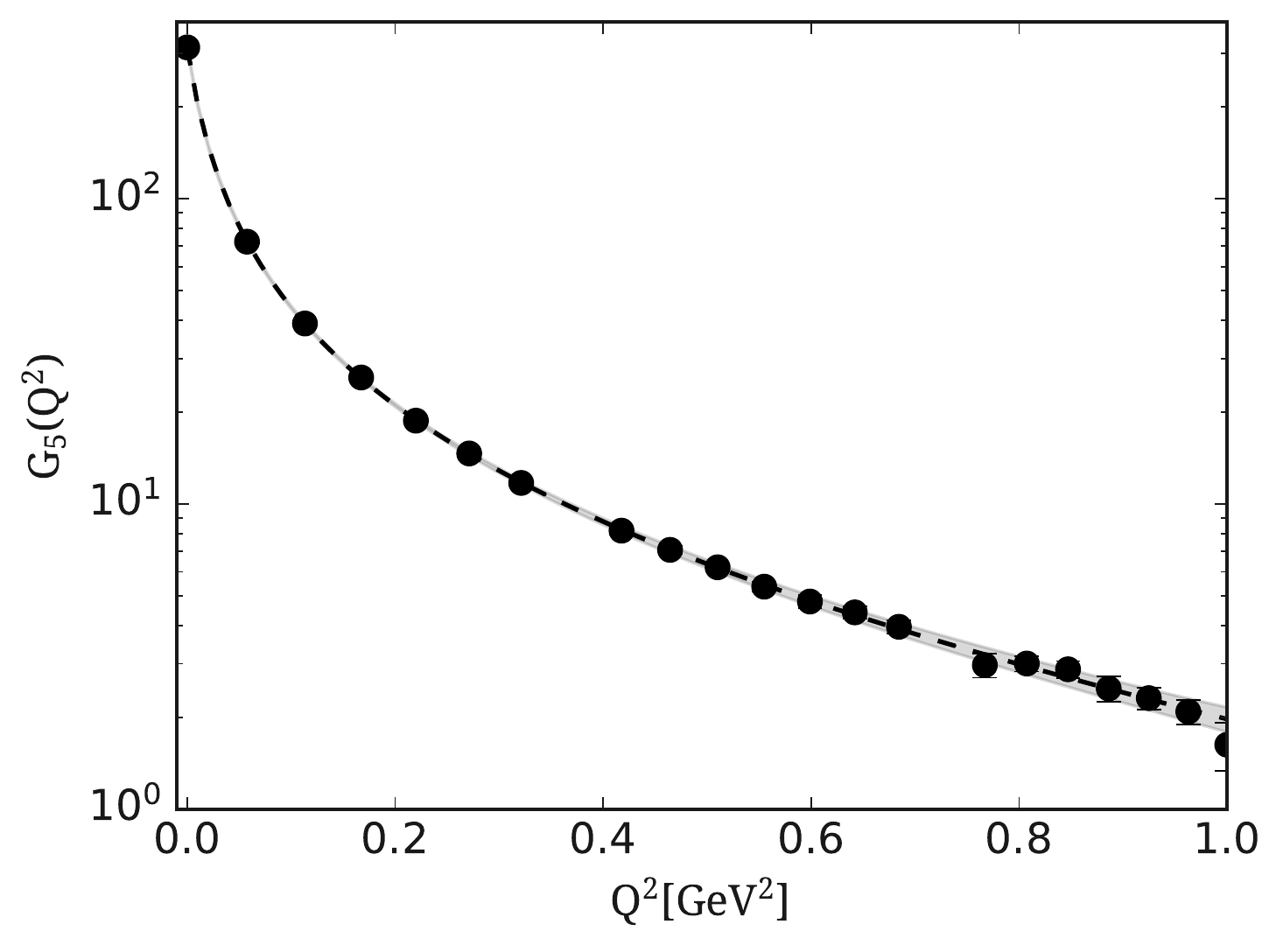}
 \caption{Results for the $Q^2$ dependence of the pseudoscalar form factor, $G_5(Q^2)$. The notation is as in Fig.~\ref{fig:GP}.}
 \label{fig:G5}
 \end{figure}

Finally,  our results for the pseudoscalar form factor $G_5(Q^2)$ extracted using Eq.~\eqref{Eq:G5eq} are shown in Fig.~\ref{fig:G5}. In principle, the pion nucleon coupling can be extracted from this form factor but since we use the PCAC and PPD relations for both $G_P(Q^2)$ and $G_5(Q^2)$,  one would obtain the  same value as the one extracted from the $G_P(Q^2)$ form factor.

We tabulate our results for the three form factors as a function of $Q^2$ in the Appendix~\ref{sec:appendix_ResFFs}.

 \begin{table}[ht!]
     \centering
     \caption{Results (from top to bottom) for the axial mass $m_A$, the square axial radius $\langle r_A^2 \rangle$, and the r.m.s $\sqrt{\langle r_A^2 \rangle}$, the induced pseudoscalar coupling determined at the muon capture~\cite{Egger:2016hcg}, the pion nucleon coupling as in Eq.~(\ref{Eq:gpiNN}), and the Goldberger-Treiman discrepancy as in Eq.~(\ref{Eq:GTD}). The first error is statistical and the second a systematic  taken as the difference in the mean values when using $w= 5$ and  $w= 20$. }
     \begin{tabular}{|c|c|}
     \hline 
          $m_A$~[GeV] & 1.169(72)(27) \\
          $\langle r_A^2 \rangle$~[fm$^2$]& 0.343(42)(16) \\
          $\sqrt{\langle r_A^2 \rangle}$~[fm] & 0.585(36)(14) \\
          $g_P^*$ & 8.69(14) \\
          $g_{\pi NN}$ & 13.48(27)(2) \\
          $\Delta_{GT}$ & 0.0276(38)(17) \\
          \hline\hline
     \end{tabular}
     \label{tab:final_res}
 \end{table}

\section{Comparison with other studies}\label{sec:CompOther}
While there are a number of lattice QCD studies  on the isovector axial and pseudoscalar form factors  using simulation with heavier than physical pion masses, we
restrict our comparison here with results obtained using   ensembles at the physical point.
We summarize below  the setup used by other groups to compute the isovector axial and pseudoscalar form factors: 
\begin{itemize}
  
\item The PNDME collaboration~\cite{Jang:2019vkm} used a hybrid action with $N_f=2+1+1$ HISQ configurations generated by the MILC collaboration with lattice spacing $a\simeq 0.0871$~fm, lattice volume $64^3\times128$ and $m_\pi=130$~MeV in the sea (referred as a09m130W) and clover improved valence quarks  with $m_\pi=138$~MeV. Three-point functions were computed from three sink-source time separations in the range of [1-1.4]~fm. They performed the two-state analysis using both the \emph{M1} and \emph{M2} fits discussed in Sec.~\ref{ssec:ExcStates}. In what follows we show their results extracted using the \emph{M2} fit since they considered them as their final values (referred in their work as $S_{A4}$ type fit). No improvement of the currents used is discussed in order to eliminate ${\cal O}(a)$ cut-off artifacts, which would imply that they have larger finite lattice spacing effects as compared to our formulation.

\item The RQCD collaboration~\cite{Bali:2019yiy}, analyzed  37 CLS ensembles,  but only two of these  were simulated using physical pion masses. The ensembles were generated using a tree-level Symanzik-improved gauge action and $N_f=2+1$ clover-improved fermions. Their axial-vector current is ${\cal O}(a)$-improved using non-perturbatively determined coefficients.  We show their results from the physical point ensemble with the finer lattice spacing of $a=0.064$~fm, volume $96^3\times 192$ and $m_\pi=130$~MeV, referred to as E250 in Ref.~\cite{Bali:2019yiy} for comparison. Four sink-source time separations are computed in the range of [0.7-1.2]~fm, which is smaller than our upper range. Final results were extracted using the  two-state \emph{M2} type fit.
    
\item The PACS collaboration~\cite{Shintani:2018ozy} used a physical point ensemble of $N_f=2+1$ with stout-smeared ${\cal O}(a)$-improved Wilson-clover fermions and Iwasaki gauge action with lattice spacing $a=0.08457(67)$~fm and volume $128^3 \times 128$. They analyzed three sink-source time  separations in the range of [1-1.36]~fm, and their final values are extracted from the plateau method. No two-state fit approaches have been attempted.  No current improvement is discussed.

\item Comparisons of our results on the form factors for the three ensembles are shown in Figs.~\ref{fig:GA_ens},~\ref{fig:GP_ens} and ~\ref{fig:G5_ens} and given in Tables~\ref{tab:res_cB211},~\ref{tab:res_cA2_48} and ~\ref{tab:res_cA2_64} of the Appendix. In this section, we restrict ourselves in comparing our results for the $N_f=2+1+1$ ensemble with the other collaborations. The derived quantities presented in Table~\ref{tab:final_res}, on the other hand,  will be done for all three ensembles.
\end{itemize}

  \begin{figure}[!ht]
     \centering
     \includegraphics[scale=0.55]{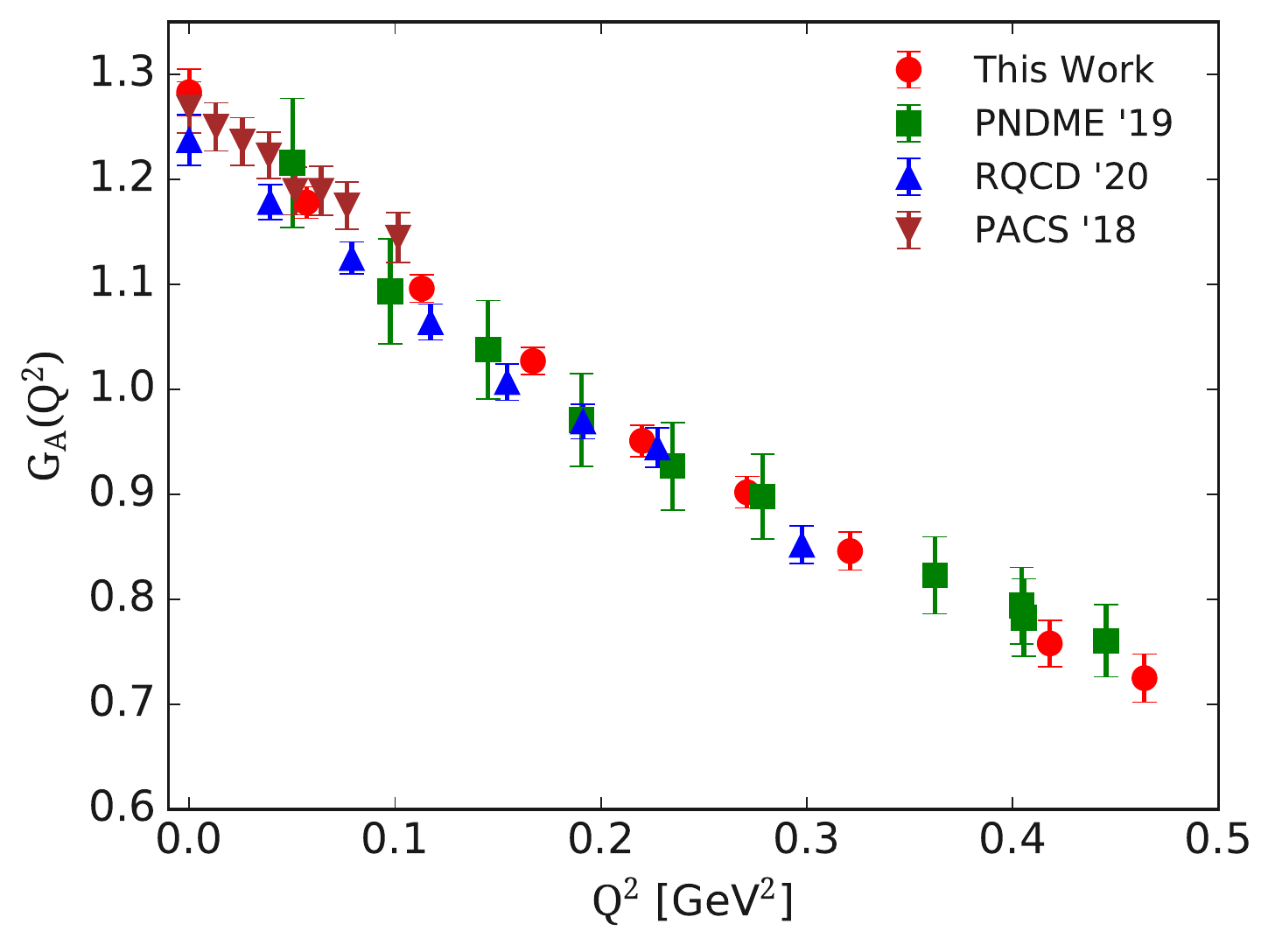}
     \caption{Lattice QCD results on the isovector axial form factor $G_A(Q^2)$ using simulations with physical pion masses. Results from this work using the cB211.072.64 ensemble are shown with red circles, from the PNDME collaboration~\cite{Jang:2019vkm} with green squares, from the RQCD collaboration~\cite{Bali:2019yiy} with blue upward-pointing triangles and from the PACS collaboration~\cite{Shintani:2018ozy} with brown down-pointing triangles.}
     \label{fig:GA_comp}
 \end{figure}
  In Fig.~\ref{fig:GA_comp}, we compare our results for $G_A(Q^2)$ using the $N_f=2+1+1$ ensemble with the aforementioned lattice QCD studies. Overall, there is a very good agreement among
  all results, which indicates that lattice artifacts are small.   PACS  results~\cite{Shintani:2018ozy} are available for very small $Q^2$ values since their lattice spatial extent  is approximately  twice as compared to the other lattices for which we show results.
 
   \begin{figure}[!ht]
     \centering
     \includegraphics[scale=0.55]{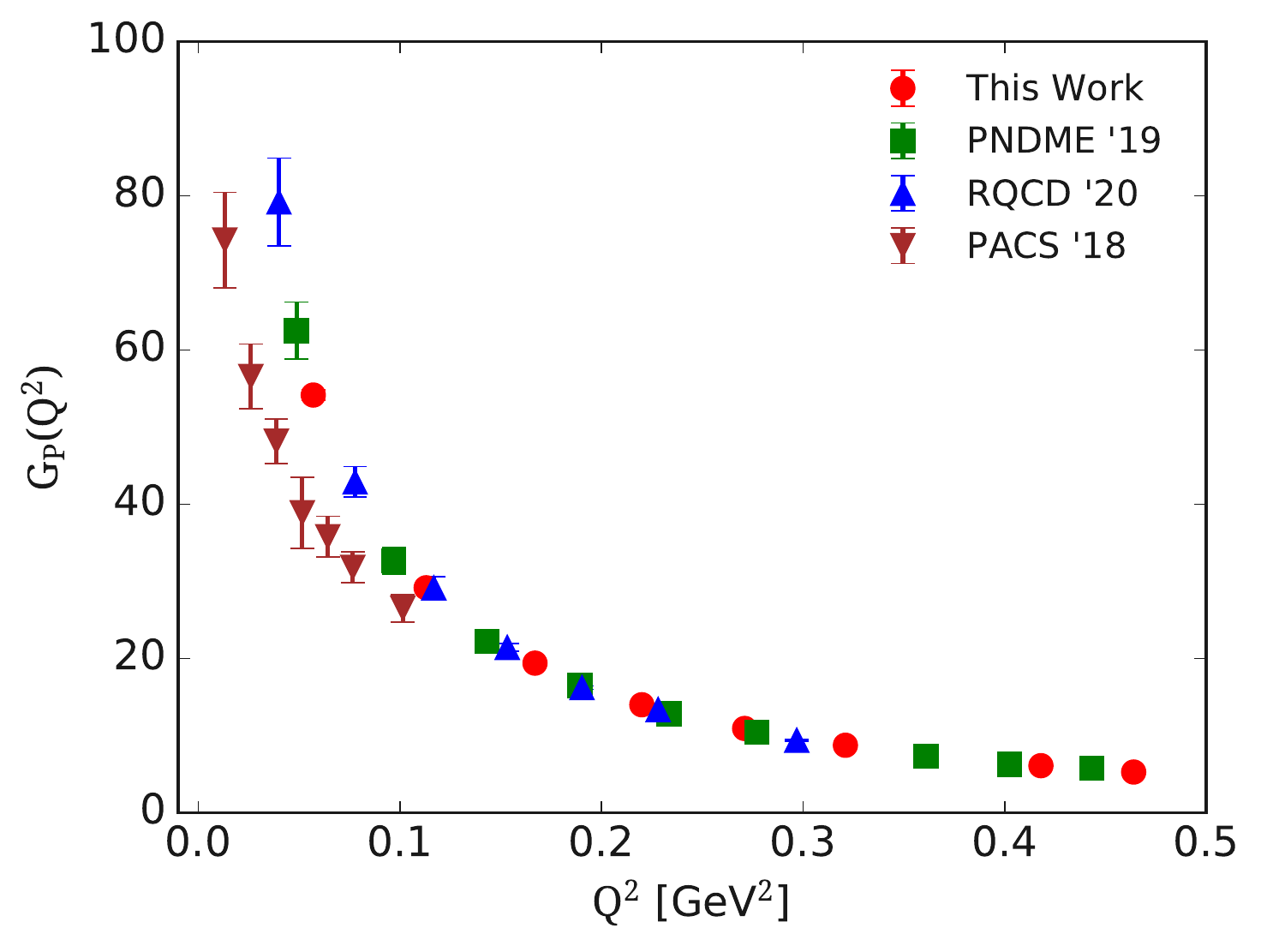}
     \caption{Comparison of lattice QCD results for the isovector induced pseudoscalar form factor $G_P(Q^2)$. The notation is the same as in Fig.~\ref{fig:GA_comp}.}
     \label{fig:GP_comp}
 \end{figure}
In Fig.~\ref{fig:GP_comp}, we compare results for the isovector induced pseudoscalar form factor. The results from PACS are extracted using the plateau method at their largest time separation. The results from the  PNDME and RQCD collaborations,  were extracted using a two-state  \emph{M2} fit. Our results are determined using $G_A(Q^2)$ and Eq.~\eqref{Eq:PPD} and are in agreement with those   from the PNDME and RQCD collaborations. While  results from PACS are lower than the others at small $Q^2$ values, their $G_P(Q^2)$ has been determined using the plateau fits at relatively small value of the the source-sink separations. Their values are higher as compared to what we find at the same time separation for the direct extraction of the $G_P(Q^2)$.  This is something that needs to be further investigated.

  \begin{figure}[!ht]
     \centering
     \includegraphics[scale=0.55]{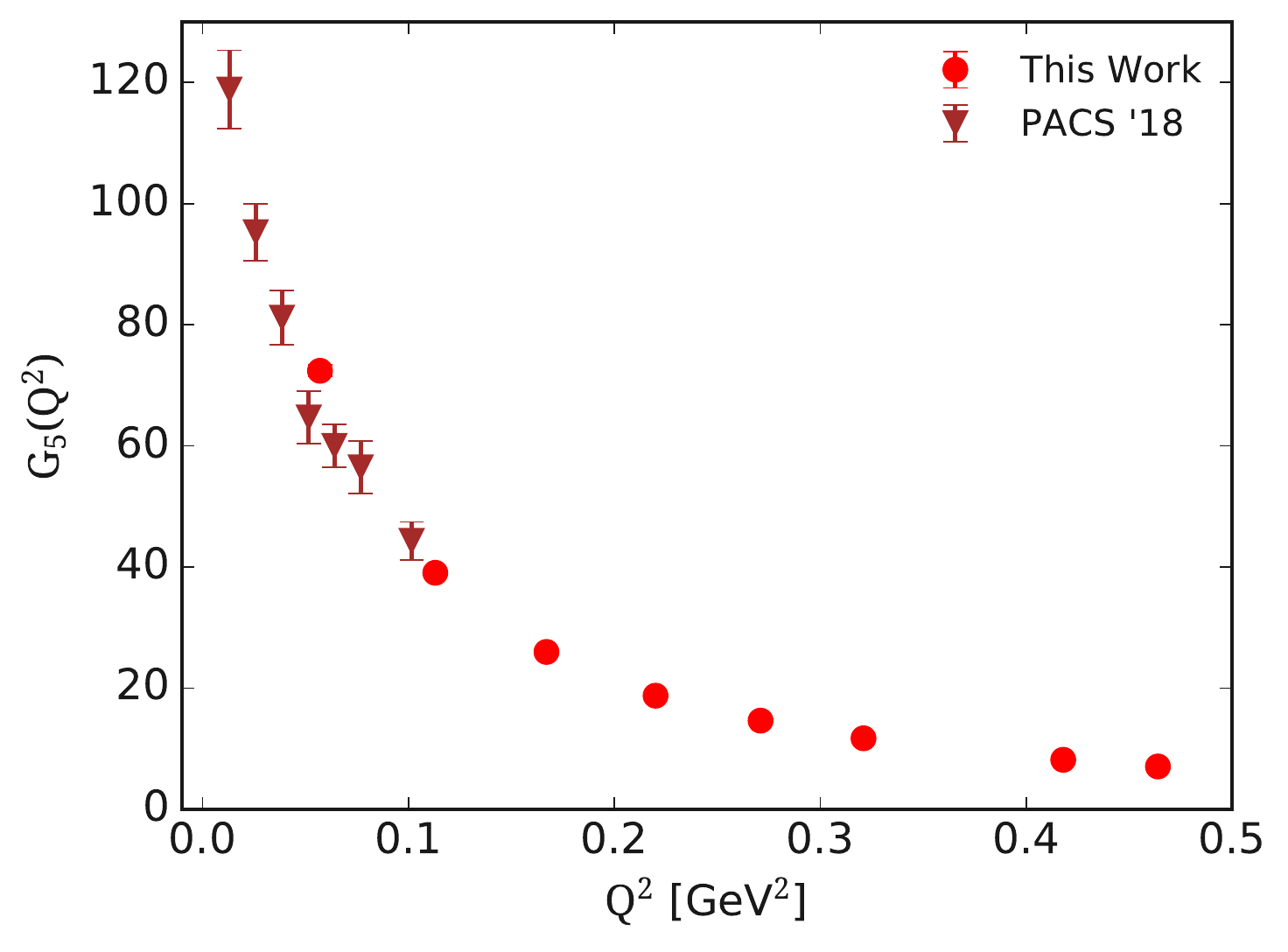}
     \caption{Lattice QCD results for the isovector  pseudoscalar form factor $G_5(Q^2)$. The notation is the same as that in Fig.~\ref{fig:GA_comp}.}
     \label{fig:G5_comp}
 \end{figure}
 In Fig.~\ref{fig:G5_comp}, we compare results for $G_5(Q^2)$. Results from PNDME are omitted since  they show only bare results and no renormalization factor  is provided. Results from RQCD are omitted because they give only results  multiplied by $m_q/m_N$ and they do not provide the renormalized value of $m_q$. Comparing our results with those  from PACS we observe agreement. This is interesting since the PACS results are extracted using the plateau method at a relatively small  source-sink time separation. However, their results, unlike what we find directly from the  three-point function of the pseudoscalar current using the  \emph{M2} fit,  show the correct pion pole behavior. Whether the reason  is because they use a large volume has to be further investigated. We plan to do such a comparison in the future when an ensemble using a large volume becomes available.

\begin{figure}[!ht]
     \centering
     \includegraphics[scale=0.46]{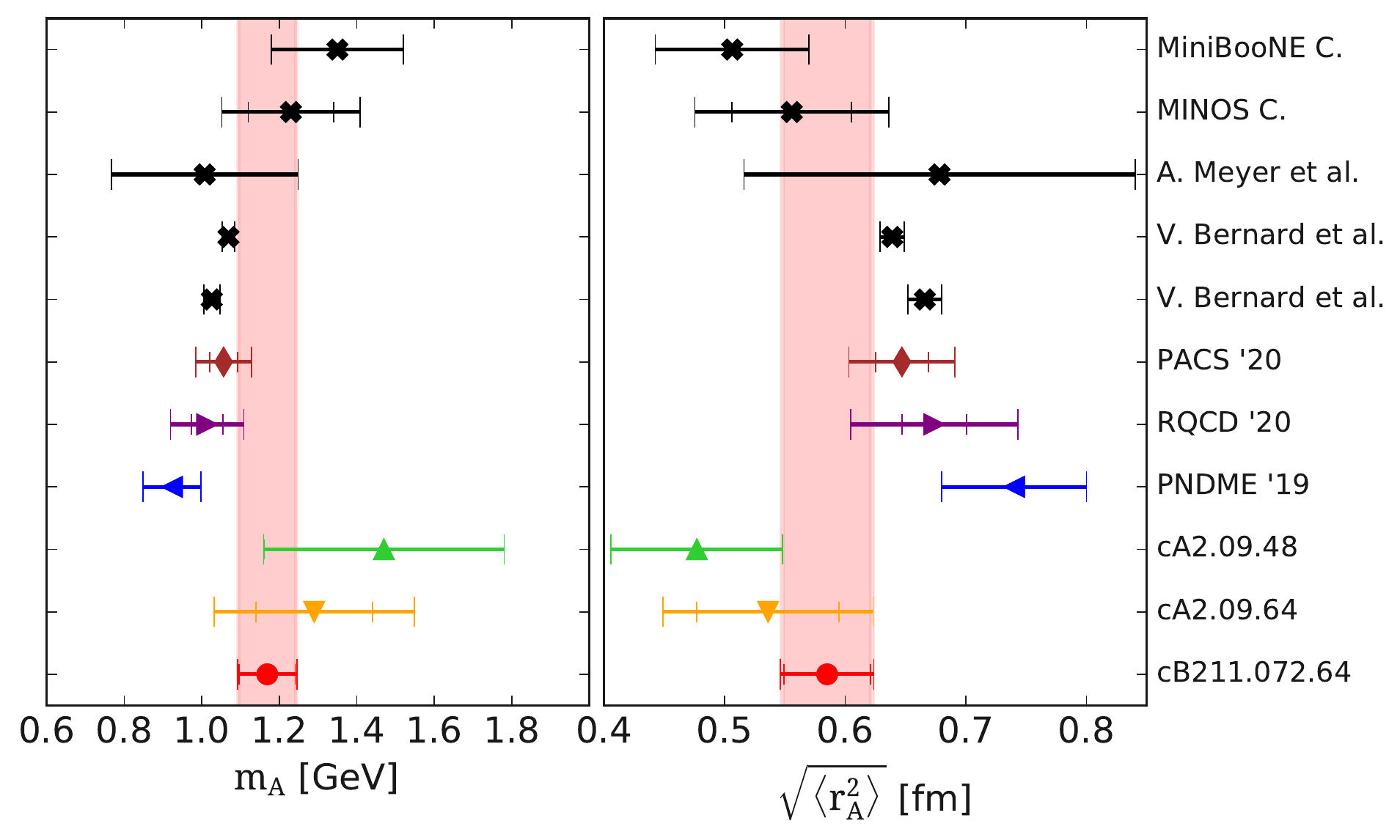}
     \caption{Results on the isovector axial mass $m_A$ (left) and the axial radius $\sqrt{\langle r_A^2 \rangle}$ (right). We show results  from our analysis of the cB211.072.64 ensemble (red circles with the associated red band), the cA2.09.64 ensemble (orange down triangle) and the cA2.09.48 ensemble (green up triangle) ensembles, from the  PNDME collaboration~\cite{Jang:2019vkm} (blue left-pointing triangle), from the RQCD collaboration~\cite{Bali:2019yiy}  (purple right-pointing triangle) when using the z-expansion, and from the PACS collaboration~\cite{Shintani:2018ozy} (brown rhombus).   Inner error bars are statistical errors while outer errors bars include systematic errors.
    The black crosses are results from phenomenology. From top to bottom we show results from the MiniBooNE experiment using charged-current muon neutrino scattering events~\cite{AguilarArevalo:2010zc}, from $\nu_\mu$-iron interactions using the MINOS Near Detector~\cite{Adamson:2014pgc}, from Ref.~\cite{Meyer:2016oeg} using  world data from neutrino-deuteron scattering and the  z-expansion for the fit, and  two very accurate results  from world averages, one is from (quasi)elastic neutrino and anti-neutrino scattering experiments~\cite{Bernard:2001rs} and the other from charged pion electroproduction experiments~\cite{Bernard:2001rs}. }
     \label{fig:mA_rA_comp}
 \end{figure}

 In Fig.~\ref{fig:mA_rA_comp}, we compare our results for the isovector $m_A$ and $\sqrt{\langle r_A^2 \rangle}$ with results from other lattice QCD studies and with phenomenological analyses using experimental data. Our results from the three ensembles are in agreement with  those using the $N_f=2+1+1$ ensemble being the most precise. That value of $m_A$ agrees with the value reported by the MiniBooNE collaboration~\cite{AguilarArevalo:2010zc} as well as  the one from the  MINOS Near detector~\cite{Adamson:2014pgc} and  Ref.~\cite{Meyer:2016oeg}. 
 Comparing with other lattice QCD results we find that our values are  compatible with the ones from the PACS and RQCD collaborations.

\begin{figure}[!ht]
     \centering
     \includegraphics[scale=0.43]{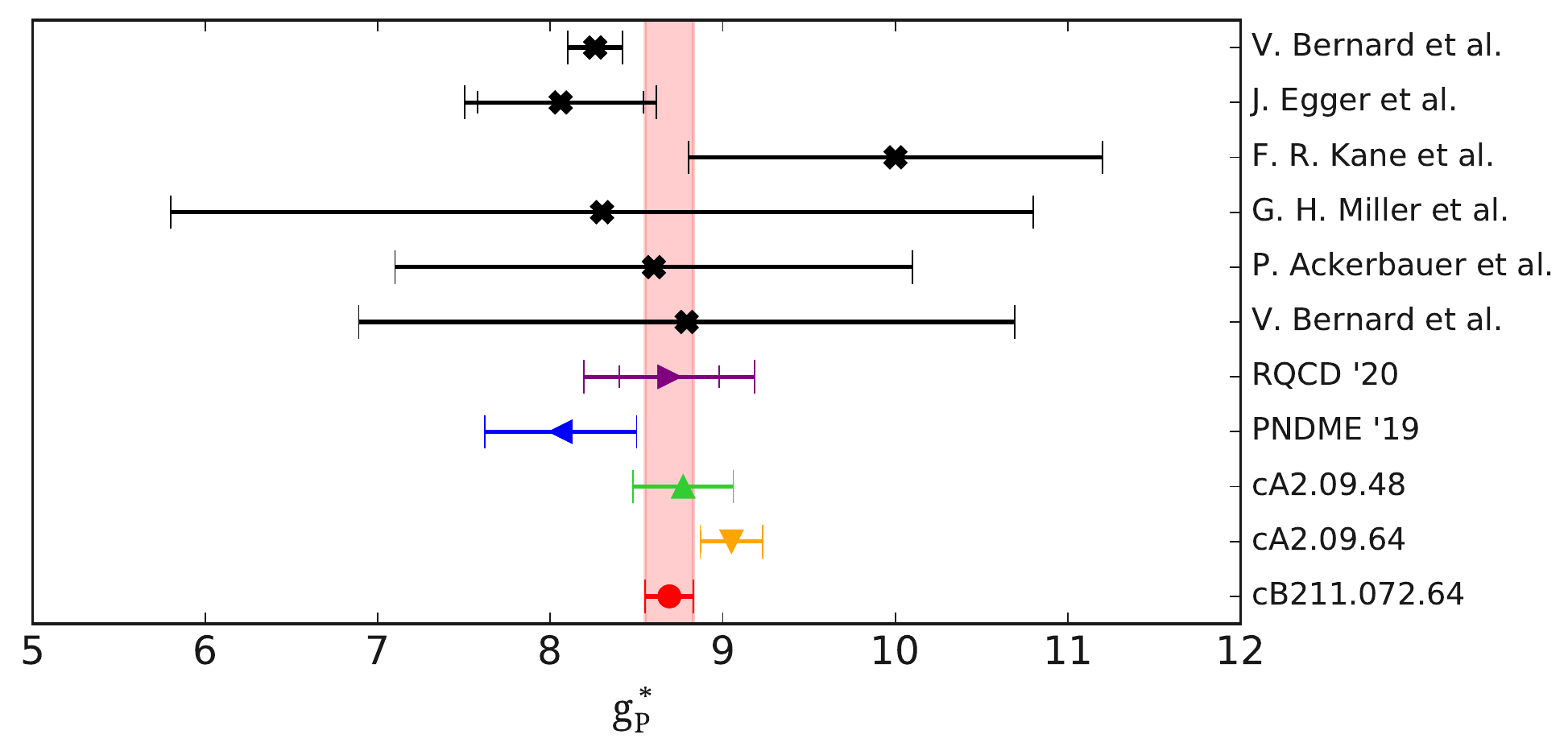}
     \caption{The results for $g_P^*$. The notation for the lattice QCD results  is the same as that in Fig.~\ref{fig:mA_rA_comp}. Black crosses are results from experimental analyses for ordinary muon capture from Refs.~\cite{Bernard:2001rs,Ackerbauer:1997rs,Miller:1972lxy,Kane:1973gg,Egger:2016hcg} and the precise result at the top of the figure is from chiral perturbation theory~\cite{Bernard:2001rs}.  }
     \label{fig:gP_comp}
 \end{figure}

\begin{figure}[!ht]
     \centering
     \hspace*{0.2cm}\includegraphics[scale=0.43]{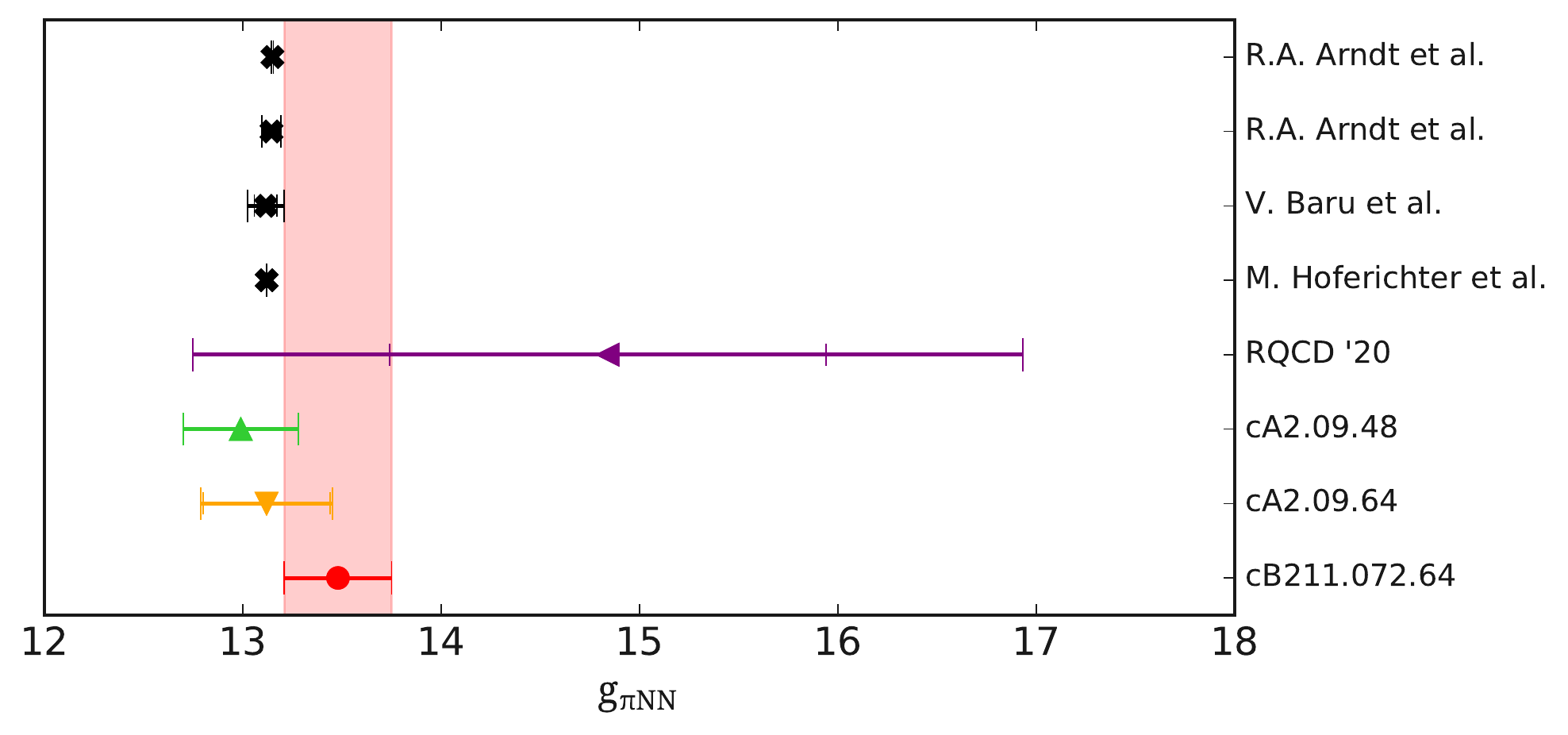}
     \includegraphics[scale=0.43]{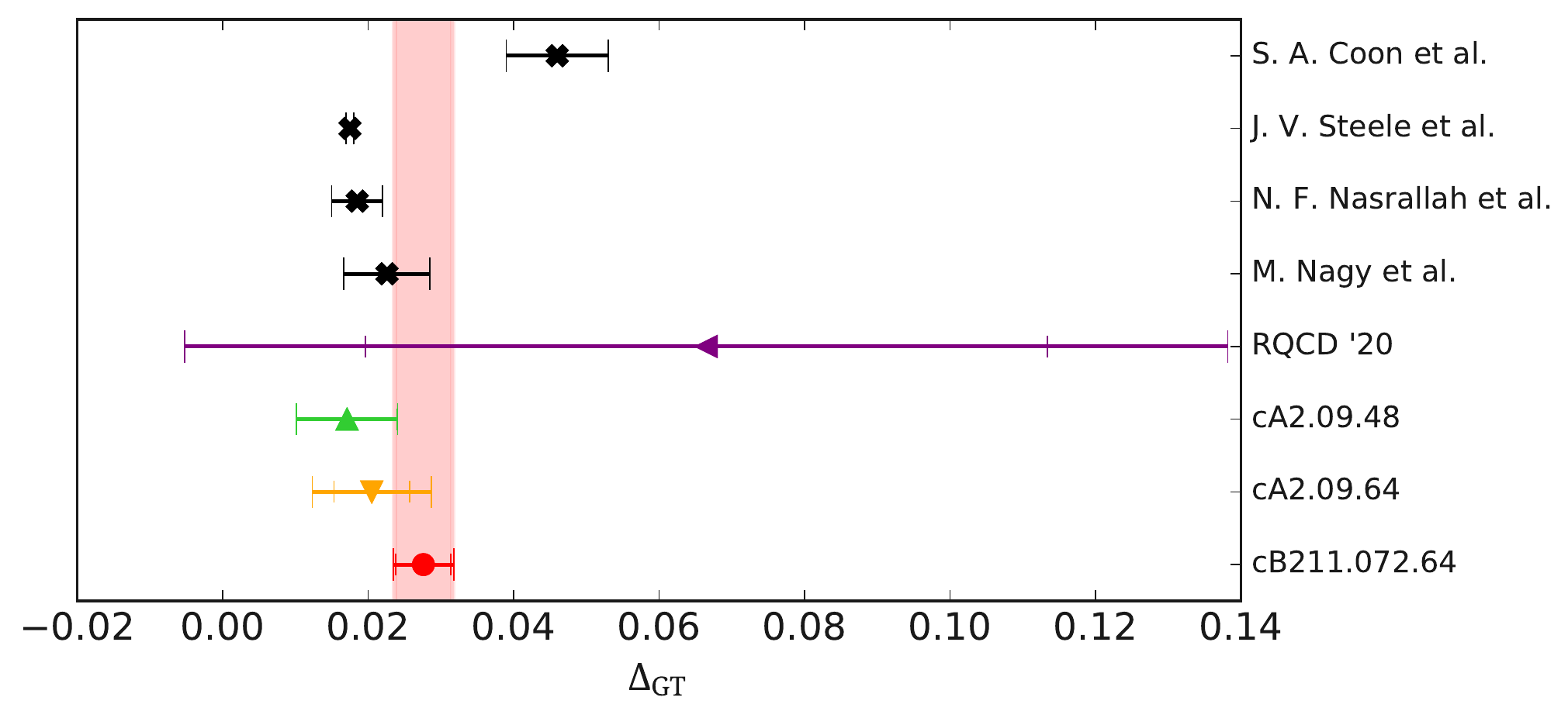}
     \caption{Results on the pion nucleon coupling constant $g_{\pi NN}$ (top) and  the Goldberger-Treiman deviation  $\Delta_{GT}$ (bottom).  The notation for the lattice QCD results  is the same as that in Fig.~\ref{fig:mA_rA_comp}.  We also show phenomenological results with the black symbols. For  $g_{\pi NN}$, these are taken from   Refs.~\cite{Hoferichter:2015hva,Baru:2011bw,Arndt:2003if,Arndt:2006bf} and are results from analyses of experimental data on pion-nucleon scattering cross-sections. For the case of $\Delta_{GT}$,  these are from Refs.~\cite{Nagy:2004tp,Coon:1990fh}  ,  from baryonic QCD sum rules~\cite{Nasrallah:1999fw}, and from heavy baryon chiral perturbation theory~\cite{Steele:1995yr}.  }
     \label{fig:gpiNN_DGT_comp}
 \end{figure}

We compare our values on muon capture coupling constant, $g_P^*$, pion-nucleon coupling  $g_{\pi NN}$ and the Goldberger-Treiman discrepancy, $\Delta_{GT}$,  with other  lattice QCD groups, experimental results and phenomenology in Figs.~\ref{fig:gP_comp} and ~\ref{fig:gpiNN_DGT_comp}.
Our results using the three ensembles are in agreement  with the values from the $N_f=2+1+1$ ensemble being the most precise.
They are also  in agreement with other lattice QCD results, although  the errors on some lattice QCD results are large. Phenomenological results are in general much more precise for $g_{\pi NN}$ and $\Delta_{GT}$.  On the other hand,  experimental results on $g_P^*$ from ordinary muon capture are compatible with lattice QCD results but carry large errors, while the result from chiral perturbation theory~\cite{Bernard:2001rs}, is as precise as our value from the cB211.072.64.

   In Fig.~\ref{fig:gpiNN_DGT_comp} we compare our results for  $g_{\pi NN}$ and $\Delta_{GT}$.  The only other lattice QCD results on $g_{\pi NN}$ and $\Delta_{GT}$ are from the RQCD collaboration~\cite{Bali:2019yiy}. As can be seen, our value for $g_{\pi NN}$ has  smaller error since it is determined from  Eq.~(\ref{Eq:GTR}) unlike the value by RQCD that does not use  $g_A$ but instead uses the $G_P(Q^2)$ form factor and Eq.(\ref{Eq:gpiNN}).  Analyses of experimental results of pion-nucleon scattering yield very  precise values. 
   We can determine $\Delta_{GT}$ precisely, extracting a value that is in agreement with the one obtained from the recent analysis of $\pi-N$ elastic scattering  data~\cite{Nagy:2004tp}. Results  using QCD sum rules~\cite{Nasrallah:1999fw},  heavy baryon chiral perturbation theory~\cite{Steele:1995yr}  and an older analysis of experimental data~\cite{Coon:1990fh}  are spread around our value.

\section{Conclusions} \label{sec:Concl}
Results on the axial and pseudoscalar form factors are presented  using an $N_f=2+1+1$ ensemble directly at the physical point avoiding chiral extrapolation that may introduce uncontrolled systematic errors in the nucleon sector. Using  $N_f=2$ ensembles with spatial extent $4.5$~fm and $6$~fm no detectable finite volume effects are observed within the range of these two volumes. Given that the analysis of the  $N_f=2+1+1$ ensemble uses more statistics and allows for a better investigation of excited states effects, we quote as our final results those obtained using the $N_f=2+1+1$ ensemble.

Our results for the axial form factor,  $G_A(Q^2)$, are the most accurate compared to those from other recent lattice QCD studies. The axial charge $G_A(0)\equiv g_A$ is in agreement with the experimental value. Fitting the  $Q^2$-dependence of $G_A(Q^2)$, we extract precisely  the axial mass $m_A$ and r.m.s radius  given  in Table~\ref{tab:final_res}. Our value for $m_A$ agrees with the value reported by the MiniBooNE collaboration~\cite{AguilarArevalo:2010zc} as well as  the one from the  MINOS Near detector experiment~\cite{Adamson:2014pgc} and  Ref.~\cite{Meyer:2016oeg}.

The analysis of the lattice data that yield the induced pseudoscalar $G_P(Q^2)$ and pseudoscalar $G_5(Q^2)$ form factors is performed  using two approaches. In  the first approach we take the excited energies extracted from the nucleon two-point function to coincide with those entering the three-point correlators, and in the second, we  allow them to be different. While we obtain different excited energies from the three-point correlators, the difference is not as large as observed in two recent studies~\cite{Jang:2019vkm, Bali:2019yiy}.   The reason is that the first excited state extracted from our two-point function is already lower as compared to what these other two studies find. The consequence is that the effect on the low $Q^2$-dependence is smaller and, thus, the $G_P(Q^2)$ and $G_5(Q^2)$ do not fulfill the PCAC and the pion-pole relations. It is interesting to note that the analysis by the PACS collaboration that uses a significantly larger volume but extracts $G_P(Q^2)$ assuming ground state dominance (via the plateau approach), finds almost agreement with pion pole dominance.    Therefore, in our view, further investigation is needed to settle the pion dominance behavior of both $G_P(Q^2)$ and $G_5(Q^2)$. In the future, we plan to perform an analysis on a larger twisted mass ensemble of spatial extent $\sim 7.7$~fm, which is currently under production by ETMC. 

Using the axial form factor $G_A(Q^2)$ and  PCAC and pion-pole dominance, we extract the values of  the pion nucleon coupling constant $g_{\pi NN}$, the Goldberger-Treiman deviation from chiral symmetry $\Delta_{GT}$ and the
muon capture coupling constant $g_p^*$, all of which are in agreement with other recent  lattice QCD studies, with our results being more accurate. These are also consistent with phenomenological extractions.
This agreement is a success of lattice QCD in being now in a good position to compute from first principles these quantities.

\begin{acknowledgements}
We would like to thank all members of ETMC for a very constructive and enjoyable collaboration and in particular V. Lubicz and  R. Frezzotti for their comments. We are also indebted to O. Baer for valuable input and discussions.
M.C. acknowledges financial support by the U.S. Department of Energy, Office of Nuclear Physics, within
the framework of the TMD Topical Collaboration, as well as, by the DOE Early Career Award under Grant No.\ DE-SC0020405. 
K.H. is financially supported by the Cyprus Research Promotion foundation under contract number POST-DOC/0718/0100.
This project has received funding from the Horizon 2020 research and innovation program
of the European Commission under the Marie Sk\l{}odowska-Curie grant agreement No 642069(HPC-LEAP)  and  under grant agreement No 765048 (STIMULATE) as well as  by the DFG as a project under the Sino-German CRC110.
S.B. and J. F. are supported by the H2020 project  PRACE 6-IP (grant agreement No 82376)  and the  COMPLEMENTARY/0916/0015 project funded by the Cyprus Research Promotion Foundation.
The authors gratefully acknowledge the Gauss Centre for Supercomputing e.V. (www.gauss-centre.eu)
for funding the project pr74yo by providing computing time on the GCS Supercomputer SuperMUC
at Leibniz Supercomputing Centre (www.lrz.de).
Results were obtained using Piz Daint at Centro Svizzero di Calcolo Scientifico (CSCS),
via the project with id s702.
We thank the staff of CSCS for access to the computational resources and for their constant support.
This work also used computational resources from Extreme Science and Engineering Discovery Environment (XSEDE),
which is supported by National Science Foundation grant number TG-PHY170022.
We acknowledge Temple University for providing computational resources, supported in part
by the National Science Foundation (Grant Nr. 1625061) and by the US
Army Research Laboratory (contract Nr. W911NF-16-2-0189).
This work used computational resources from the John von Neumann-Institute for Computing on the Jureca system~\cite{jureca} at the research center
in J\"ulich, under the project with id ECY00 and HCH02.
\end{acknowledgements}

\clearpage
\bibliography{refs}

\clearpage
\appendix
\widetext

\section{Results for the axial, induced pseudoscalar and pseudoscalar form factors}
\label{sec:appendix_ResFFs}

In Tables~\ref{tab:res_cB211},~\ref{tab:res_cA2_48} and ~\ref{tab:res_cA2_64} we give our results on the axial form factors $G_A(Q^2)$, $G_P(Q^2)$ and the pseudoscalar form factor $G_5(Q^2)$ as a function of the $Q^2$ values for the cB211.072.64, cA2.09.48 and cA2.09.64 ensembles, respectively.  
\begin{table}[ht!]
    \centering
    \bgroup
    \def\arraystretch{1.5}
    \begin{tabular}{c|c|c|c}
    \hline
    $Q^2$ [GeV$^2$] & $G_A(Q^2)$ & $G_P(Q^2)$ & $G_5(Q^2)$ \\
    \hline 
0.000 & 1.283(22) & 237.0(4.0) & 313.4(5.4)  \\
0.057 & 1.178(15) & 54.17(69)  & 72.40(92)   \\
0.113 & 1.096(13) & 29.18(33)  & 39.06(45)   \\
0.167 & 1.027(13) & 19.40(24)  & 25.99(32)   \\
0.220 & 0.951(15) & 14.01(22)  & 18.77(29)   \\
0.271 & 0.902(15) & 10.93(18)  & 14.65(24)   \\
0.321 & 0.846(18) & 8.75(18)   & 11.73(25)   \\
0.418 & 0.758(22) & 6.11(18)   & 8.19(24)    \\
0.464 & 0.725(23) & 5.28(16)   & 7.08(22)    \\
0.510 & 0.696(23) & 4.63(15)   & 6.21(20)    \\
0.555 & 0.653(25) & 4.00(16)   & 5.37(21)    \\
0.599 & 0.628(32) & 3.58(18)   & 4.80(24)    \\
0.642 & 0.619(28) & 3.30(15)   & 4.42(20)    \\
0.684 & 0.591(28) & 2.96(14)   & 3.96(19)    \\
0.767 & 0.494(45) & 2.21(20)   & 2.97(27)    \\
0.807 & 0.526(31) & 2.24(13)   & 3.00(18)    \\
0.847 & 0.528(34) & 2.15(14)   & 2.88(19)    \\
0.886 & 0.477(45) & 1.86(17)   & 2.49(23)    \\
0.925 & 0.463(38) & 1.73(14)   & 2.31(19)    \\
0.963 & 0.435(41) & 1.56(15)   & 2.09(20)    \\
1.000 & 0.352(63) & 1.21(22)   & 1.63(29)    
    \end{tabular}
    \egroup
    \caption{Results for the axial (second column), induced pseudoscalar (third column) and pseudoscalar (forth column) form factors as a function of $Q^2$ for the $N_f=2+1+1$  cB221.072.64 ensemble.}
    \label{tab:res_cB211}
\end{table}

\begin{table}[ht!]
    \centering
    \bgroup
    \def\arraystretch{1.5}
    \begin{tabular}{c|c|c|c}
    \hline
    $Q^2$ [GeV$^2$] & $G_A(Q^2)$ & $G_P(Q^2)$ & $G_5(Q^2)$ \\
    \hline 
    0.000 &1.258(28) &259.1(5.7) &310.0(6.8) \\
    0.074 &1.109(17) &42.36(64) &50.66(77)\\
    0.146 &1.023(15) &21.92(33) &26.22(40)\\
    0.214 &0.961(18) &14.48(27) &17.32(32)\\
    0.281 &0.893(19) &10.45(23) &12.50(27)\\
    0.345 &0.833(18) &8.02(17) &9.59(21)\\
    0.407 &0.783(19) &6.43(16) &7.69(19)\\
    0.527 &0.675(27) &4.32(17) &5.17(20)\\
    0.584 &0.669(26) &3.88(15) &4.64(18)\\
    0.640 &0.639(30) &3.38(16) &4.05(19)\\
    0.695 &0.610(31) &2.98(15) &3.57(18)\\
    0.749 &0.590(48) &2.68(22) &3.21(26)\\
    0.801 &0.530(42) &2.26(18) &2.70(21)\\
    0.853 &0.523(46) &2.09(18) &2.51(22)
    \end{tabular}
    \egroup
    \caption{Results using the cA.09.48 ensemble using the same  notation as in Table~\ref{tab:res_cB211}.}
    \label{tab:res_cA2_48}
\end{table}

\begin{table}[ht!]
    \centering
    \bgroup
    \def\arraystretch{1.5}
    \begin{tabular}{c|c|c|c}
    \hline
    $Q^2$ [GeV$^2$] & $G_A(Q^2)$ & $G_P(Q^2)$ & $G_5(Q^2)$ \\
    \hline 
    0.000 &1.240(26) &255.6(5.4) &305.7(6.5) \\
    0.042 &1.185(21) &69.9(1.3) &83.6(1.5) \\
    0.083 &1.122(19) &39.01(67) &46.65(81)\\
    0.123 &1.062(19) &26.35(46) &31.51(56)\\
    0.163 &1.019(18) &19.74(35) &23.61(42)\\
    0.201 &0.963(17) &15.37(28) &18.38(33)\\
    0.239 &0.930(18) &12.64(25) &15.12(30)\\
    0.313 &0.863(21) &9.12(22) &10.91(26)\\
    0.348 &0.830(21) &7.91(20) &9.47(24)\\
    0.384 &0.780(21) &6.79(18) &8.12(22)\\
    0.418 &0.761(22) &6.09(18) &7.29(21)\\
    0.452 &0.787(37) &5.84(27) &6.99(33)\\
    0.486 &0.732(26) &5.07(18) &6.06(22)\\
    0.519 &0.715(27) &4.65(18) &5.56(21)\\
    0.583 &0.603(48) &3.50(28) &4.18(33)\\
    0.615 &0.646(33) &3.56(18) &4.26(22)\\
    0.646 &0.607(36) &3.19(19) &3.81(23)\\
    0.677 &0.624(49) &3.13(25) &3.75(29)\\
    0.707 &0.590(40) &2.84(19) &3.40(23)\\
    0.737 &0.543(43) &2.51(20) &3.00(24)\\
    0.825 &0.452(76) &1.87(32) &2.24(38)
    \end{tabular}
    \egroup
    \caption{Results using the cA2.09.64 ensemble using the same notation as in Table~\ref{tab:res_cB211}.}
    \label{tab:res_cA2_64}
\end{table}

\section{Expressions for the axial and pseudoscalar form factors}
\label{sec:appendix_FFs}
The following expressions are provided in Euclidean space. 
In the case of the axial matrix element we have
\begin{eqnarray}
\Pi_i^A(\Gamma_k,\vec{q}) &=& \frac{i C}{4 m_N} [ \frac{q_k q_i}{2 m_N} G_P - (E+m_N) G_A \delta_{i,k}]
\label{Eq:Aik_decomp}
\end{eqnarray}
for the case that the current is in the $i$-direction. For the temporal direction the corresponding expression is
\begin{eqnarray}
\Pi_0^A(\Gamma_k,\vec{q}) &=& C \frac{-q_k}{2m_N} [G_A  + G_P \frac{(m_N-E)}{2 m_N}]. 
\label{Eq:A0k_decomp}
\end{eqnarray}
 The matrix  of kinematical coefficients then becomes
\begin{equation}
{\cal G}_{\mu}(\Gamma_k;\vec{q}) = 
\begin{pmatrix}
    \frac{- q_k C}{2 m_N} & \frac{- q_k C (m_N-E)}{4 m_N^2} \\
  \frac{-i C (E+m_N) \delta_{i,k}}{4 m_N}  &   \frac{i C q_k q_i}{8 m_N^2}
\end{pmatrix}
\label{Eq:coeffs}
\end{equation}
where the first row is for $\mu=0$, the second row for $\mu=i$, the first column the kinematic coefficients for $G_A$ and the second column those for $G_P$.

For the case of the pseudoscalar matrix element we have
\begin{eqnarray}
\Pi^5(\Gamma_k,\vec{q}) = \frac{-i C q_k}{2 m} G_5.
\label{Eq:g5_decomp}
\end{eqnarray}
In the above expressions,  $E$ is the energy and $m$ the mass of the nucleon. The kinematic factor $C$ is given by
\begin{equation}
  C = \sqrt{\frac{2 m_N^2}{E (E +m)}}.
  \label{Eq:C_coeff}
\end{equation}

\end{document}